\documentclass[12pt]{article}

\usepackage{graphicx}
\usepackage{amssymb}

\catcode`\@=11
\def\citenum#1{{\def\@cite##1##2{##1}\cite{#1}}}
\catcode`\@=12
\def\9{\phantom 0}     
\renewcommand\linebreak{\unskip\break} 

\newlength{\captsize} \let\captsize=\small 
\def\tautau{\tau^+\tau^-}

\def\be{\begin{equation}}
\def\ee{\end{equation}}
\def\ba{\begin{array}}
\def\ea{\end{array}}
\def\bea{\begin{eqnarray}}
\def\eea{\end{eqnarray}}
\def\t{\tau}

\begin{document}

\begin{titlepage}
\def\thepage {}        
\title{\bf The Meaning of Higgs:\\
$\tau^+ \tau^-$ and $\gamma \gamma$ \\
at the Tevatron and the LHC\footnote{MSUHEP-050608}}

\author{Alexander Belyaev, Alexander Blum, \\
R. Sekhar Chivukula, and Elizabeth  H. Simmons \\ \\
{\sl Department of Physics and Astronomy}\\{\sl Michigan State University}\\ 
{\sl East Lansing, MI 48824, USA}\\ \\
{\sl e-mail: belyaev@pa.msu.edu, blumalex@msu.edu,}\\ {\sl sekhar@msu.edu, esimmons@msu.edu}}

\date{June 8, 2005}

\maketitle


\begin{abstract}
{In this paper we discuss how to extract  information about physics beyond the Standard Model (SM) from searches
for a light SM
Higgs at Tevatron Run II and CERN LHC.   We demonstrate that new (pseudo)scalar states
predicted in both supersymmetric and dynamical models can have enhanced visibility in
standard Higgs search channels, making them potentially discoverable at Tevatron Run
II and CERN LHC.
We discuss the likely sizes of the enhancements in the various search channels for
each model and identify the model features having the largest influence on the degree
of enhancement.   We compare the
key signals for the non-standard scalars across models and also with expectations in
the SM, to show how one could start to identify which state has actually been found.
In particular, we suggest the likely mass reach of the Higgs search in $p\bar{p}/pp \to {\cal
H} \to \tau^+\tau^-$ for each kind of non-standard scalar state and we demonstrate that 
$p\bar{p}/pp \to {\cal H} \to \gamma\gamma$ may cleanly distinguish the scalars of supersymmetric models from those of dynamical models.}

\end{abstract}

\end{titlepage}

\setcounter{footnote}{0}
\setcounter{page}{1}
\setcounter{section}{0}
\setcounter{subsection}{0}
\setcounter{subsubsection}{0}

\newpage


\section{Introduction}

The origin of electroweak symmetry breaking remains unknown.  While the
Standard Model (SM) of particle physics is consistent with existing data,
theoretical considerations suggest that this theory is only a low-energy effective
theory and must be supplanted by a more complete description of the underlying
physics at energies above those reached so far by experiment.

The CDF and D\O\ experiments at the Fermilab Tevatron are currently searching for
the Higgs boson of the Standard Model.  The production cross-section and decay
branching fractions for this state have been predicted in great detail for the
mass range accessible to Tevatron Run II.  Search strategies have been
carefully planned and optimized.  

However, if the Tevatron does find evidence for a new scalar state, it may not
necessarily be the Standard Higgs. Many alternative models of electroweak
symmetry breaking have spectra that include new scalar or pseudoscalar states
whose masses could easily lie in the range to which Run II is sensitive.  The
new scalars tend to have cross-sections and branching fractions that differ
from those of the SM Higgs.  The potential exists for one of these scalars to
be more visible in a standard search than the SM Higgs would be.

In this paper we discuss how to extract  information about non-Standard theories of electroweak symmetry breaking from searches
for a light SM
Higgs at Tevatron Run II 
and CERN LHC.

The idea of using standard Higgs searches to place limits on new scalar states
associated with electroweak symmetry breaking beyond the Standard Model has
been applied to LEP results (see e.g. 
Refs.~\cite{Rupak:1995kg,Lubicz:1995xi,Lynch:2000hi,higgs-lep1,higgs-lep2,higgs-lep3,higgs-lep4,higgs-lep5}).   
The Tevatron and LHC can 
potentially access significantly
heavier scalars than those to which LEP was sensitive, particularly in models
of dynamical symmetry breaking. Ref.~\cite{Belyaev:2002zz} studied the potential
of Tevatron Run II to augment its search for the SM Higgs boson by considering the
process $gg \to h_{SM} \to \tau^+\tau^-$.   While this channel would not suffice as
a sole discovery mode,\footnote{The authors established that discovery of  $h_{SM}$
in this channel alone (assuming a mass in the range 120 - 140 GeV) would require an
integrated luminosity of 14-32 fb$^{-1}$, which is unlikely to be achieved.} the
authors found that it could usefully be combined with other channels such as $h_{SM}
\to W^+W^-$ or associated Higgs production to enhance the overall visibility of the
Higgs.  At the same time, the authors determined what additional enhancement of
scalar production and branching rate, such as might be provided in a non-standard
model like the MSSM, would enable a scalar to become visible in the $\tau^+\tau^-$
channel alone at Tevatron Run II.   Similar work has been done for $gg\to h_{MSSM}\to \tau^+\tau^-$ at the LHC ~\cite{Cavalli:2002vs} and for $gg\to h_{SM} \to \gamma\gamma$ at the Tevatron \cite{Mrenna:2000qh} and LHC \cite{Kinnunen:2005aq}. 

Our work builds on these results, considering an additional production
mechanism (b-quark annihilation), more decay channels ($b\bar{b}$, $W^+W^-$, $ZZ$,
and $\gamma\gamma$), and a wider range of non-standard physics (supersymmetry
and dynamical electroweak symmetry breaking) from which rate enhancement may
derive.  We discuss the possible sizes of the enhancements in the various search
channels for each model and pinpoint the model features having the largest
influence on the degree of enhancement.   We suggest the mass reach of the
standard Higgs searches for each kind of non-standard scalar state.   We also
compare the key signals for the non-standard scalars across models and also
with expectations in the SM, to show how one could start to identify which
state has actually been found.

Much of our discussion will focus on the degree to which certain  standard Higgs search channels are enhanced in non-standard models due to changes
in the production rate or branching fractions of the non-standard scalar  $({\cal H})$ relative to the values for the standard Higgs boson
$(h_{SM})$.   We define the enhancement factor for the process $yy \to {\cal H} \to xx$ as the ratio of the products of the width of the (exclusive)
production mechanism  and the branching ratio of the decay: 
\begin{equation}
\kappa_{yy/xx}^{\cal H} = \frac{ \Gamma({\cal H} \to yy) \times BR({\cal H} \to x x)}
         { \Gamma(h_{SM} \to yy) \times BR(h_{SM} \to  x x)}.
\label{eq:kappa}
\end{equation}
Analytic formulas for the decay widths of the SM Higgs boson are taken
from \cite{Gunion:1989we}, \cite{Gunion:1992hs} and numerical values are calculated using the HDECAY program
\cite{Djouadi:1997yw}.

  In Section 2, we introduce supersymmetric and dynamical  models of electroweak symmetry breaking and indicate which model features will be
particularly relevant to our analysis.   In Section 3, we discuss the production and decay of the scalar states of the various models at the Tevatron and LHC  and present our results for the enhancement factors.   In Section 4, we compare the different models to one another and
to the SM.  Section 5 holds our conclusions.

\section{Models of Electroweak Symmetry Breaking}

\subsection{General Remarks}

The Standard Higgs Model of particle physics, based on the
gauge group $SU(3)_c \times SU(2)_W \times U(1)_Y$, accommodates electroweak 
symmetry breaking by including a fundamental weak doublet of scalar
(``Higgs'') bosons ${\phi = {\phi^+ \choose \phi^0}}$ with potential
function $V(\phi) = \lambda \left({\phi^\dagger \phi - \frac12
v^2}\right)^2$.  However the SM does not explain the dynamics
responsible for the generation of mass.  Furthermore, the scalar
sector suffers from two serious problems.  The scalar mass is
unnaturally sensitive to the presence of physics at any higher scale
(e.g. the Planck scale), through contributions of loops of
SM particles to the Higgs self-energy.  
This is known as the gauge hierarchy problem \cite{'tHooft:1980xb,Witten:1981nf,Dimopoulos:1981zb}.  In addition, if the
scalar must provide a good description of physics up to arbitrarily
high scale (i.e., be fundamental), the scalar's self-coupling
($\lambda$) is driven to zero at finite energy scales.  That is, the scalar field theory is free (or
``trivial'') \cite{Wilson:1971bg,Wilson:1973jj} . Then the scalar cannot fill its intended role: if
$\lambda = 0$, the electroweak symmetry is not spontaneously broken.
The scalars involved in electroweak symmetry breaking must therefore
be a party to new physics at some finite energy scale -- e.g., they
may be composite or may be part of a larger theory with a UV fixed
point.  The SM is merely a low-energy effective field theory, and the
dynamics responsible for generating mass must lie in physics outside
the SM.

In this section, we briefly introduce two classes of physics beyond the standard model that may carry
the answer to the puzzle of electroweak symmetry breaking. For a review of supersymmetric models, see
\cite{Dawson:1996cq},\cite{Murayama:2000fm};  for an introduction to dynamical electroweak symmetry
breaking, see  \cite{Hill:2002ap}.  In the meantime, we will summarize the aspects of these models
which are most germane to our analysis.  

\subsection{Supersymmetry}

One interesting possibility for addressing the hierarchy and triviality problems is to introduce supersymmetry. 
The gauge structure of the minimal supersymmetric SM (MSSM) is identical to
that of the SM, but each ordinary fermion (boson) is paired with a new
boson (fermion), called its ``superpartner,'' and two Higgs doublets 
provide mass to all the ordinary fermions.  Each loop of ordinary particles contributing to the Higgs boson's mass is now countered by a loop of superpartners.  If the masses of the
ordinary particles and superpartners are close enough, the gauge hierarchy
can be stabilized \cite{Witten:1981nf,Dimopoulos:1981zb,Anderson:1996ew,Murayama:1996ec}.  Supersymmetry relates the scalar
self-coupling to gauge couplings, so that triviality is not a concern.

In order to provide masses to both 
up-type and down-type quarks, and to ensure anomaly cancellation,
the minimal supersymmetric Standard Model (MSSM) contains two
Higgs complex-doublet superfields:  $\Phi_d=(\Phi_d^0,\Phi_d^-)$ and 
$\Phi_u=(\Phi_u^+,\Phi_u^0)$.
When electroweak symmetry breaking occurs,  the
neutral components of the Higgs doublets acquire independent 
vacuum expectation values (vevs):
\be
\label{eq-higgs-sector}
\langle {\Phi_d} \rangle={1\over\sqrt{2}} \left( \ba{c} v_d\\ 0   \ea \right), \ \ 
\langle {\Phi_u} \rangle={1\over\sqrt{2}} \left( \ba{c}  0 \\ v_u \ea \right),
\ee
 where $\sqrt{v_d^2+v_u^2}= 2 M_W/g = 246$~GeV.
Out of the original 8 degrees of freedom, 3 serve as Goldstone  bosons, absorbed
into longitudinal components of the $W^\pm$ and $Z$, making them massive.
The other 5 degrees of freedom remain in the spectrum as distinct scalar states, namely 
two neutral, CP-even states
\bea
h &=& -(\sqrt{2}  \mbox{\rm Re } \Phi_d^0-v_d)\sin\alpha+
     (\sqrt{2}  \mbox{\rm Re } \Phi_u^0-v_u)\cos\alpha~,     \\
H &=& (\sqrt{2} \mbox{\rm Re } \Phi_d^0-v_d)\cos\alpha+
  (\sqrt{2}   \mbox{\rm Re } \Phi_u^0-v_u)\sin\alpha~, 	  
\eea
one neutral, CP-odd state
\begin{equation}
A = \sqrt{2} ( \mbox{\rm Im } \Phi_d^0\sin\beta 
           +  \mbox{\rm Im } \Phi_u^0\cos\beta )~, 	
\end{equation}
and a charged pair
\begin{equation}
H^\pm = \Phi_d^\pm\sin\beta+\Phi_u^\pm\cos\beta~. 
\end{equation}
Here $\alpha$ is the mixing angle between $h$ and $H$ 
which diagonalizes the  neutral boson mass-squared matrix:
\be
{\cal M}_0^2 =    \left(
\matrix{M_A^2 \sin^2\beta + M^2_Z \cos^2\beta&
           -(M_A^2+M^2_Z)\sin\beta\cos\beta \cr
  -(M_A^2+M^2_Z)\sin\beta\cos\beta&
    M_A^2\cos^2\beta+ M^2_Z \sin^2\beta }
   \right)\ ,
\ee
 and $\beta$ is defined through the ratio  $v_u/v_d$
(sometimes denoted as $v_2/v_1$ )
\be
\tan\beta=v_u/v_d\ .
\ee
It is conventional to choose $\tan\beta$ and 
\be
M_A=\sqrt{M_{H^\pm}^2-M_W^2}~
\ee
to define the SUSY Higgs sector. From the above equations one may derive the relations 
\bea
&& M^2_{h,H} = {1\over 2}\left[ (M^2_A + M^2_Z) \mp 
\sqrt{(M^2_A+M^2_Z)^2-4 M^2_A M^2_Z \cos^2 2\beta}\right]~,
\label{eq:mha}\\
&& \cos^2(\beta-\alpha) = {M_h^2(M_Z^2-M_h^2)~,\over
M_A^2(M_H^2-M_h^2)}
\label{eq:cosbetal}
\eea
which will be useful for determining when Higgs boson interactions with fermions are enhanced. 

The Yukawa interactions of the Higgs fields with the quarks and leptons are given by:
\bea
\label{eq-lagr}
-{\cal L}_{\rm Yukawa}
=&
 h_u\left[
\bar u P_L u \Phi^0_u-\bar u P_L d \Phi^+_u
\right]
 + 
 h_d\left[
 \bar d P_L d \Phi^0_d-\bar d P_L u\Phi^-_d\right] \nonumber\\
 & +h_\ell\left[
 \bar \ell P_L \ell \Phi^0_d-\bar \ell P_L \nu \Phi^-_d\right] 
 + {\rm h.c.}\
\eea
Using Eq.~(\ref{eq-higgs-sector}) and Eq.~(\ref{eq-lagr})
we find, for example, for the 3rd generation:
\bea
h_t    =& {\sqrt{2}\,m_t   \over v_u}={\sqrt{2}\, m_t   \over v\sin\beta},\\
h_{b,\,\tau}    =& {\sqrt{2}\,m_{b,\,\tau}   \over v_d}={\sqrt{2}\, m_{b,\,\tau}   \over v\cos\beta}\ .
\label{htdef}
\eea
To display this in terms of the interactions of the mass eigenstate Higgs bosons with the fermions ($Y_{{\cal H}f\bar{f}}$) we may write\footnote{Note that the interactions of the $A$ are pseudoscalar,
{\it i.e.} it couples to $\bar{\psi} \gamma_5 \psi$.}
\bea 
Y_{ht\bar{t}} /  Y^{SM}_{ht\bar{t}}  & =  &{ \cos\alpha/\sin\beta}  \qquad\qquad Y_{hb\bar{b}} / Y^{SM}_{hb\bar{b}}  = {-\sin\alpha/\cos\beta} 	 \nonumber	\\
Y_{Ht\bar{t}} /  Y^{SM}_{ht\bar{t}} & =  &{ \sin\alpha/\sin\beta}  \qquad\qquad Y_{Hb\bar{b}} / Y^{SM}_{hb\bar{b}}  = {\  \cos\alpha/\cos\beta}	 \label{eq-yukawas} 
	\\
Y_{At\bar{t}}/   Y^{SM}_{ht\bar{t}} & =  &{ \cot\beta}   \qquad\qquad\qquad Y_{Ab\bar{b}}/ 
 Y^{SM}_{hb\bar{b}}  =  { \tan\beta}     \nonumber
\eea
relative to the Yukawa couplings of the Standard Model ($Y^{SM}_{hf\bar{f}} = m_f/v)$.  Once again, the same pattern holds for the tau lepton's  Yukawa couplings as for those of the $b$ quark.

There are several circumstances under which various Yukawa couplings are enhanced relative to 
Standard Model values.
For high $\tan\beta$ (small $\cos\beta$), eqns. (\ref{eq-yukawas}) show that the
interactions of all neutral Higgs bosons with the down-type fermions
are enhanced by a factor of $1/\cos\beta$.  In the decoupling limit, where $M_A\to\infty$, 
applying eqns. (\ref{eq:mha}) and (\ref{eq:cosbetal}) 
to eqns. (\ref{eq-yukawas}) shows that the $H$ and $A$
Yukawa couplings to down-type fermions are enhanced by a factor of $\tan\beta$
\bea
Y_{Hb\bar{b}}  /Y^{SM}_{hb\bar{b}} =  
Y_{H\t\bar{\t}}/Y^{SM}_{h\t\bar{\t}}\simeq \tan\beta,
\label{eq:bhi}
\eea
Conversely, for low $m_A\simeq m_h$, one can check that
\bea
Y_{hb\bar{b}}  /Y^{SM}_{hb\bar{b}} =  
Y_{h\t\bar{\t}}/Y^{SM}_{h\t\bar{\t}}\simeq \tan\beta
\label{eq:bhii}
\eea
that $h$ and $A$ Yukawas are enhanced instead.
For further details we refer 
to Ref.~\cite{Boos:2002ze}
where issues of mass-degenerate Higgs bosons in MSSM
at large $\tan\beta$ have been studied in great detail.

\subsection{Technicolor}

Another intriguing class of theories, dynamical electroweak symmetry
breaking (DEWSB), supposes that the scalar states involved in
electroweak symmetry breaking could be manifestly composite at scales
not much above the electroweak scale $v \sim 250$ GeV.  In these
theories, a new asymptotically free strong gauge interaction  (technicolor 
\cite{Susskind:1978ms,Weinberg:1975gm, Weinberg:1979bn}) breaks the chiral symmetries of massless fermions $f$ at
a scale $\Lambda \sim 1$ TeV.  If the fermions carry appropriate
electroweak quantum numbers (e.g. left-hand (LH) weak doublets and right-hand (RH) weak
singlets), the resulting condensate $\langle \bar f_L f_R \rangle \neq
0$ breaks the electroweak symmetry as desired.  Three of the Nambu-Goldstone Bosons
(technipions) of the chiral symmetry breaking become the
longitudinal modes of the $W$ and $Z$. The logarithmic running of the
strong gauge coupling renders the low value of the electroweak scale
natural.  The absence of fundamental
scalars obviates concerns about triviality.

Many models of DEWSB have additional light neutral pseudo Nambu-Goldstone bosons which could potentially be accessible to a standard Higgs search; these are called ``technipions" in technicolor models.  
There is not one particular DEWSB model that has been singled out 
as a benchmark, in the manner of the MSSM among supersymmetric theories.
Rather, several different classes of models have been proposed to address various 
challenges within the DEWSB paradigm of the origins of mass.  In this paper, we look at several
representative technicolor models.  We both evaluate the potential of standard Higgs searches to
discover the lightest Pseudo Nambu-Goldstone Boson (PNGB) of each of these models, and also draw some inferences about the
characteristics of technicolor models that have the greatest impact on this search potential.

Our analysis will assume, for simplicity, that the lightest PNGB state is significantly lighter than other
neutral (pseudo) scalar technipions, so as to heighten the comparison to the SM Higgs boson.  The
precise spectrum of any technicolor model generally depends on a number of parameters, particularly
those related to whatever ``extended technicolor" \cite{Dimopoulos:1979es,Eichten:1979ah} interaction transmits electroweak symmetry breaking
to the ordinary quarks and leptons.  Models in which several light neutral PNGBs were nearly degenerate
would produce even larger signals than those discussed here.  

The specific models we examine are: 1) the traditional one-family model \cite{Farhi:1980xs} with a full family of
techniquarks and technileptons, 2) a variant on the one-family model \cite{Casalbuoni:1998fs} in which the lightest
technipion contains only down-type technifermions and is significantly lighter than the other pseudo Nambu-Goldstone
bosons,  3) a multiscale walking technicolor model \cite{Lane:1991qh} designed to reduce flavor-changing neutral currents,
and  4) a low-scale technciolor model (the Technicolor Straw Man model) \cite{Lane:1999uh} with many weak doublets of technifermions, in which the second-lightest technipion $P'$ is the state relevant for our study (the lightest, being composed of technileptons, lacks the anomalous coupling to gluons required for $gg \to P$ production).  For simplicity the lightest relevant neutral technipion of each model will be generically denoted $P$; where a specific model is meant, a superscript will be used.

One of the key differences among these models is the value of the technipion decay constant $F_P$, which is related to the
number $N_D$ of weak doublets of technifermions that contribute to electroweak symmetry breaking.  In a theory like model 2, in
which  only a single technifermion condensate breaks the electroweak symmetry, the value of $F_P$ is simply the weak
scale: $F_P^{(2)} = v = 246$ GeV. In models where more than one technifermion condensate breaks the EW
symmetry, one finds $v^2 = f_P^2 + f_2^2 + f_3^2 + ...$ For example, in the one-family model (model 1), all four technidoublets
corresponding to a technifermion ``generation" condense, so that the decay constant is fixed to be $F_P^{(1)} =
\frac{v}{2}.$ In the lowscale model (model 4), the number of condensing technidoublets is much higher, of order 10; setting $N_D
= 10$ yields $F_P^{(4)} = \frac{v}{ \sqrt{10}}$.  In the  multiscale model (model 3), the scales at which various
technicondensates form are assumed to be significantly different, so that the lowest scale is simply bounded from above.
In keeping with \cite{Lane:1991qh}  and to ensure that the technipion mass will be in the range to which the standard
Tevatron Higgs searches are sensitive, we set $F_P^{(3)}$ = $ \frac{v}{4}$.

In section 3, we study the enhancement factors for several production and decay modes of the lightest PNGBs of each
technicolor model. Then in section 4, we compare the signatures of these PNGBs to those of a SM Higgs and the Higgs bosons
of the MSSM in order to determine how the standard search modes (or additional channels) can help tell these states
apart.

\section{Results For Each Model}

In this section, we examine the single production of SUSY Higgses and technicolor PNGBs via the two dominant
methods at the  Tevatron and LHC:  gluon fusion and $b\bar{b}$ annihilation.    We determine the degree to which these
production channels are enhanced relative to production of a SM Higgs, and find which channel dominates for
each scalar state.  We likewise study the major decay modes:  $b\bar{b}$, $\tau^+\tau^-$, $\gamma\gamma$, and $W^+W^-$
in order to determine the branching fractions relative to those of an SM Higgs.  We then combine this information to obtain the overall enhancement factor in each channel and the estimated cross-section at each collider.

\subsection{Supersymmetry}

\subsubsection{Factors affecting signal strength}

\begin{figure}[tb]
\includegraphics[width=7cm]{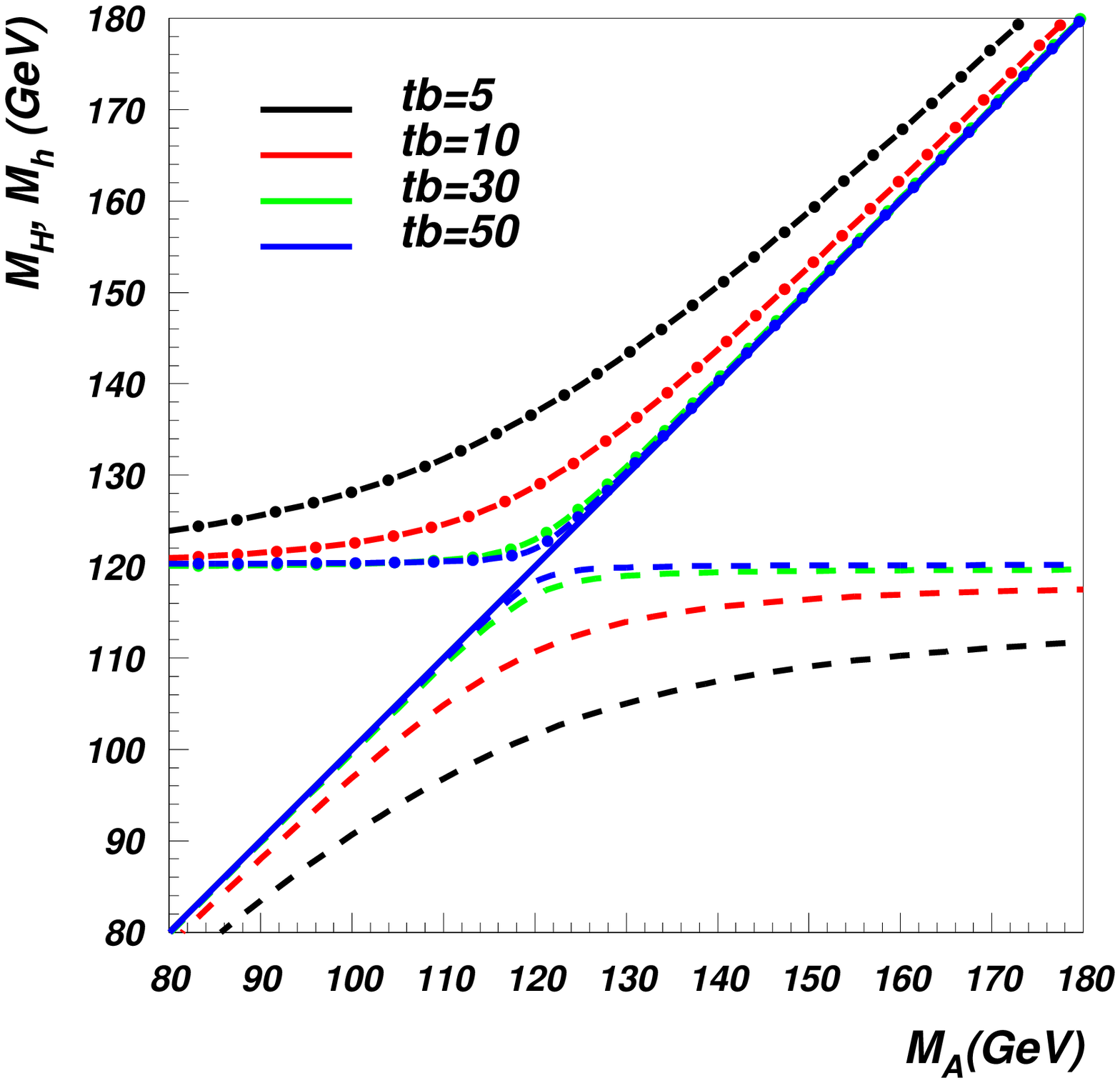}
\includegraphics[width=7cm]{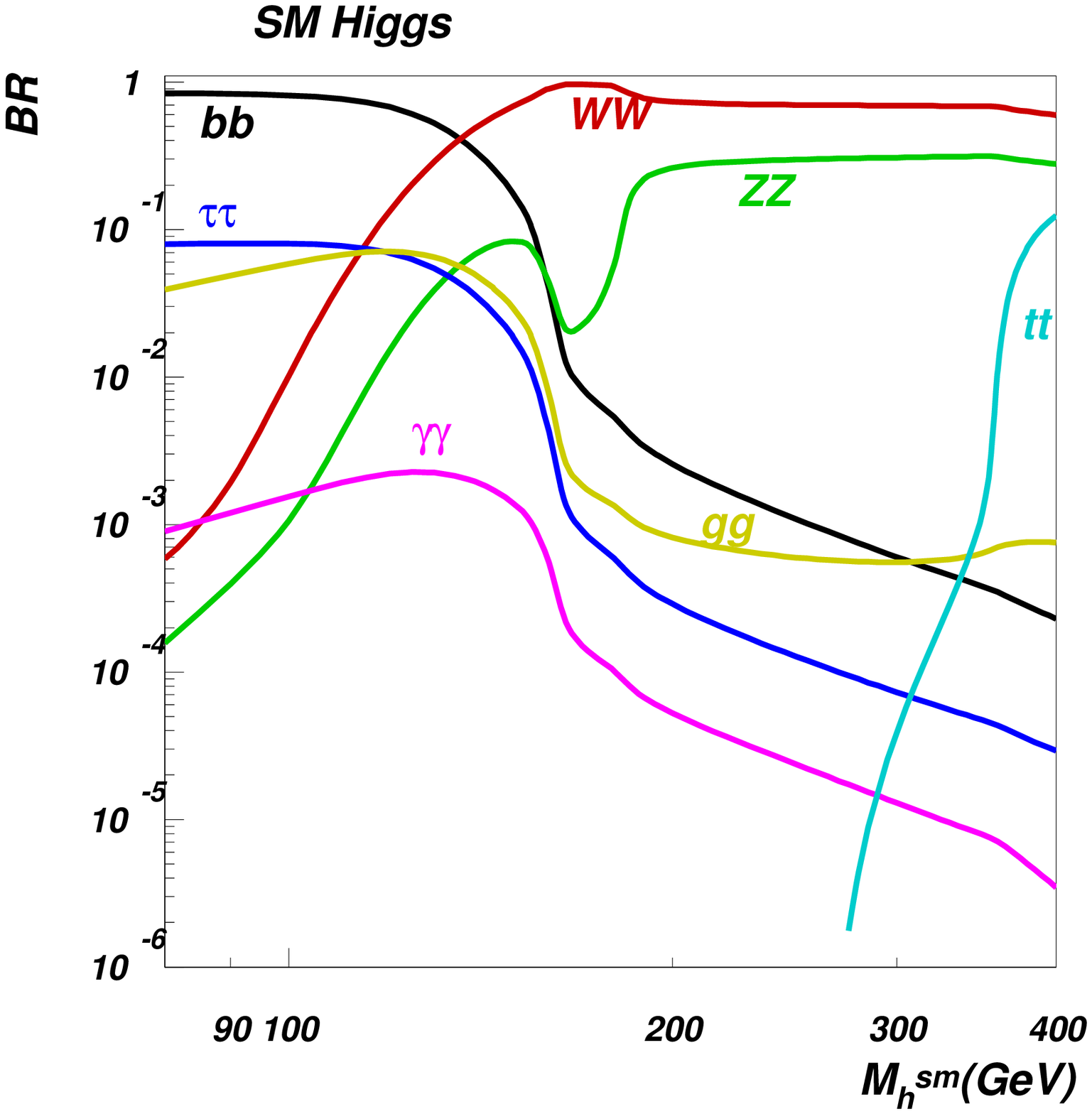}
\caption[sm-higg-br]{
Left frame: Higgs spectrum in the MSSM.   Lower curves indicate \protect{$M_h\ vs.\  M_A$} for indicated values of \protect{$\tan\beta$}.  Upper curves do likewise for \protect{$M_H\ vs.\ M_A$}.  Note the potential for degenerate Higgs masses near 120 GeV.  Right frame:  Branching ratios of the dominant decay
modes of the SM\ Higgs boson. Results
have been obtained with the program HDECAY \protect{\cite{Djouadi:1997yw}}
for \protect{$\alpha_s(M_Z^2) = 0.120$}, \protect{$\overline{m}_b(m_b) = 4.22$}~GeV,
\protect{$m_t =178$}~GeV.}
\label{fig:degenspec}
\end{figure}

Let us consider how the signal of a light Higgs boson could be changed in the MSSM,
compared to expectations in the SM.   
There are several important sources of alterations in the predicted signal, some of which are
interconnected.

First, the MSSM includes three neutral Higgs bosons ${\cal H}=(h,H,A)$ states. The
apparent signal of a single light Higgs could be enhanced if two or three neutral
Higgs species are nearly degenerate so that more than one Higgs is actually
contributing to the final state being studied.    The left-hand frame of 
Figure~\ref{fig:degenspec} illustrates that for  Higgs masses around 120 GeV it is
possible for several Higgs states to be close in mass.   We take advantage of this
near-degeneracy by combining the signals of the different neutral Higgs bosons when
their masses are closer than the experimental resolution.  Specifically, when
combining the signal from  $A$, $h$, and $H$,   we require $|M_A-M_h|$ and/or
$|M_A-M_H|$ to be less than 0.3$\sqrt{M_A/GeV}$ GeV, as compared to the approximate
experimental resolution for the Higgs mass of $\sqrt{M_A/GeV}$ GeV
 for $\tau^+\tau^-$ or $b\bar{b}$ channels.
For the Higgs
mass range studied here, 0.3$\sqrt{M_A/GeV}$ would correspond to a fairly small
mass gap of order $\sim 3-5$ GeV. 
For the $\gamma\gamma$ channel we do not combine the $(h,H,A)$ states but use just  one, 
the $A\to \gamma\gamma$
process, since the experimental mass resolution for 
this final state could be of the order of one GeV.
\begin{figure}
\includegraphics[width=7cm]{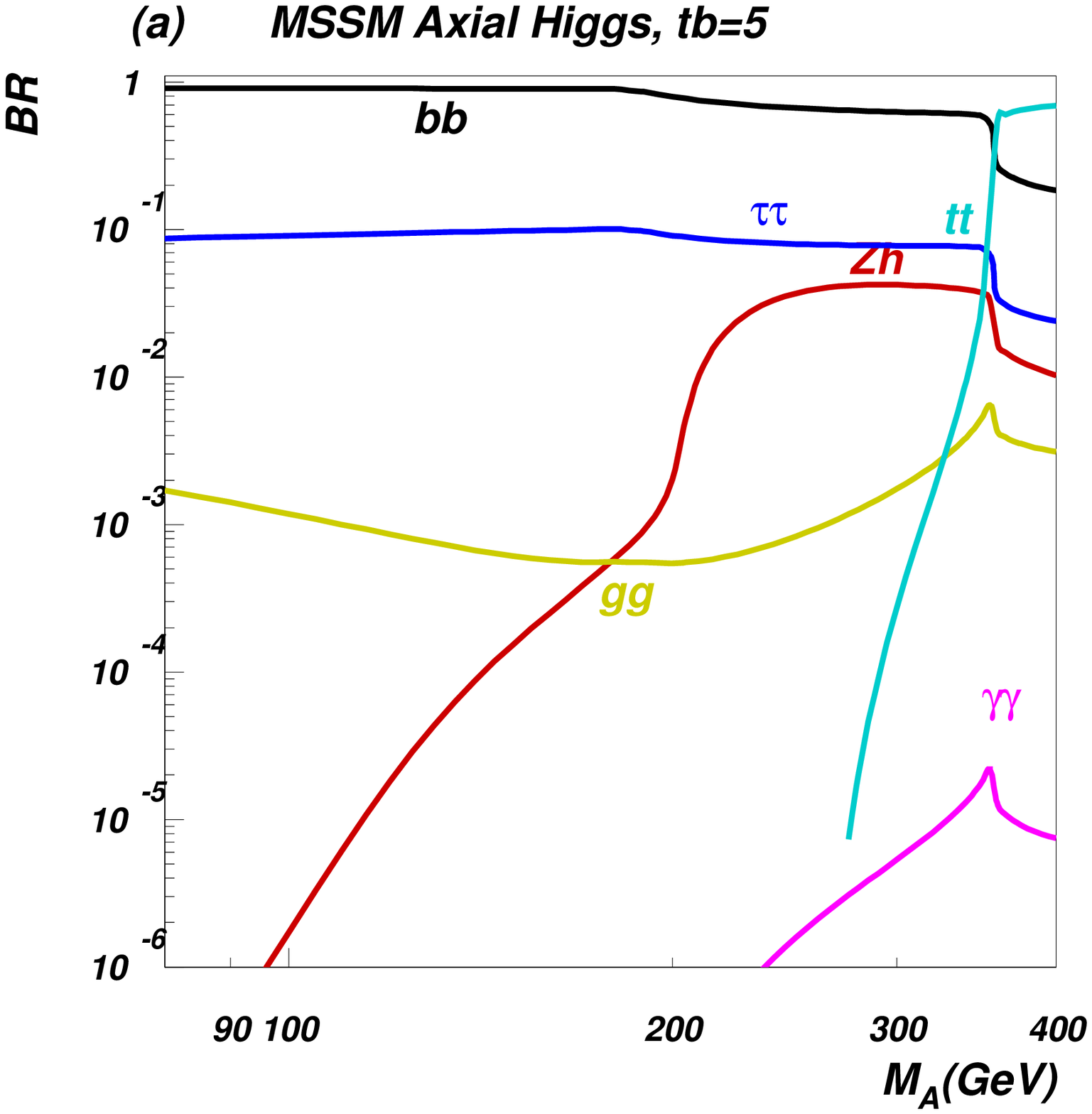}
\includegraphics[width=7cm]{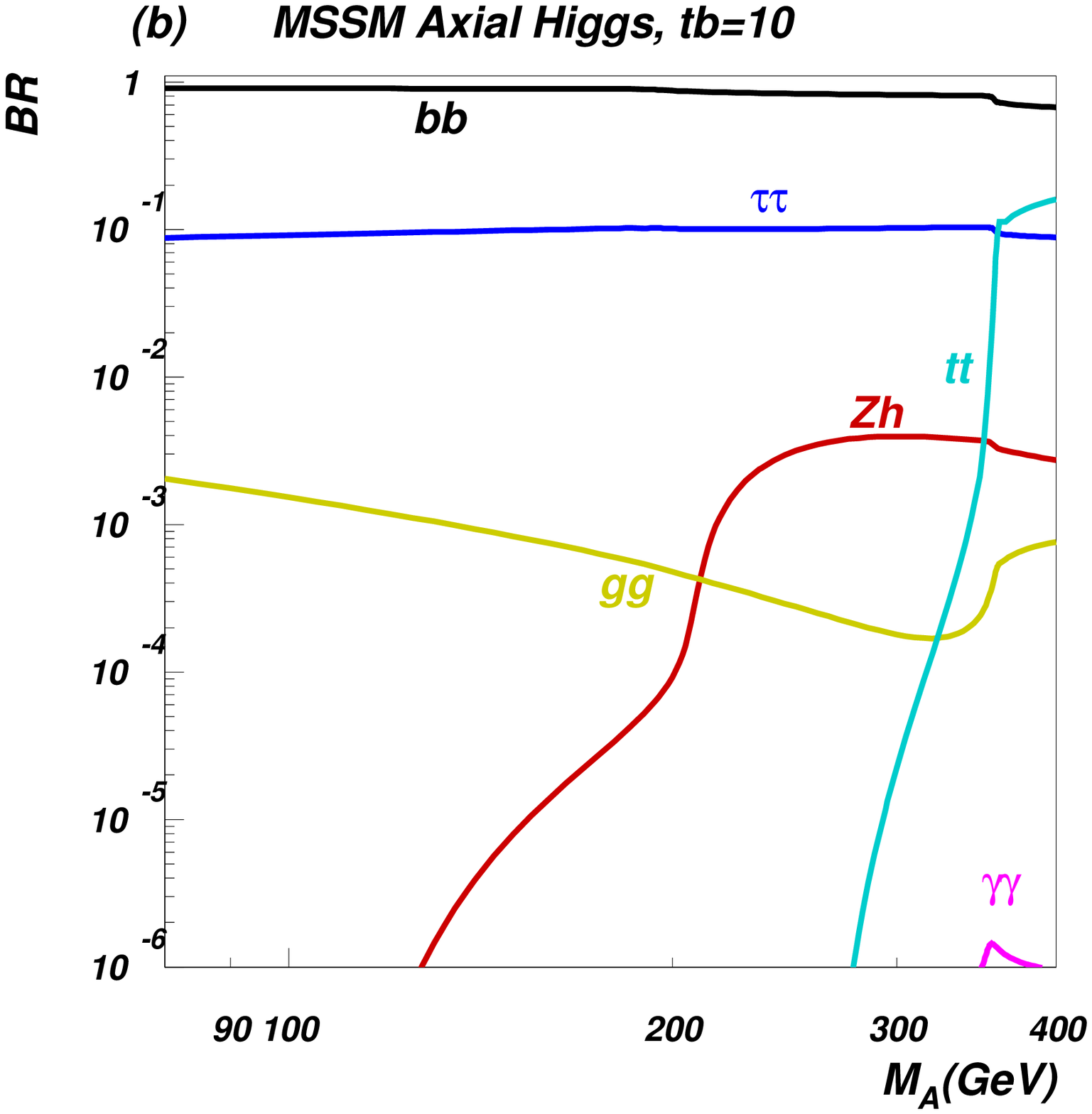}\\
\includegraphics[width=7cm]{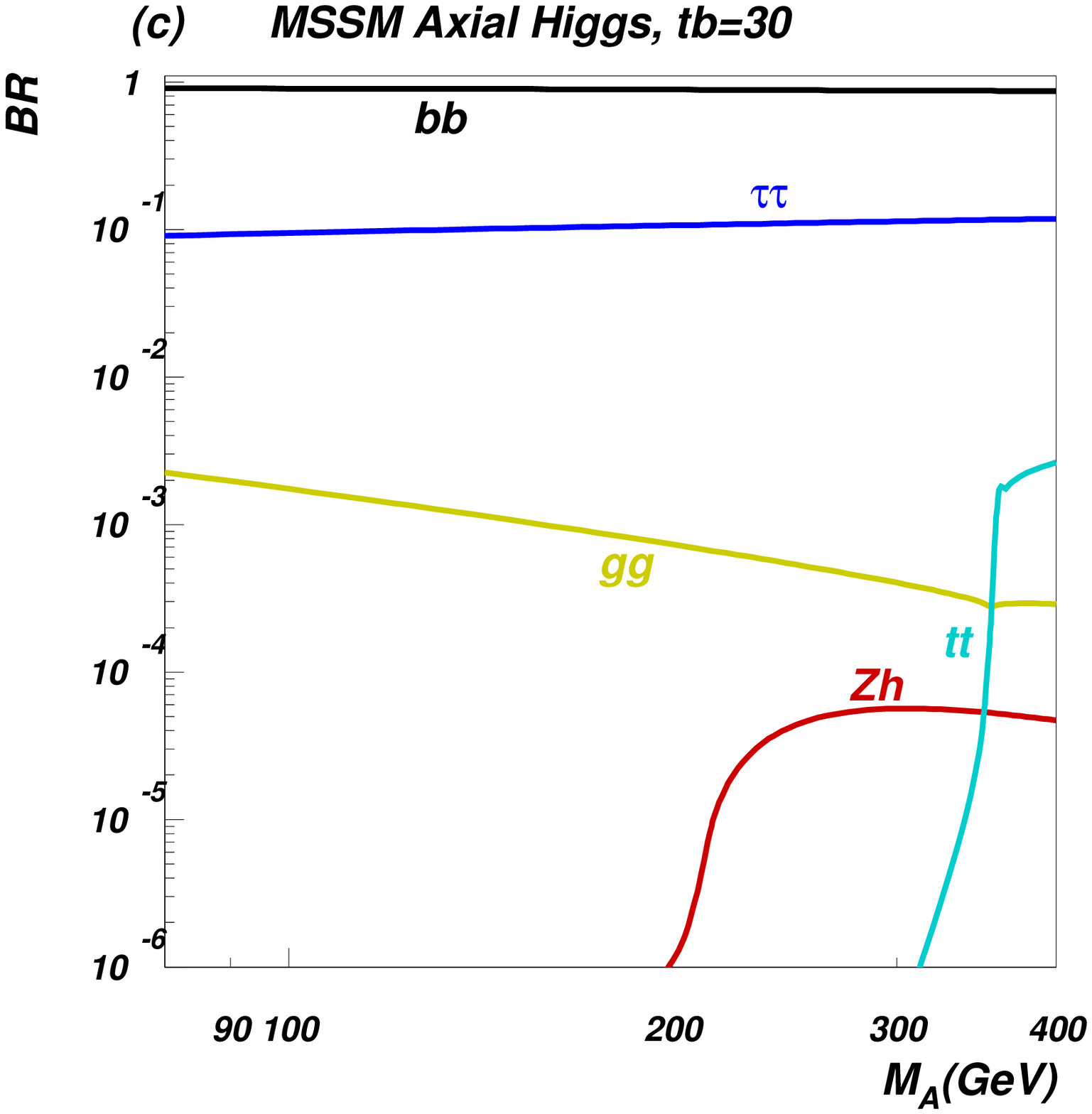}
\includegraphics[width=7cm]{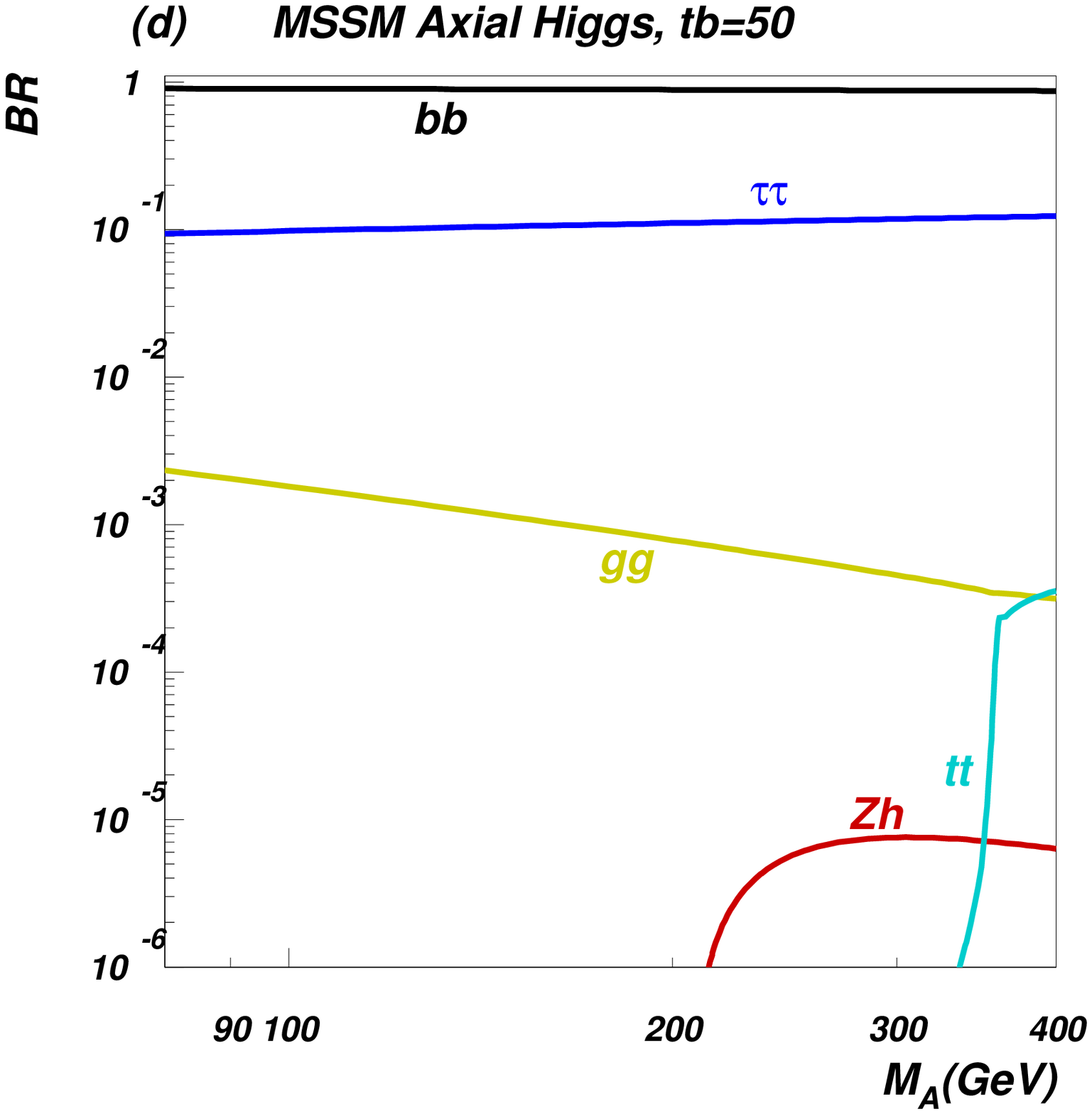}
\caption[mssm-a-br]{
Branching ratios of the dominant decay
modes of the MSSM CP-odd Higgs boson. Results
have been obtained with the program HDECAY \cite{Djouadi:1997yw}
for
$\alpha_s(M_Z^2) = 0.120$, $\overline{m}_b(m_b) = 4.22$~GeV,
$m_t =178$~GeV. Frames (a), (b), (c), and (d) correspond
to $\tan\beta=5,10,30$ and 50, respectively.
}
\label{fig:mssm-a-br}
\end{figure}

\begin{figure}
\includegraphics[width=7cm]{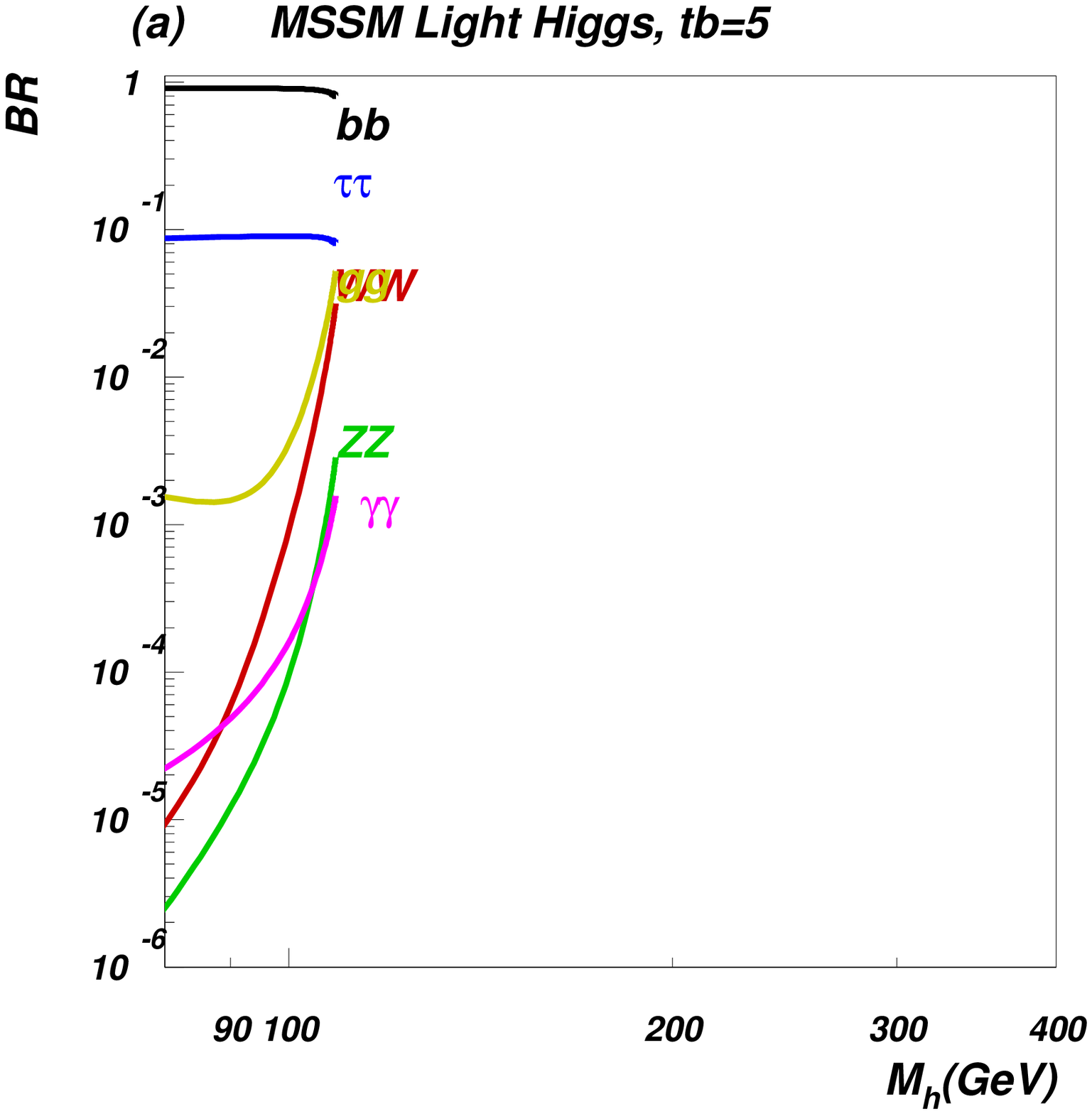}
\includegraphics[width=7cm]{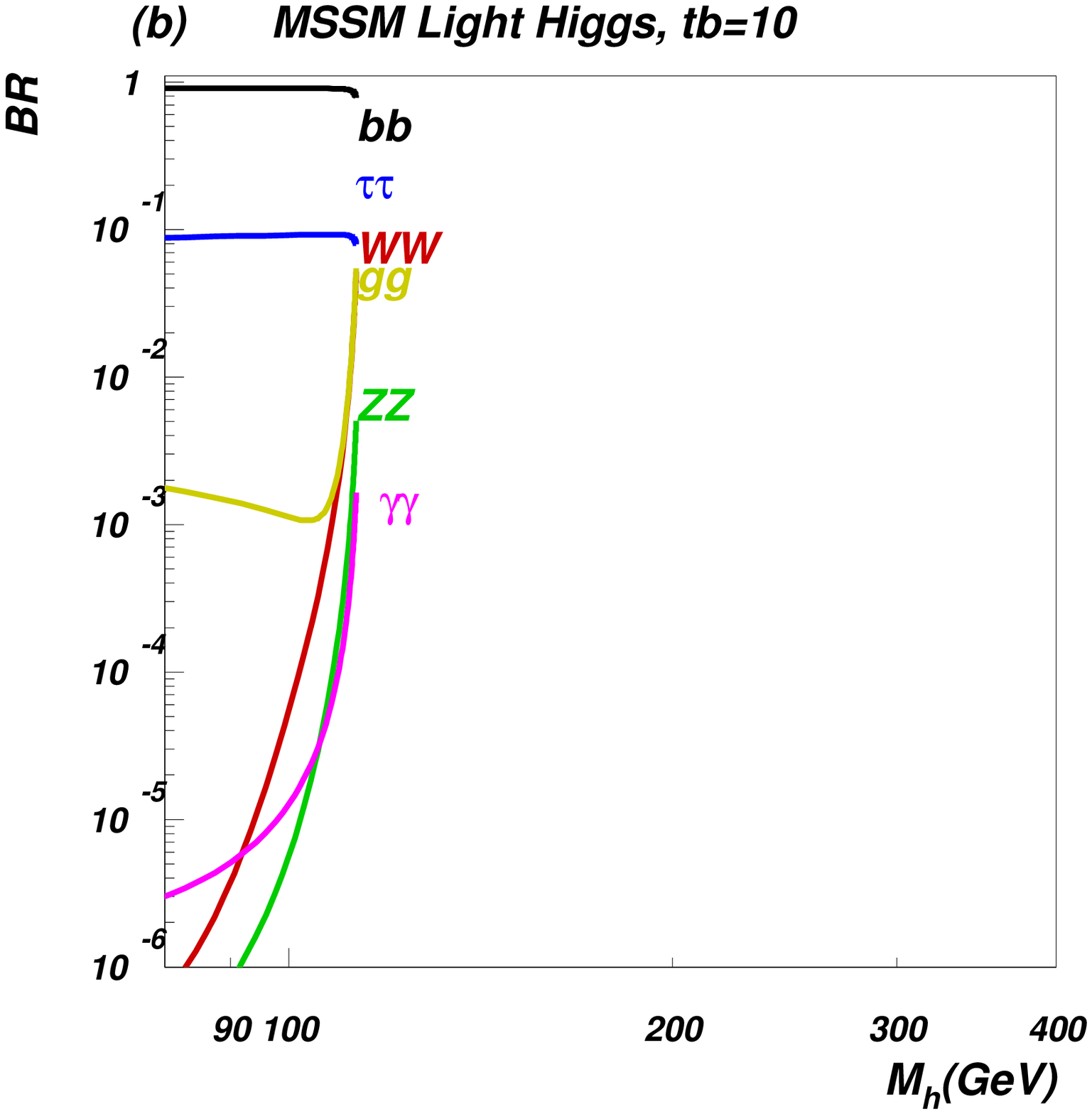}\\
\includegraphics[width=7cm]{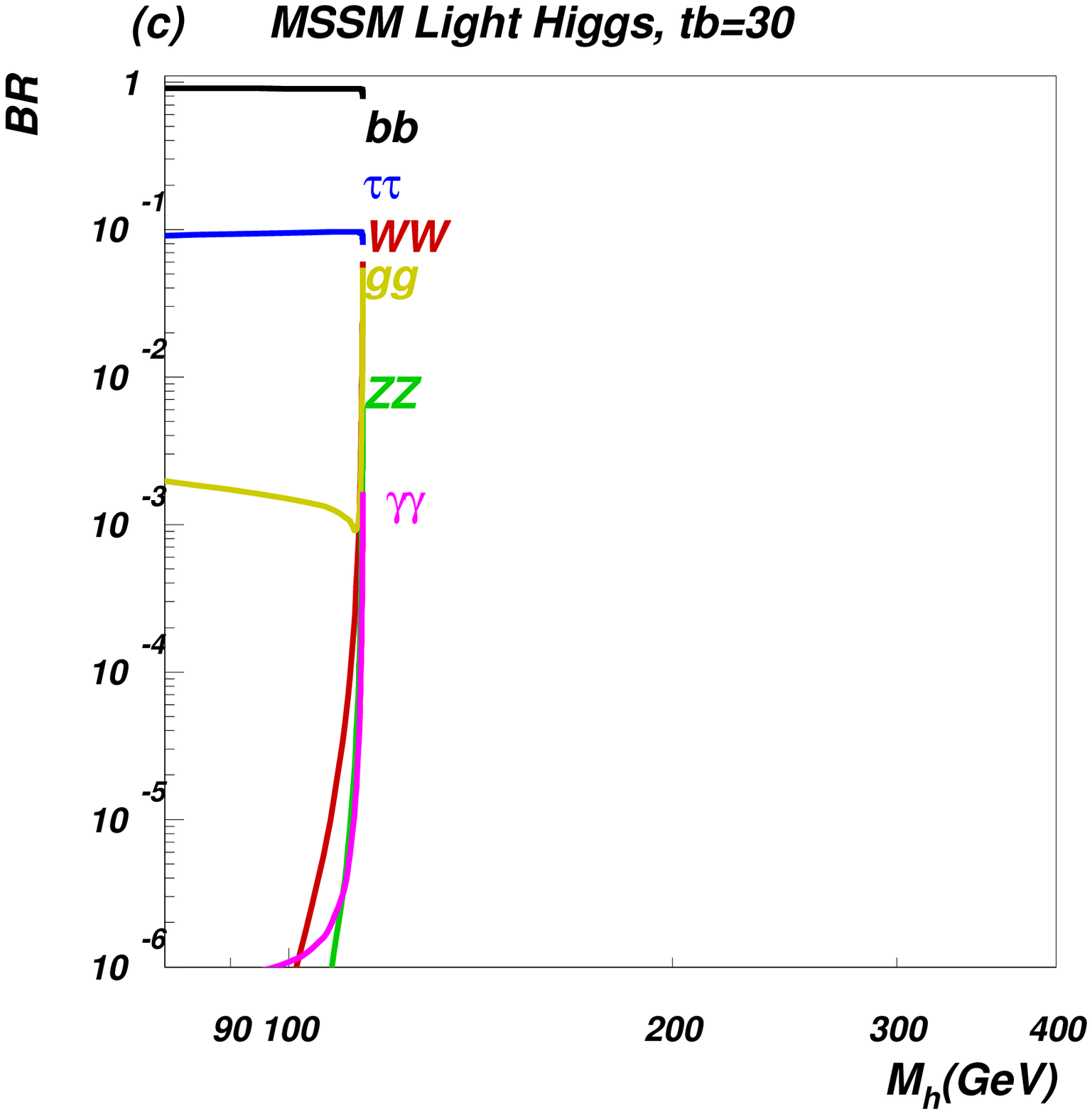}
\includegraphics[width=7cm]{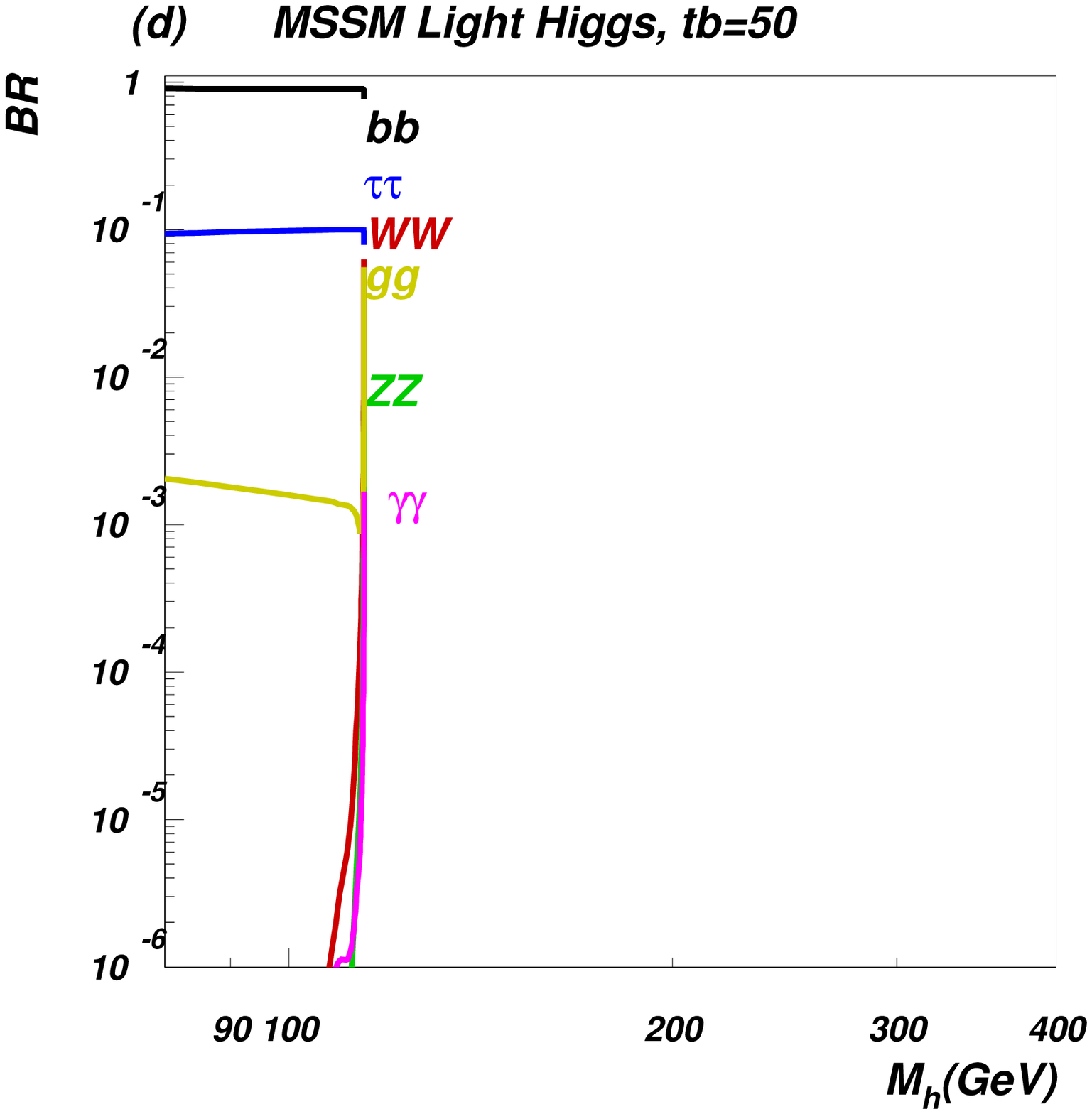}
\caption[mssm-h-br]{  
Branching ratios of the dominant decay
modes of the MSSM Light Higgs boson. Results
have been obtained with the program HDECAY \cite{Djouadi:1997yw}
for
$\alpha_s(M_Z^2) = 0.120$, $\overline{m}_b(m_b) = 4.22$~GeV,
$m_t =178$~GeV. Frames (a), (b), (c), and (d) correspond
to $\tan\beta=5,10,30$ and 50, respectively.
}
\label{fig:mssm-h-br}
\end{figure}

\begin{figure}
\includegraphics[width=7cm]{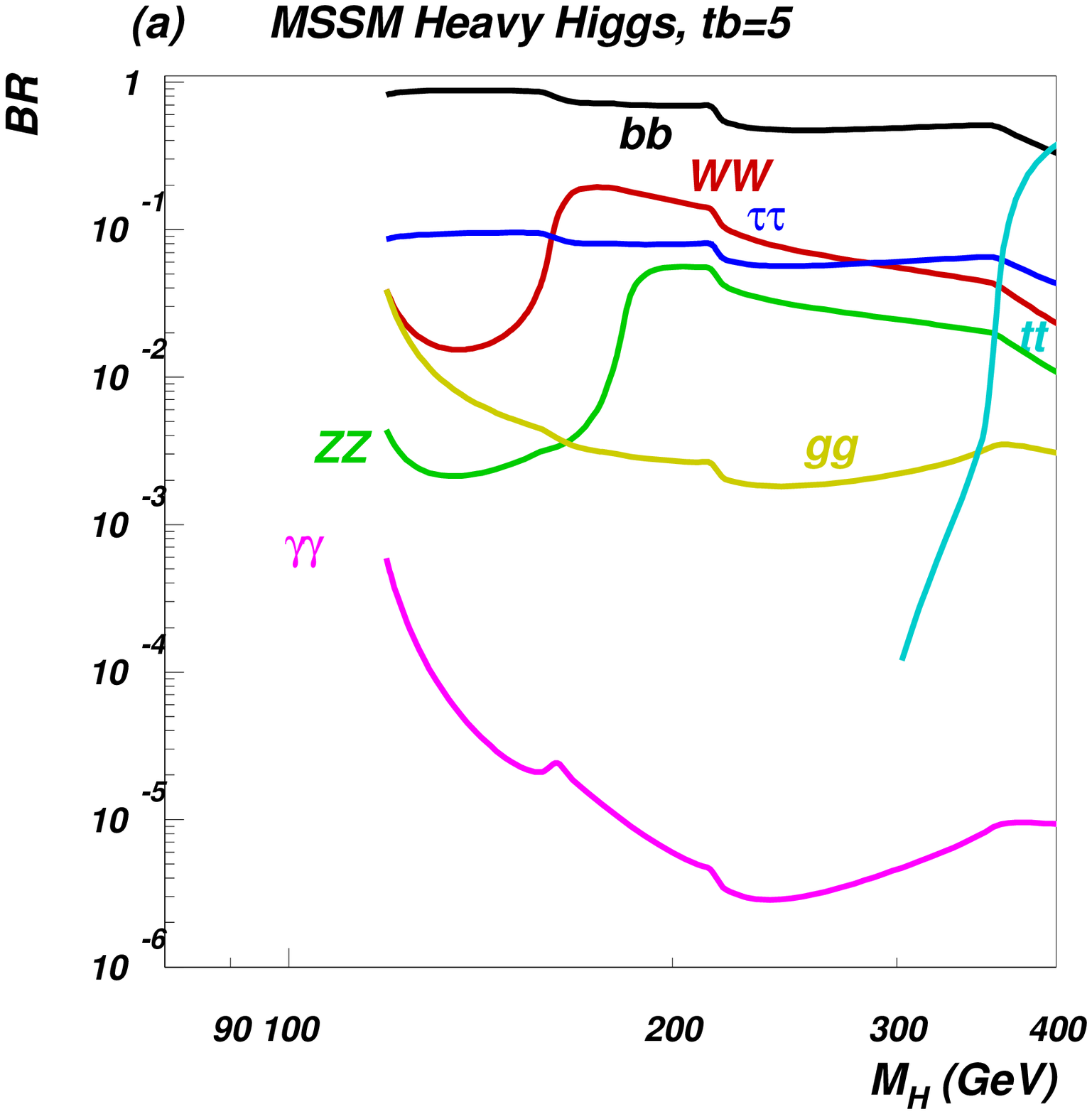}
\includegraphics[width=7cm]{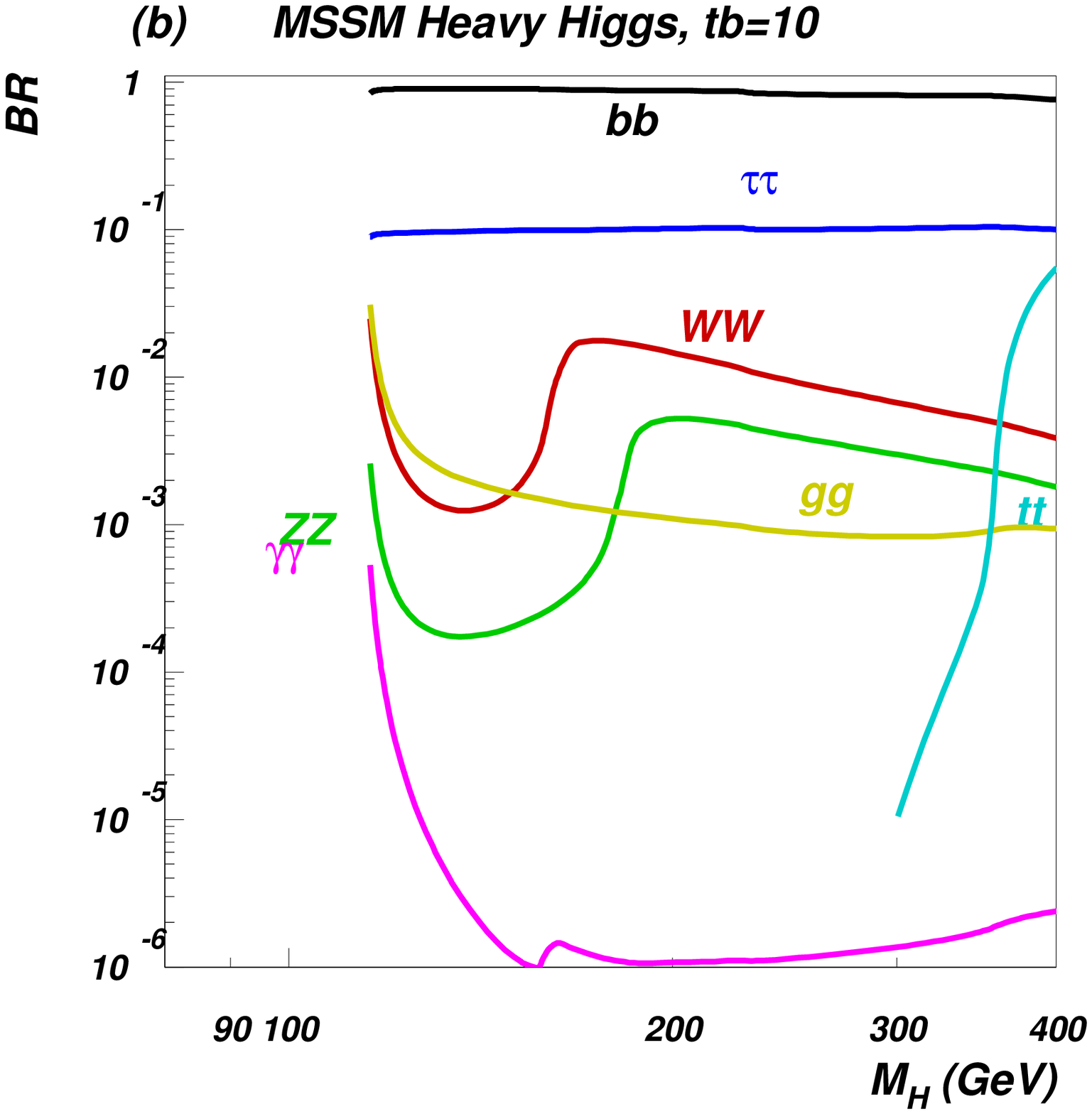}\\
\includegraphics[width=7cm]{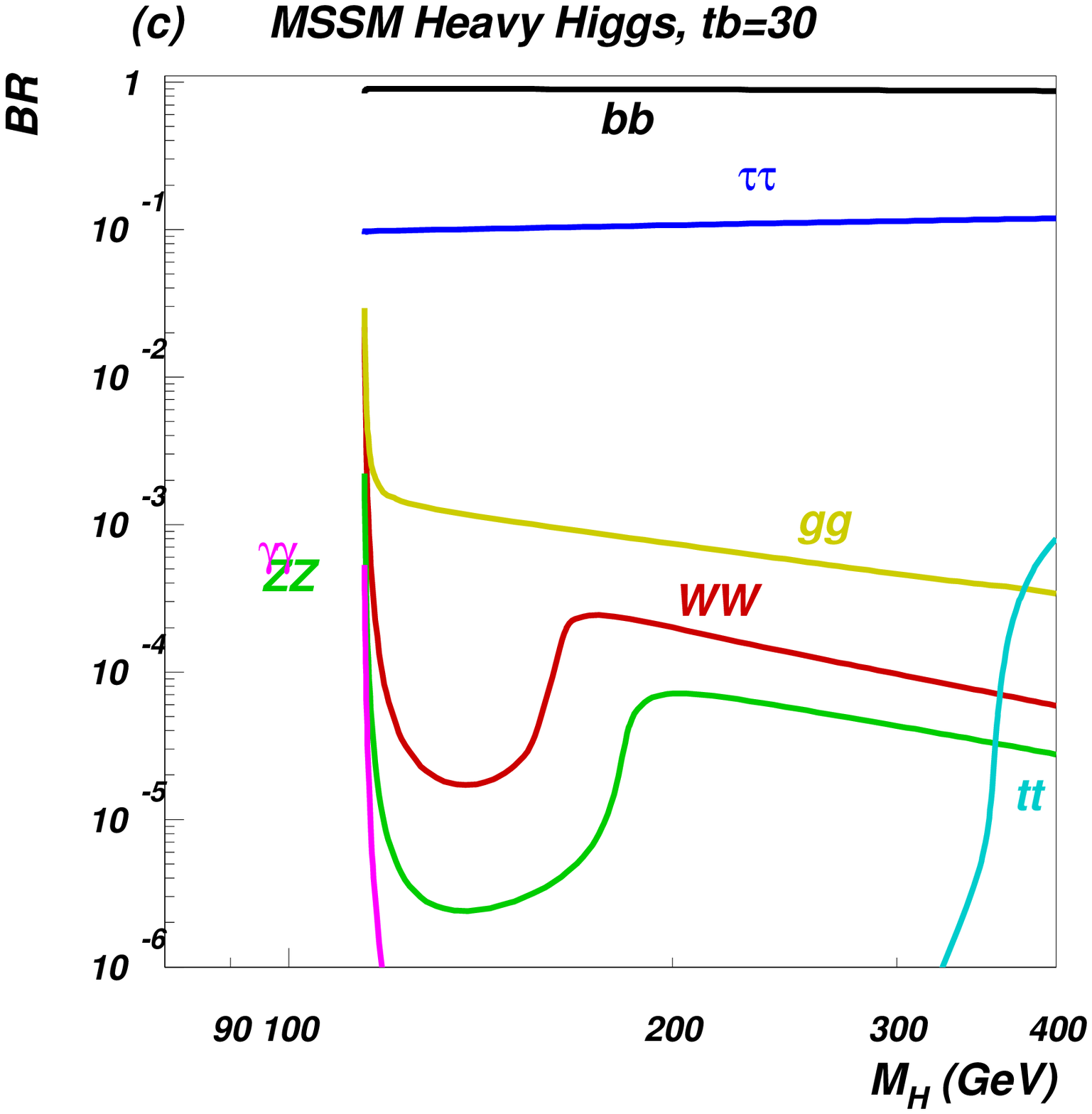}
\includegraphics[width=7cm]{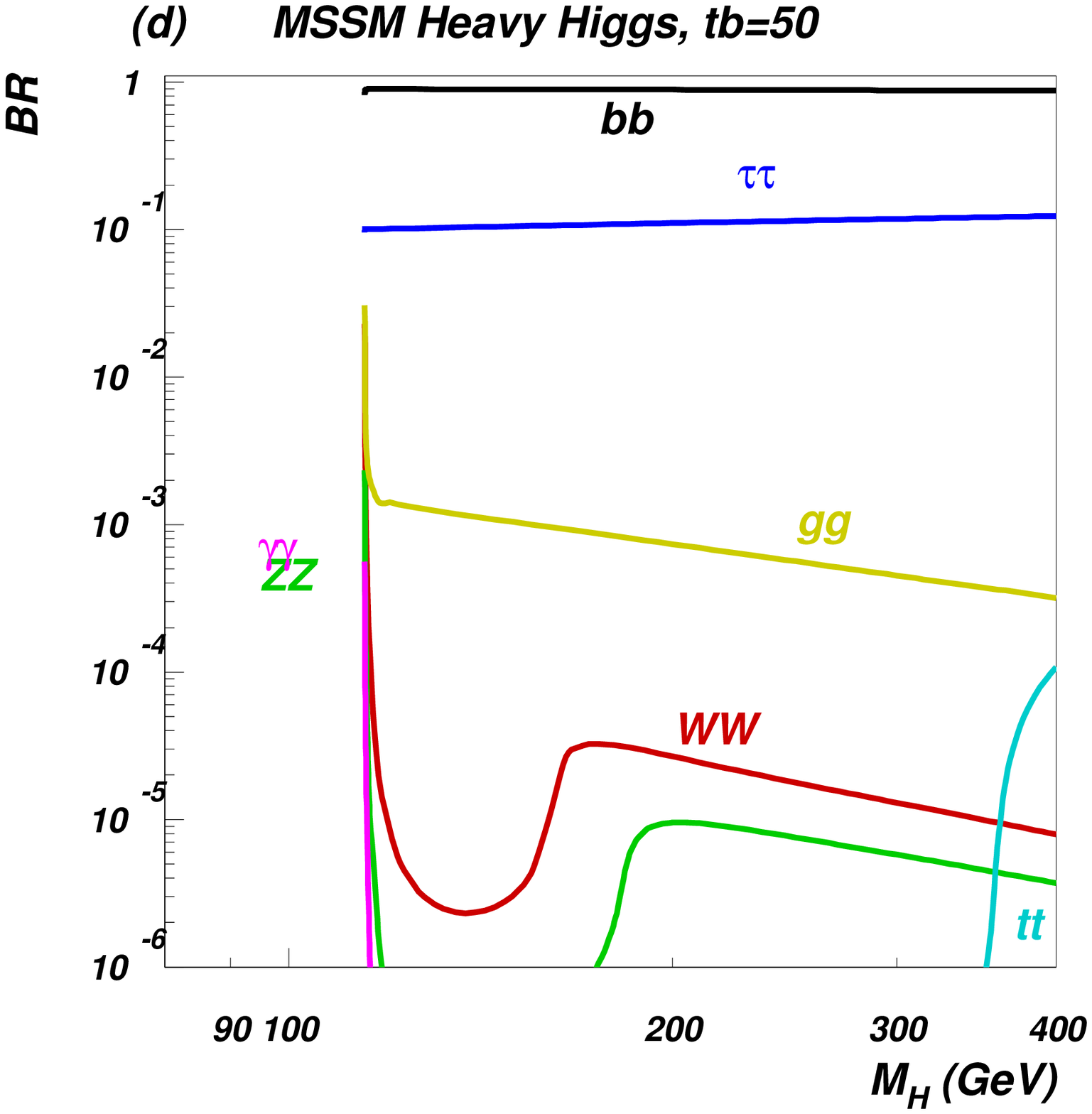}
\caption[mssm-H-br]{ 
Branching ratios of the dominant decay
modes of the MSSM Heavy Higgs boson. Results
have been obtained with the program HDECAY \cite{Djouadi:1997yw}
for
$\alpha_s(M_Z^2) = 0.120$, $\overline{m}_b(m_b) = 4.22$~GeV,
$m_t =178$~GeV. Frames (a), (b), (c), and (d) correspond
to $\tan\beta=5,10,30$ and 50, respectively.
}
\label{fig:mssm-H-br}
\end{figure}

Second, the alterations of the couplings between Higgs bosons and ordinary fermions
in the MSSM, which were discussed in section 2.2, can change the Higgs decay widths
and branching ratios relative to those in the SM.  The SM branching fractions are
pictured in the right-hand frame of Figure~\ref{fig:degenspec} and those in the
MSSM (as calculated with the HDECAY\footnote{ For SUSY HDECAY input we have chosen 
    the squark masses to be 1~TeV, while the trilinear \protect{$A_{t,b,\tau}$}
    and \protect{$\mu$} parameters were taken equal to 200 GeV.} program~\cite{Djouadi:1997yw}) are in
Figures~\ref{fig:mssm-a-br}, \ref{fig:mssm-h-br}, \ref{fig:mssm-H-br}, and the
relevant branching ratios for a 130 GeV CP-odd Higgs are given for various
$\tan\beta$ in Table~1.  These changes directly effect the enhancement factor for a
given process, as in equation (\ref{eq:kappa}). 
 When radiative effects on the
masses and couplings are included, the Higgs boson production rate as well as the
decay branching fractions can be substantially altered, in a non-universal way. 
For instance, $B(h\to\tautau)$ could be enhanced by up to an order of magnitude due
to the suppression of $B(h\to b\bar{b})$ in certain regions of parameter space
\cite{Carena:1998gk,Carena:1999bh}.  However, this gain in branching  fraction
would be offset\footnote{
    There can be a suppression of $BR(H\to b\bar{b})$ 
    and $BR(H\to \tau\tau)$ 
    in the parameter region where all Higgs bosons are nearly
    degenerate~\cite{Boos:2002ze}.
}
 to some degree by a reduction in Higgs production through channels
involving  $Y_{{\cal H}b\bar{b}}$.%

\begin{table}[tb]
\label{tab:cpoddbr}
\begin{center}
\let\normalsize=\captsize
\caption{Branching ratios for a CP-odd MSSM Higgs of mass 130 GeV.}
\vskip .5pc
\small
\renewcommand\tabcolsep{9pt}
\begin{tabular}{|c||c||c||c||c||}\hline
Decay 			& $\tan\beta=5$ 	&  $\tan\beta=10$ 	& $\tan\beta=30$ 	&   $\tan\beta=50$\\Channel & & & &  \\
\hline
$b \overline{b}$ 	& 0.90 			& 0.90  		& 0.90  		& 0.89  		\\
$c \overline{c}$ 	& $ 6.5 \times 10^{-5}$	& $ 4.1 \times 10^{-6}$ & $ 5.2 \times 10^{-8}$ & $ 6.9 \times 10^{-9}$ \\
$ \tau^+ \tau^-$ 	& 0.095 		& 0.096 		& 0.099 		& 0.10  		\\
$gg$ 			& $ 7.5 \times 10^{-4}$	& $ 1.0 \times 10^{-3}$ & $ 1.3 \times 10^{-3}$ & $ 1.3 \times 10^{-3}$ \\
$ \gamma \gamma $ 	& $ 2.7 \times 10^{-7}$ & $ 5.4 \times 10^{-7}$ & $ 5.9 \times 10^{-7}$	& $ 5.9 \times 10^{-7}$ \\
$ W^+ W^-$ 		& 0 			& 0			& 0			& 0			\\
\hline
\end{tabular}
\end{center}
\end{table}

Third, recall that SM production of the light Higgs via gluon fusion is dominated
by a top-quark loop; the large top quark mass both increases the top-Higgs coupling
and suppresses the loop.  In the MSSM, a large value of $\tan\beta$ enhances the
bottom-Higgs coupling (eqns. (\ref{eq:bhi}) and (\ref{eq:bhii})), making gluon fusion through a 
$b$-quark loop significant, and possibly even dominant over the top-quark loop contribution.

Fourth, the presence of superpartners in the MSSM gives rise to new squark-loop
contributions  to Higgs boson production through gluon fusion. 
Light squarks with masses of order 100 GeV have been argued to lead
to a considerable universal enhancement (as much as a factor of five)
\cite{Dawson:1996xz,Harlander:2003bb,Harlander:2003kf,Harlander:2004tp}
for MSSM Higgs production compared to the SM.

Finally, enhancement of the $Y_{{\cal H}b\bar{b}}$ coupling at moderate to large $\tan\beta$
makes  $b\bar{b} \to {\cal H}$ a significant means of Higgs production in the
MSSM -- in contrast to the SM where it is negligible.   To include both production
channels when looking for a Higgs decaying as ${\cal H} \to xx$, we define
a combined enhancement factor 
\begin{eqnarray}
\kappa_{total/xx}^{\cal H} 
&=&
\sigma(gg\to{\cal H}\to xx)+\sigma(bb\to{\cal H} \to xx)\over
{\sigma(gg\to h_{SM} \to xx)+\sigma(bb\to h_{SM} \to xx)} \nonumber\\
&=&
\kappa_{gg/xx}^{\cal H}+\sigma(bb\to {\cal H}\to xx)/\sigma(gg\to h_{SM} \to xx)\over
{1+\sigma(bb\to h_{SM} \to xx)/\sigma(gg\to h_{SM} \to xx)} \nonumber\\
&=&
\kappa_{gg/xx}^{\cal H}+\kappa_{bb/xx}^{\cal H}\sigma(bb\to h_{SM} \to xx)/\sigma(gg\to h_{SM} \to xx)\over
{1+\sigma(bb\to h_{SM} \to xx)/\sigma(gg\to h_{SM} \to xx)} \nonumber\\
&\equiv&
[\kappa_{gg/xx}^{\cal H}+\kappa_{bb/xx}^{\cal H} R_{bb:gg}]/
[{1+  R_{bb:gg}}]  .
\label{kappab}
\end{eqnarray}
Here 
 $R_{bb:gg}$ is the ratio of $b\bar{b}$ and $gg$ initiated Higgs boson production in the 
 Standard Model, which can be calculated using HDECAY.  
\begin{figure}
\includegraphics[width=7cm]{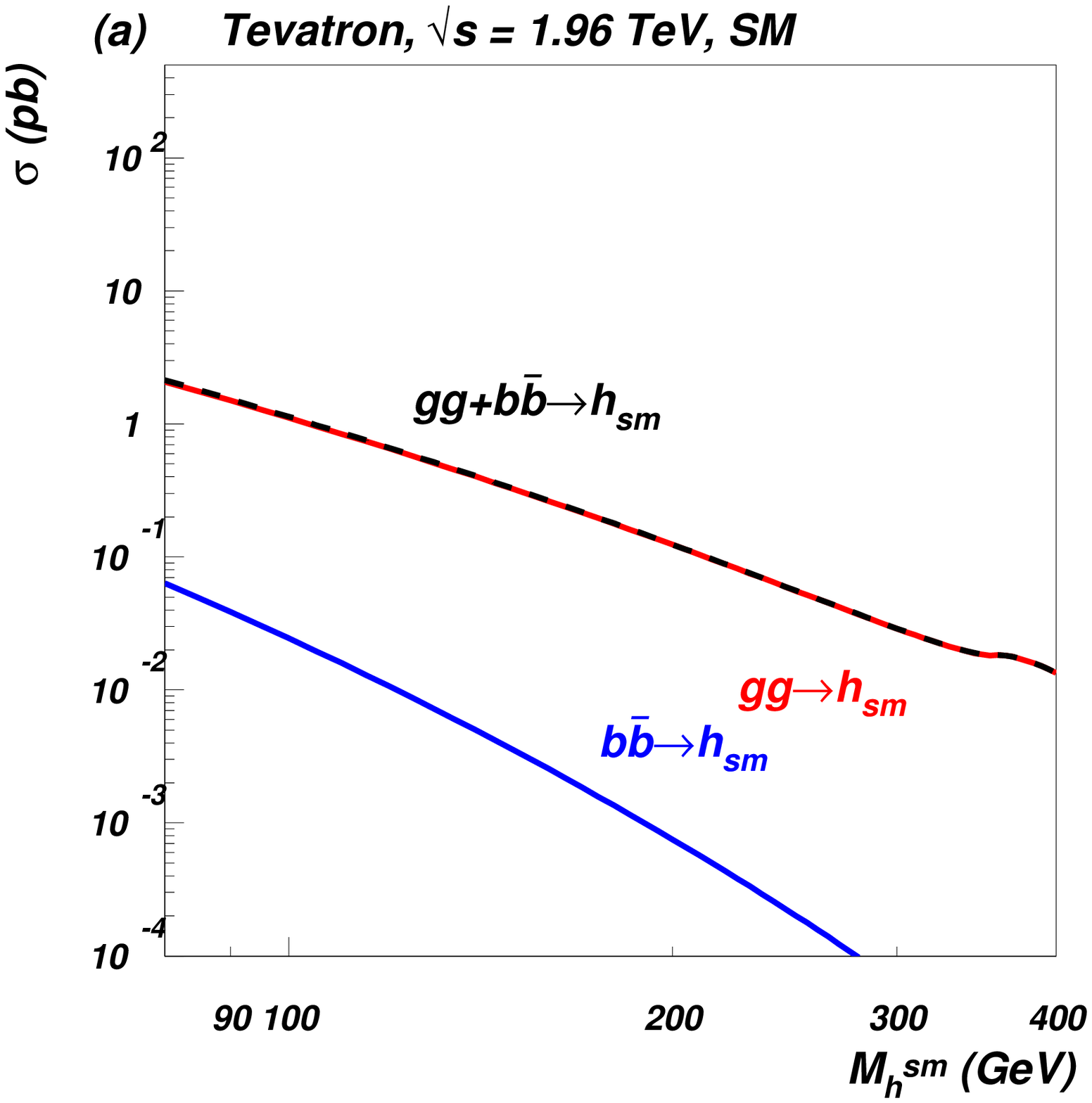}
\includegraphics[width=7cm]{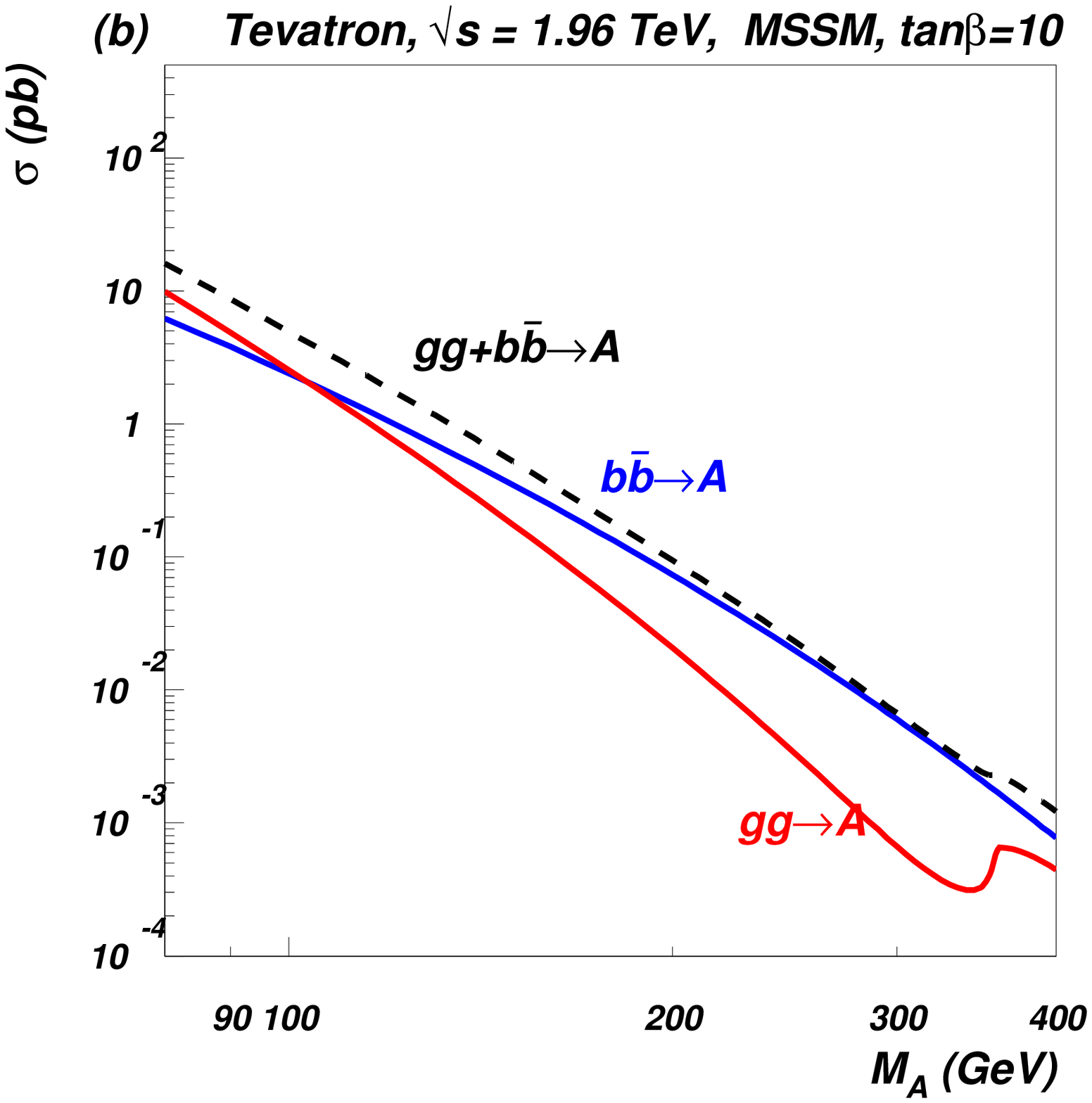}\\
\includegraphics[width=7cm]{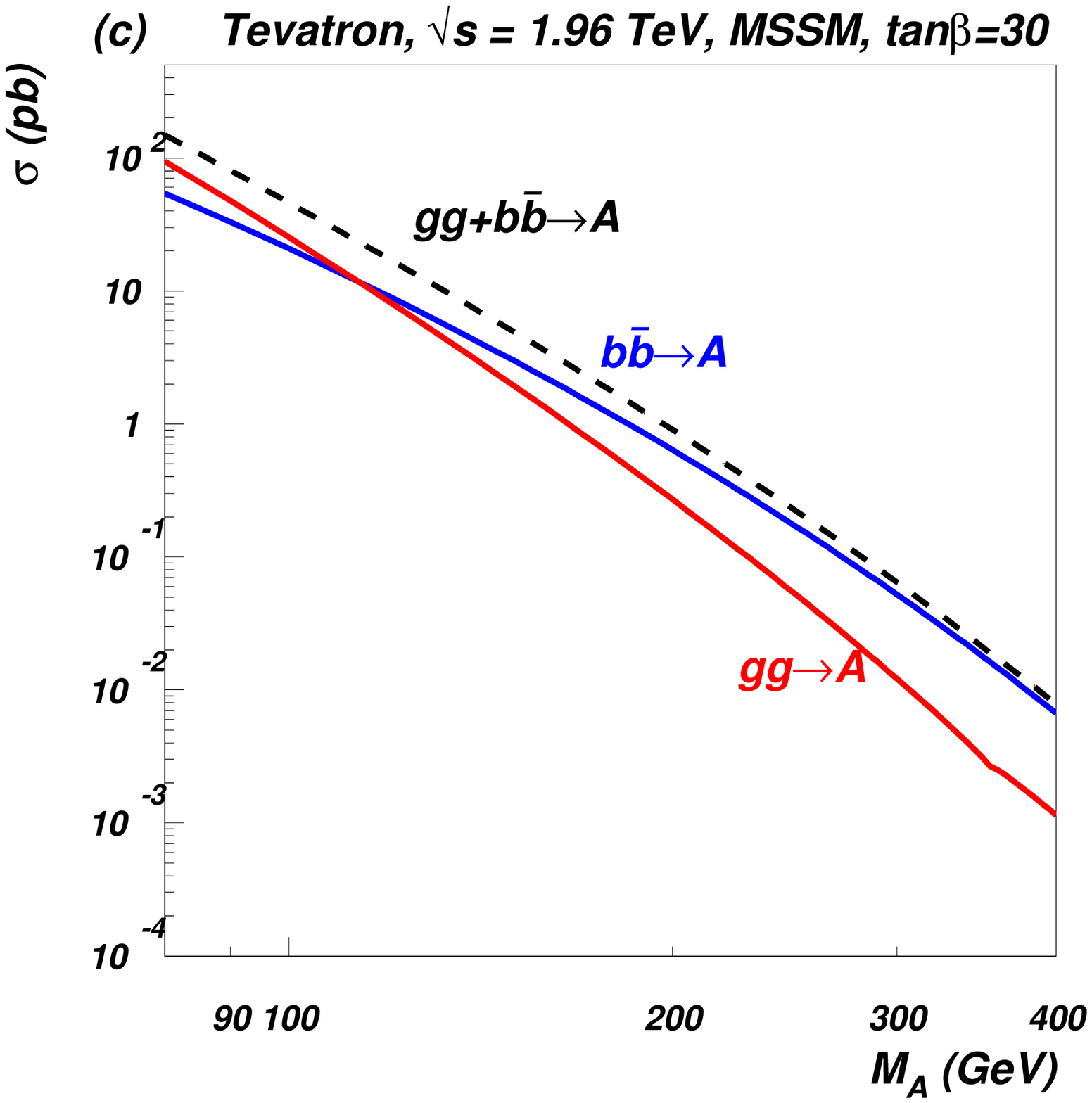}
\includegraphics[width=7cm]{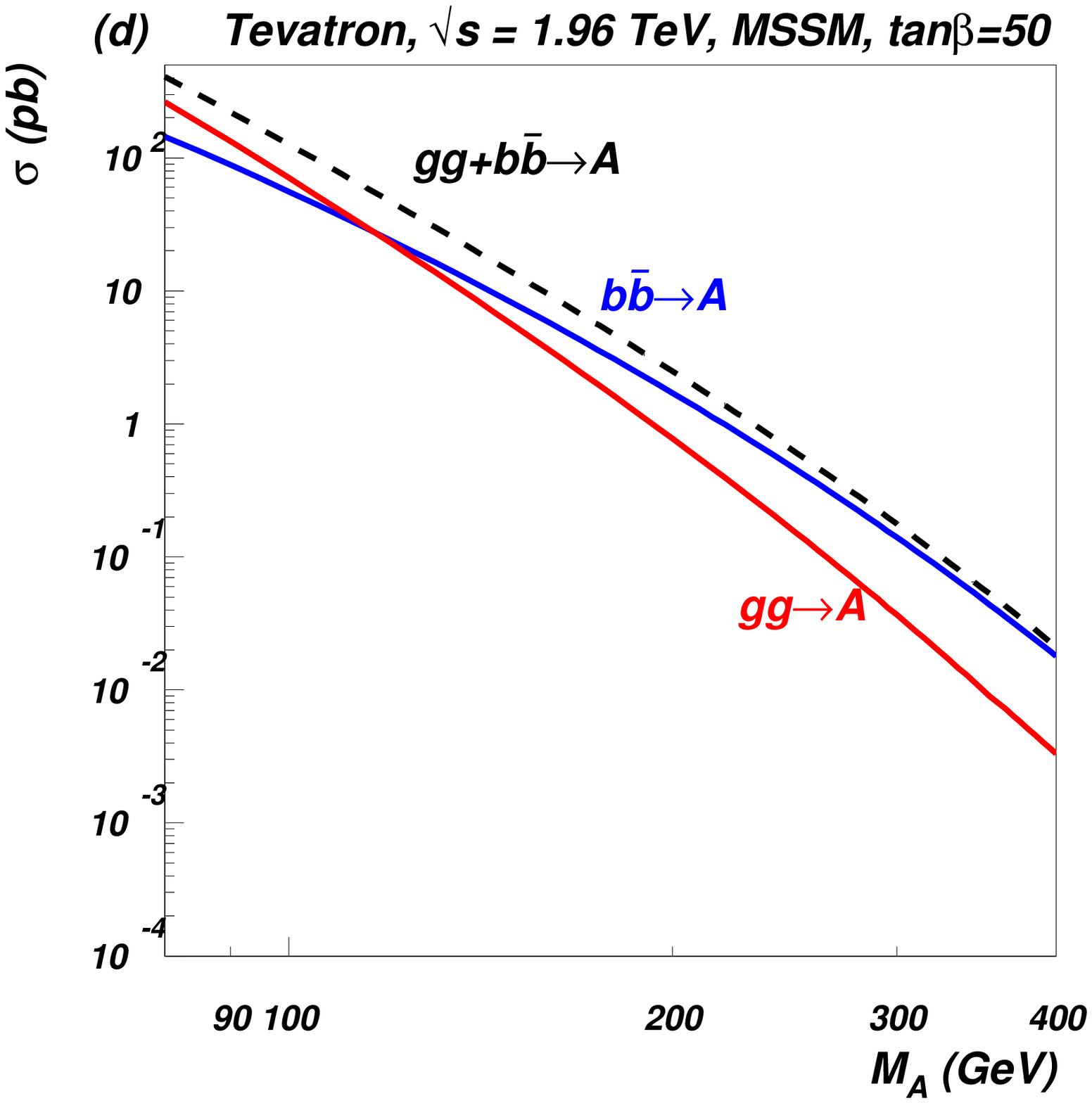}
\caption[mssm-tot-prod]{
NLO cross sections for Higgs production via the $b\bar{b}\to {\cal H}$ and $gg\to {\cal H}$
processes (as well as their sum) for (a) the SM Higgs, and (b)-(d) the Supersymmetric  axial Higgs
boson with $\tan\beta=10,30$ and 50, respectively, at the Tevatron.}
\label{fig:mssm-tot-prod}
\end{figure}

\begin{figure}
\includegraphics[width=7cm]{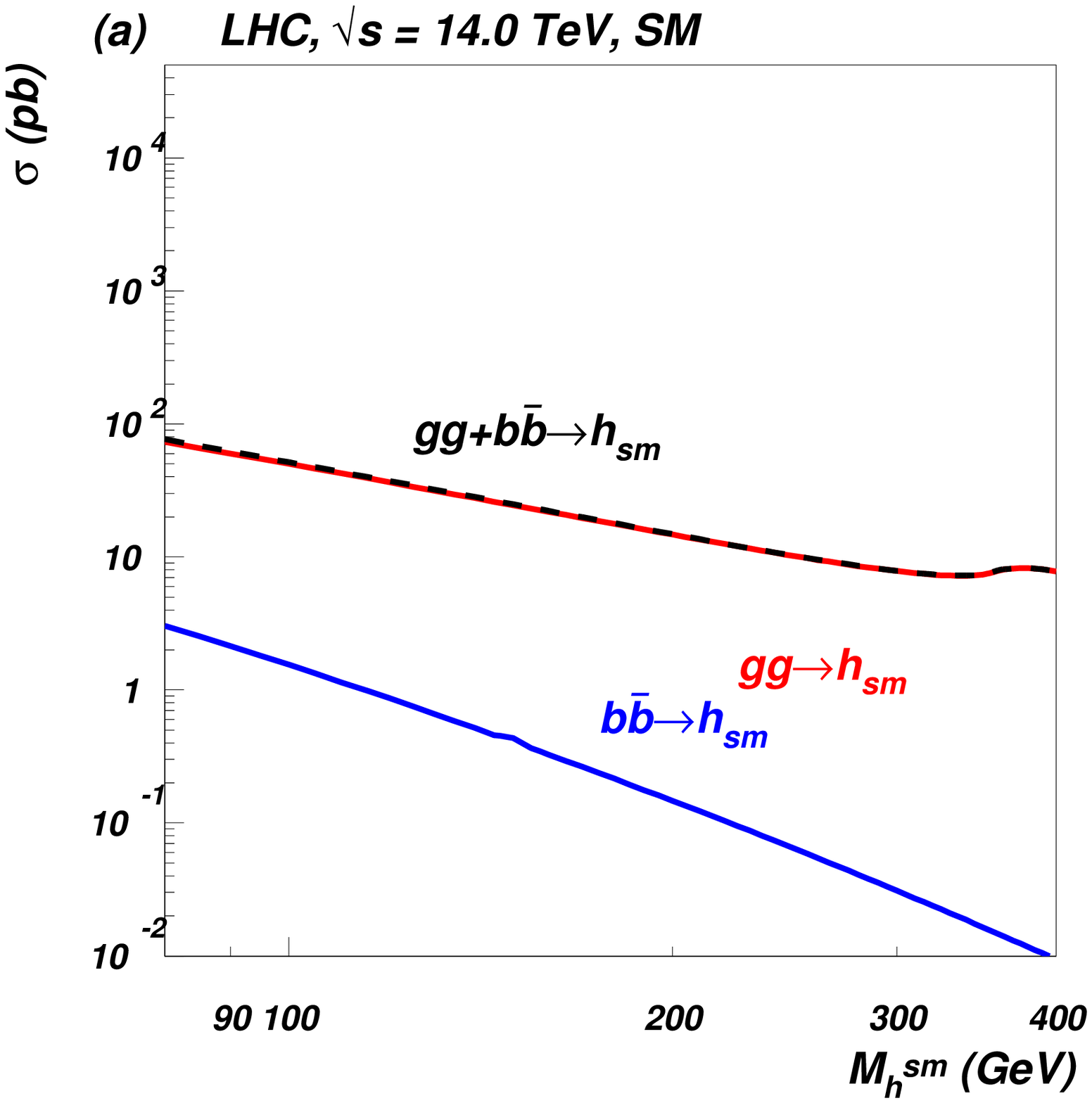}
\includegraphics[width=7cm]{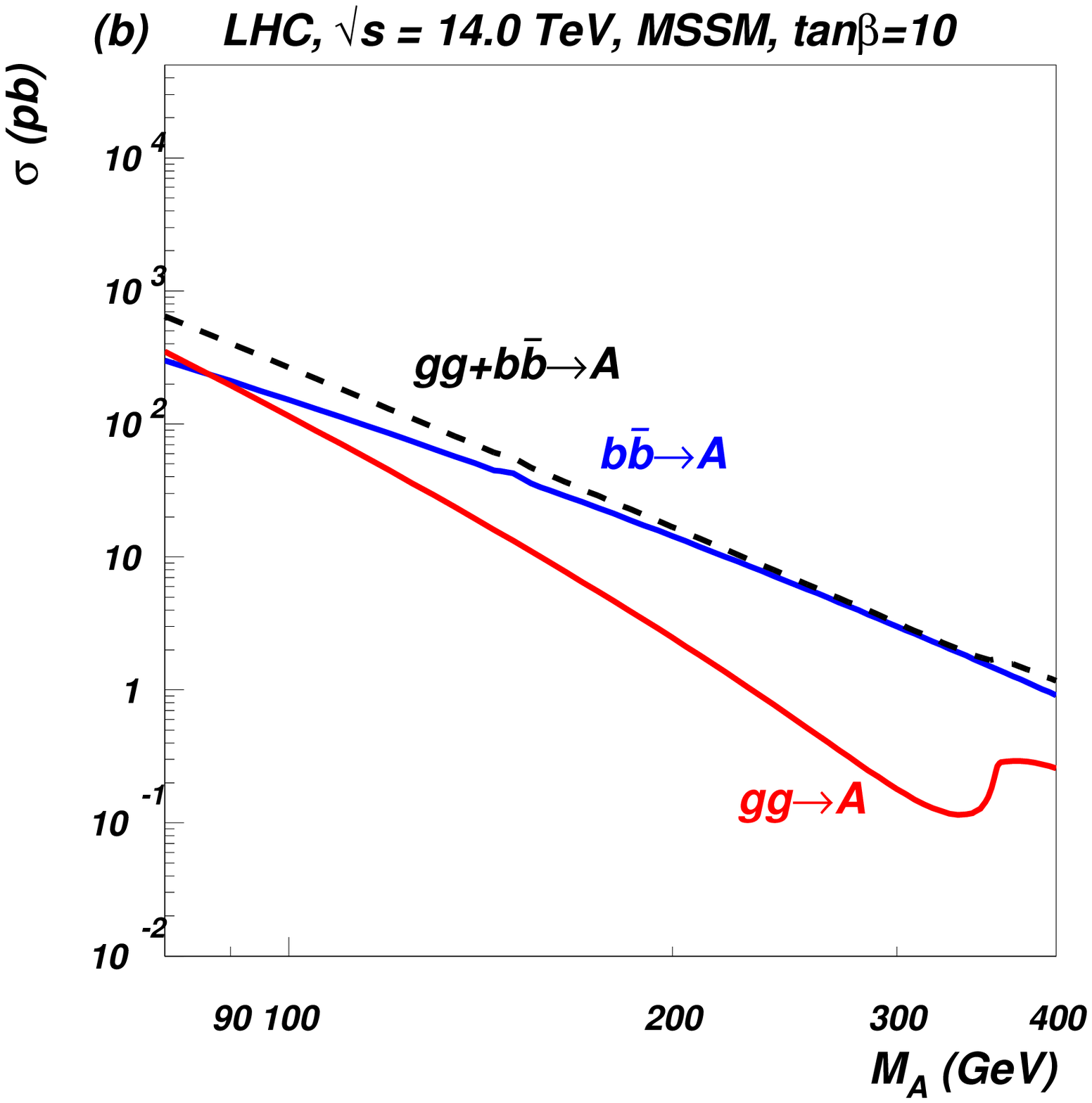}\\
\includegraphics[width=7cm]{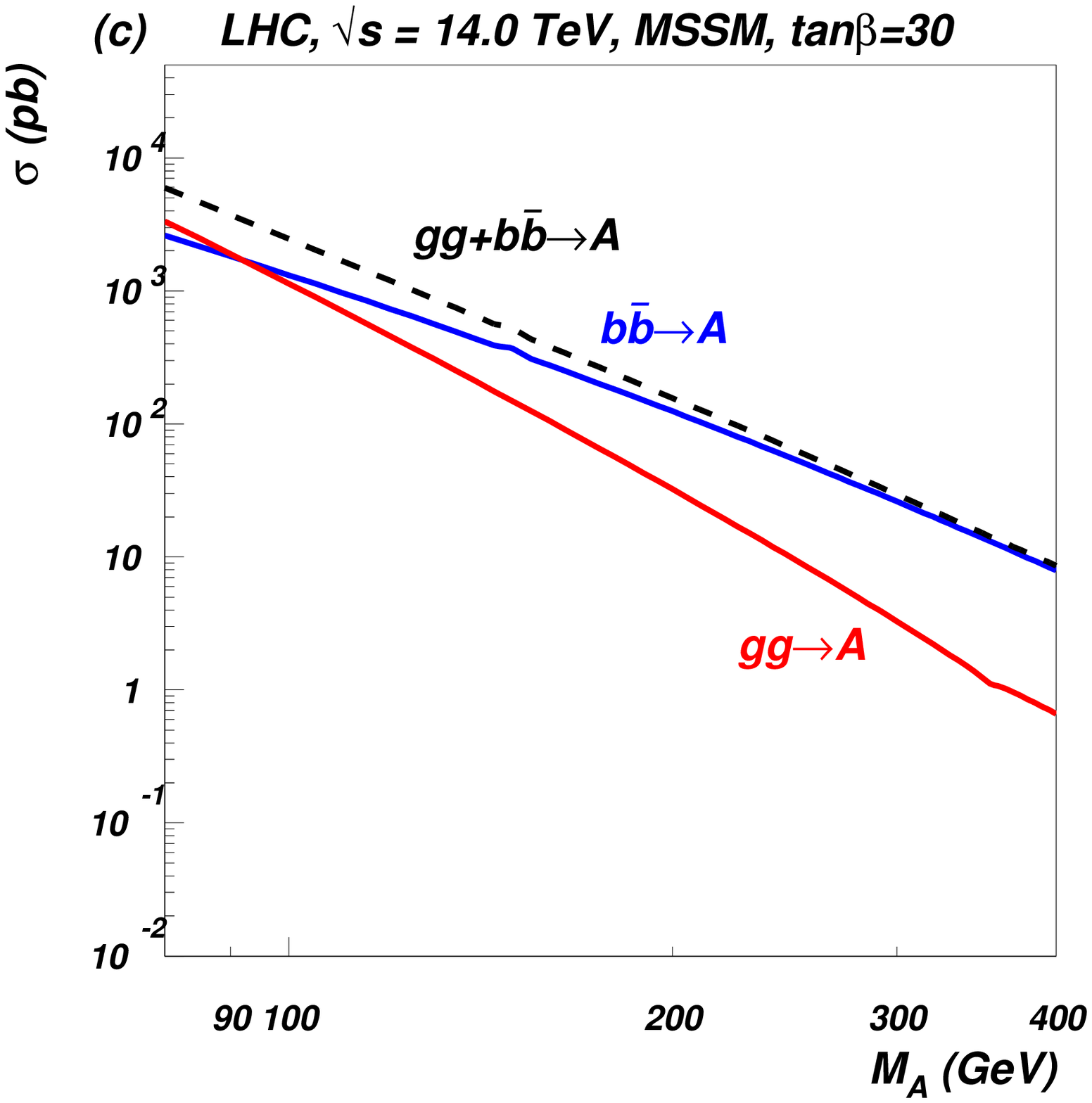}
\includegraphics[width=7cm]{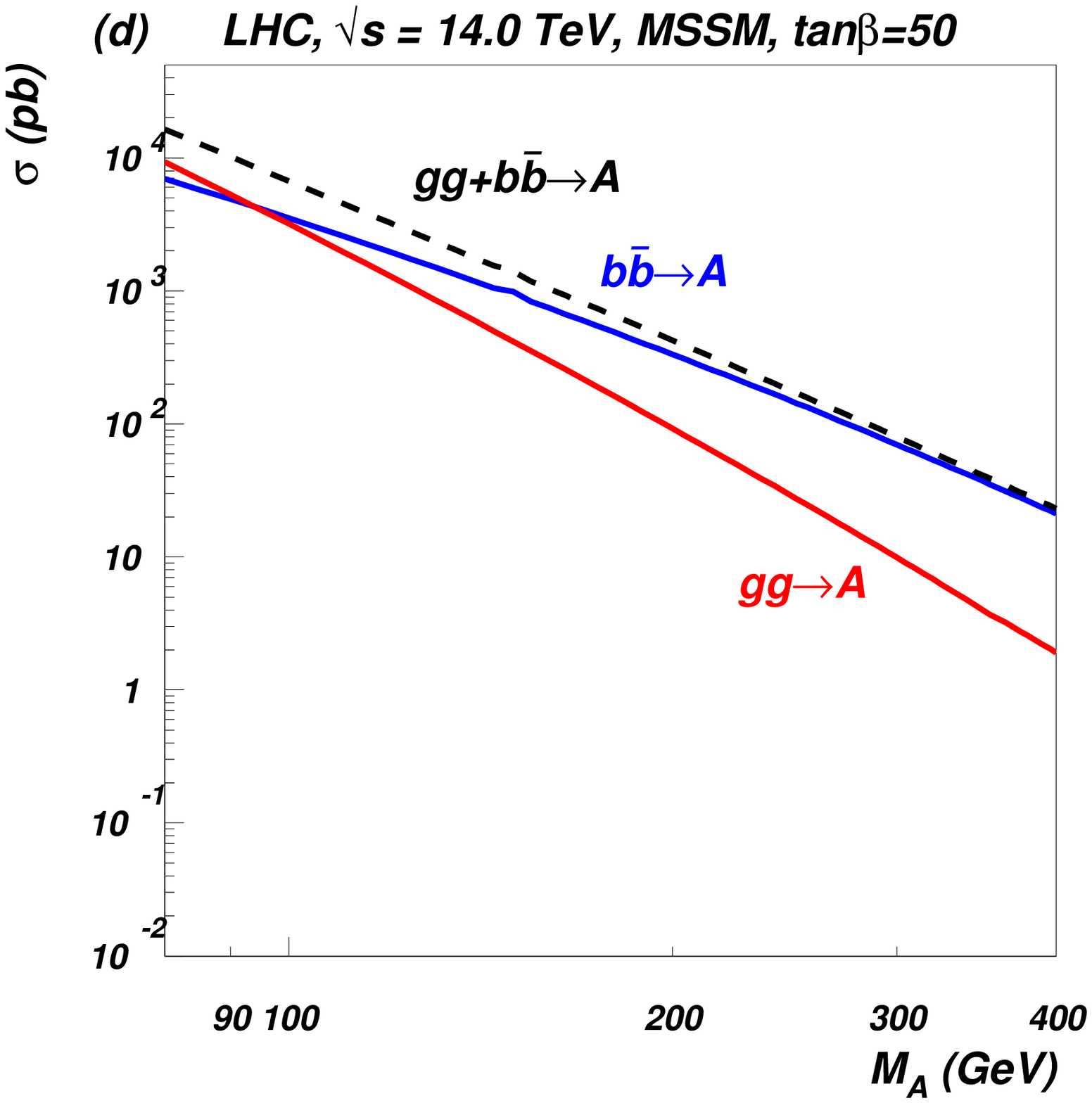}
\caption[mssm-tot-prod]{
NLO cross sections for Higgs production via the $b\bar{b}\to {\cal H}$ and $gg\to {\cal H}$
processes (as well as their sum) for (a) the SM Higgs, and (b)-(d) the Supersymmetric  axial Higgs
boson with $\tan\beta=10,30$ and 50, respectively, at the LHC.}
\label{fig:mssm-tot-prod-lhc}
\end{figure}

\begin{figure}
\includegraphics[width=7cm]{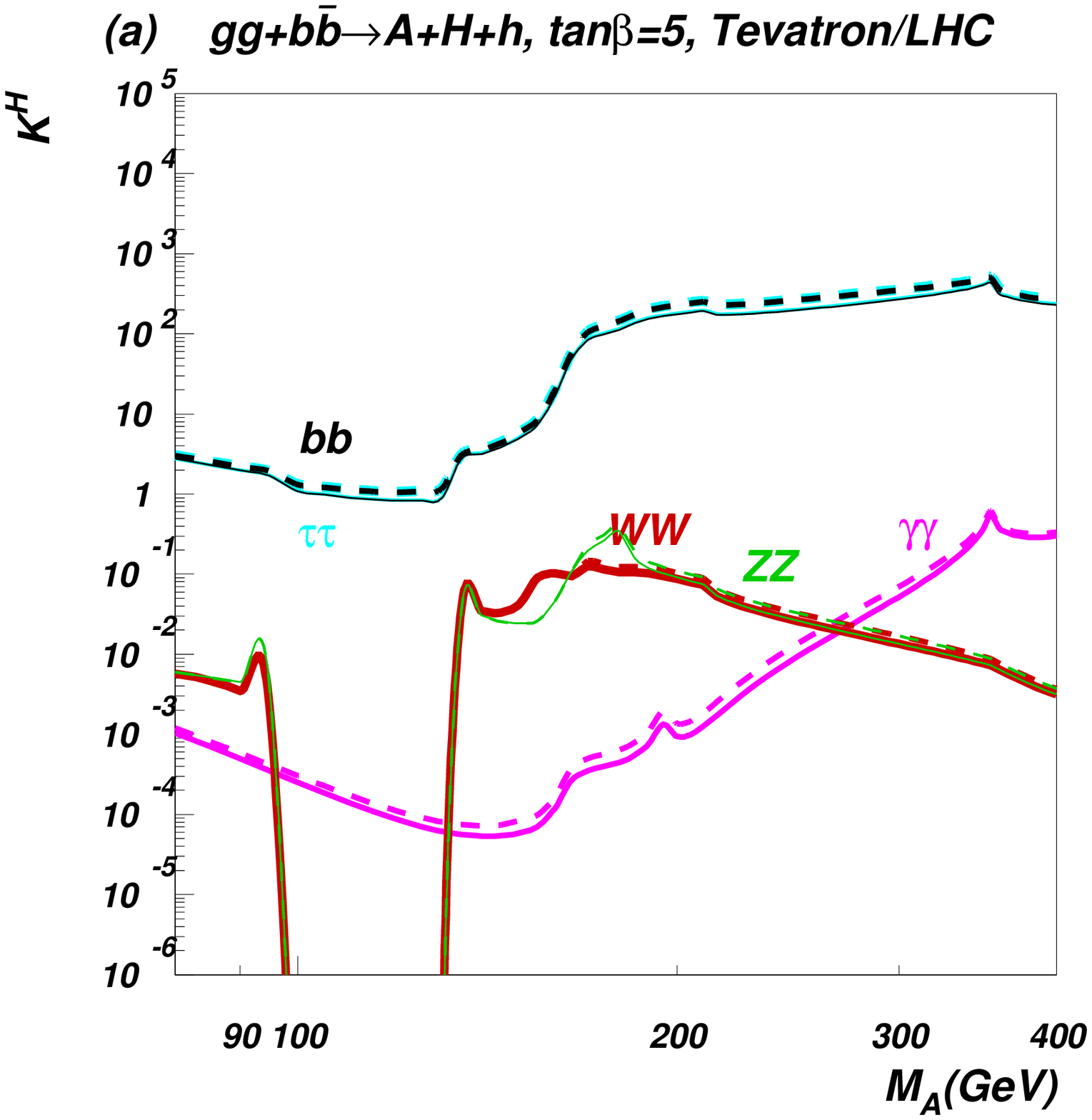}
\includegraphics[width=7cm]{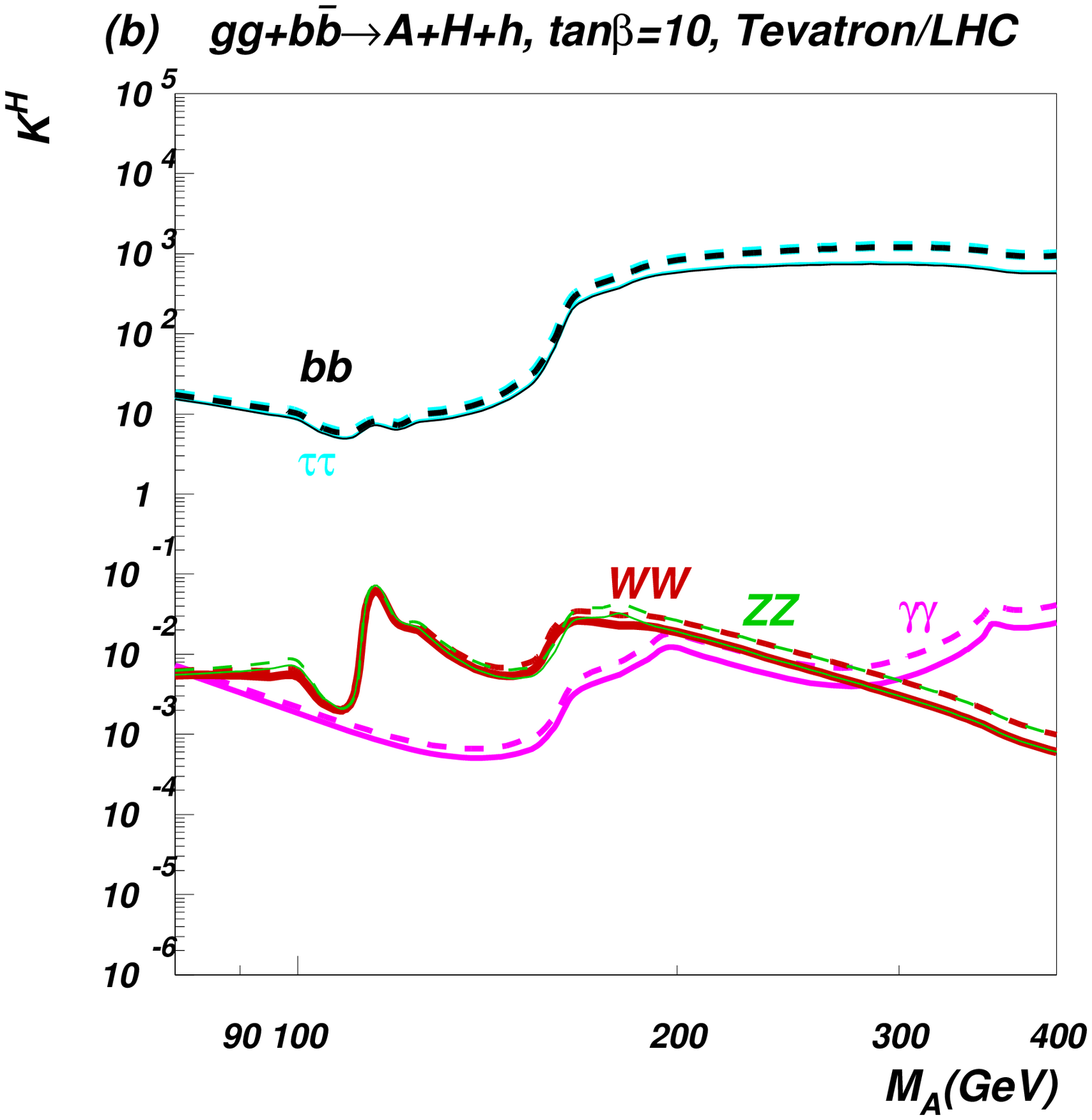}\\
\includegraphics[width=7cm]{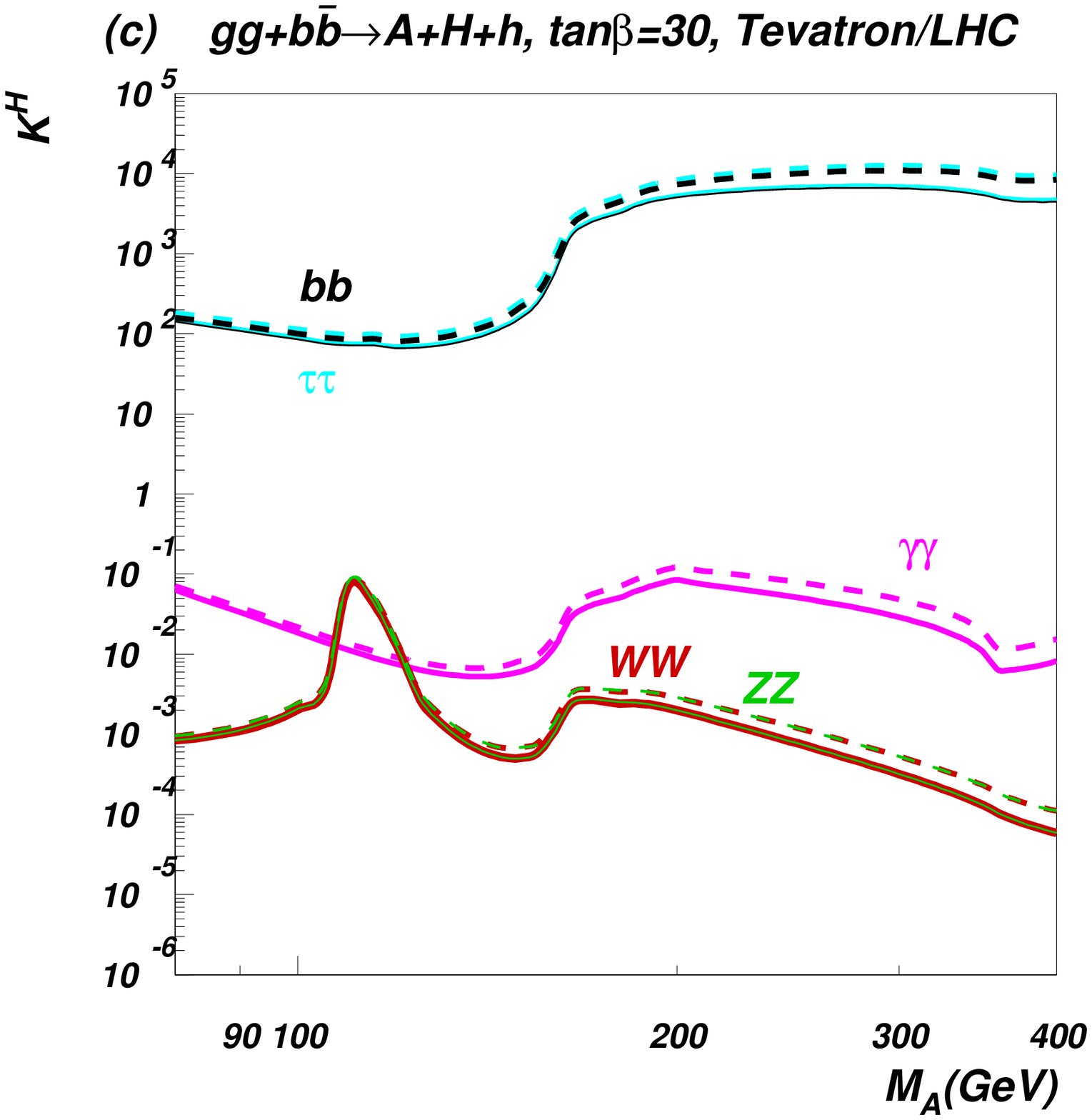}
\includegraphics[width=7cm]{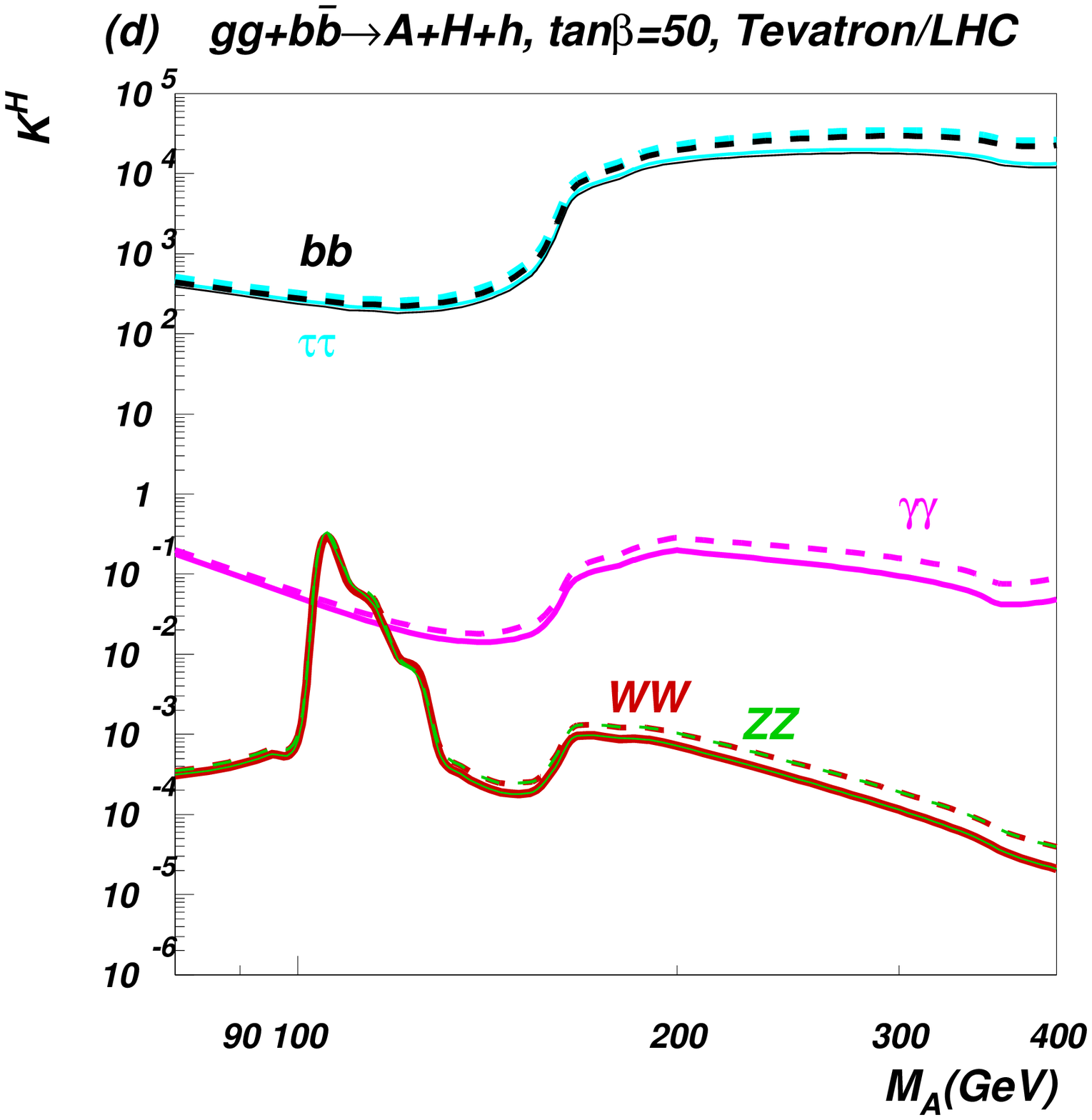}
\caption[fig:kthtot]{
Enhancement factor $\kappa^{\cal H}_{tot/xx}$ for final states $xx = b\bar{b},\ \tau^+\tau^-,\ WW,\  ZZ,\
\gamma\gamma$ when both $gg\to {\cal H}$ and $b\bar{b}\to {\cal H}$ are included and the signals of all
three MSSM Higgs states are combined.  Frames (a), (b), (c), and (d) correspond to $\tan\beta=5,10,30$
and 50, respectively,
at the Tevatron (solid lines) and at the LHC (dashed lines).
}
\label{fig:kthtot}
\end{figure}

\subsubsection{Enhancement Factors and Cross-sections}

Figure~\ref{fig:mssm-tot-prod} (\ref{fig:mssm-tot-prod-lhc}) 
presents NLO cross sections at the Tevatron (LHC).
For $b\bar{b}\to {\cal H}$ we are using the 
code of Ref.~\cite{Balazs:1998sb}, \footnote{
Note that  $b\bar{b}\to {\cal H}$
has been recently calculated at NNLO
in \protect\cite{Harlander:2003ai}.}
while for $gg\to {\cal H}$ we use HIGLU \cite{Spira:1996if}
and HDECAY \cite{Djouadi:1997yw} .\footnote{
Specifically, we use the HIGLU package to calculate 
the $gg\to h_{sm}$ cross section. We then use
the ratio of the Higgs decay widths from HDECAY
(which includes a more complete set of one-loop MSSM corrections
 than HIGLU)
to get the MSSM $gg\to {\cal H}$ cross section:
$\sigma^{MSSM}= 
\sigma^{SM}\times \Gamma({\cal H}\to gg)/ \Gamma(h_{SM}\to gg)$.
}
Frame (a) shows production of $h_{SM}$; frames (b)-(d) show
production of  the MSSM axial Higgs for several values of $\tan\beta$.  One can see
that in the MSSM  the contribution from  $b\bar{b}\to {\cal H}$ becomes  important even
for moderate values of $\tan\beta\sim 10$. For $M_{\cal H}<110-115$~GeV the contribution
from  $gg\to {\cal H}$ process is a bit bigger than that from $b\bar{b}\to {\cal H}$,  while for
$M_{\cal H}>115$~GeV $b$-quark-initiated production begins to outweigh gluon-initiated
production.

\begin{figure}[tb]
\includegraphics[width=7cm]{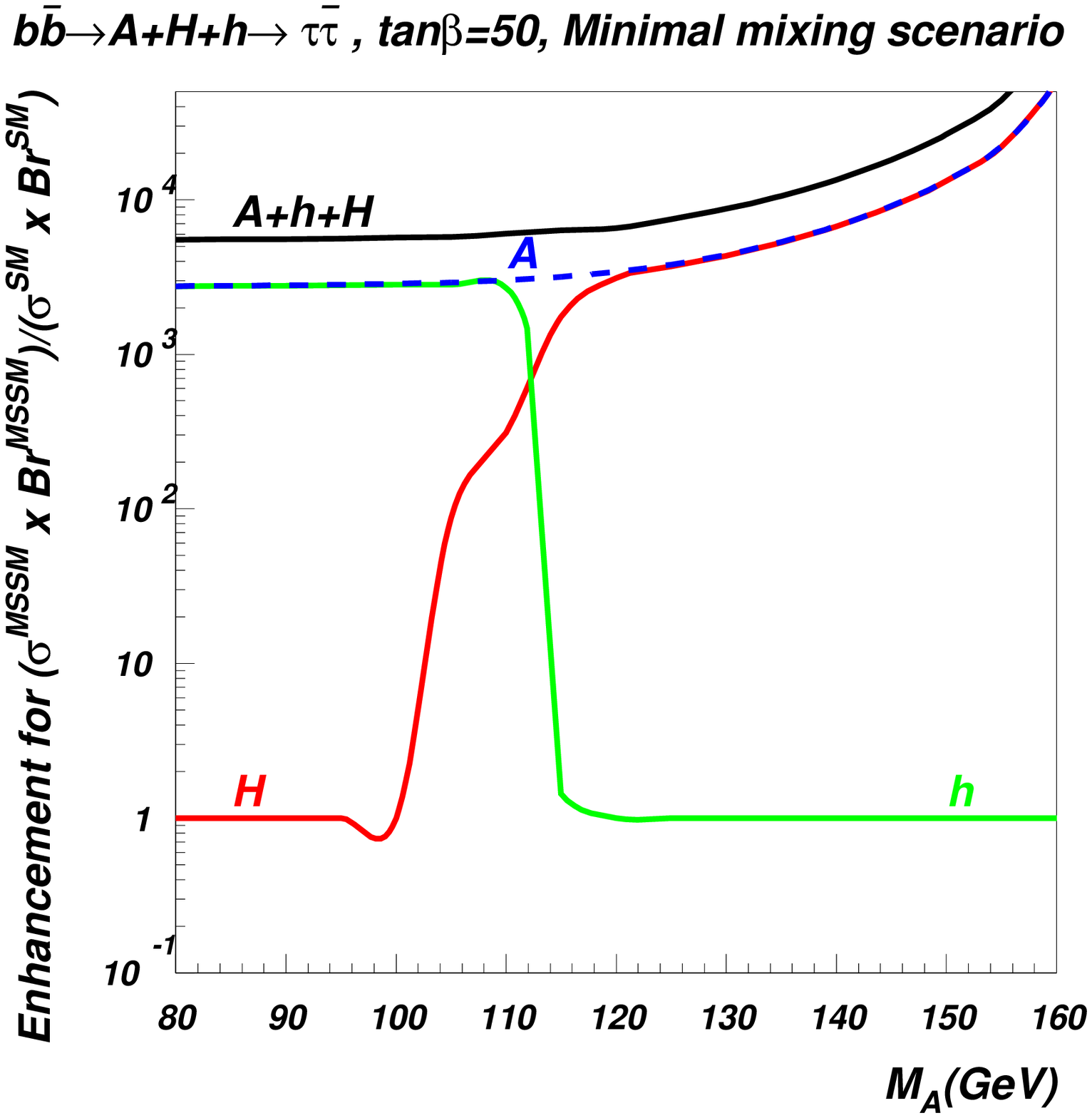}
\includegraphics[width=7cm]{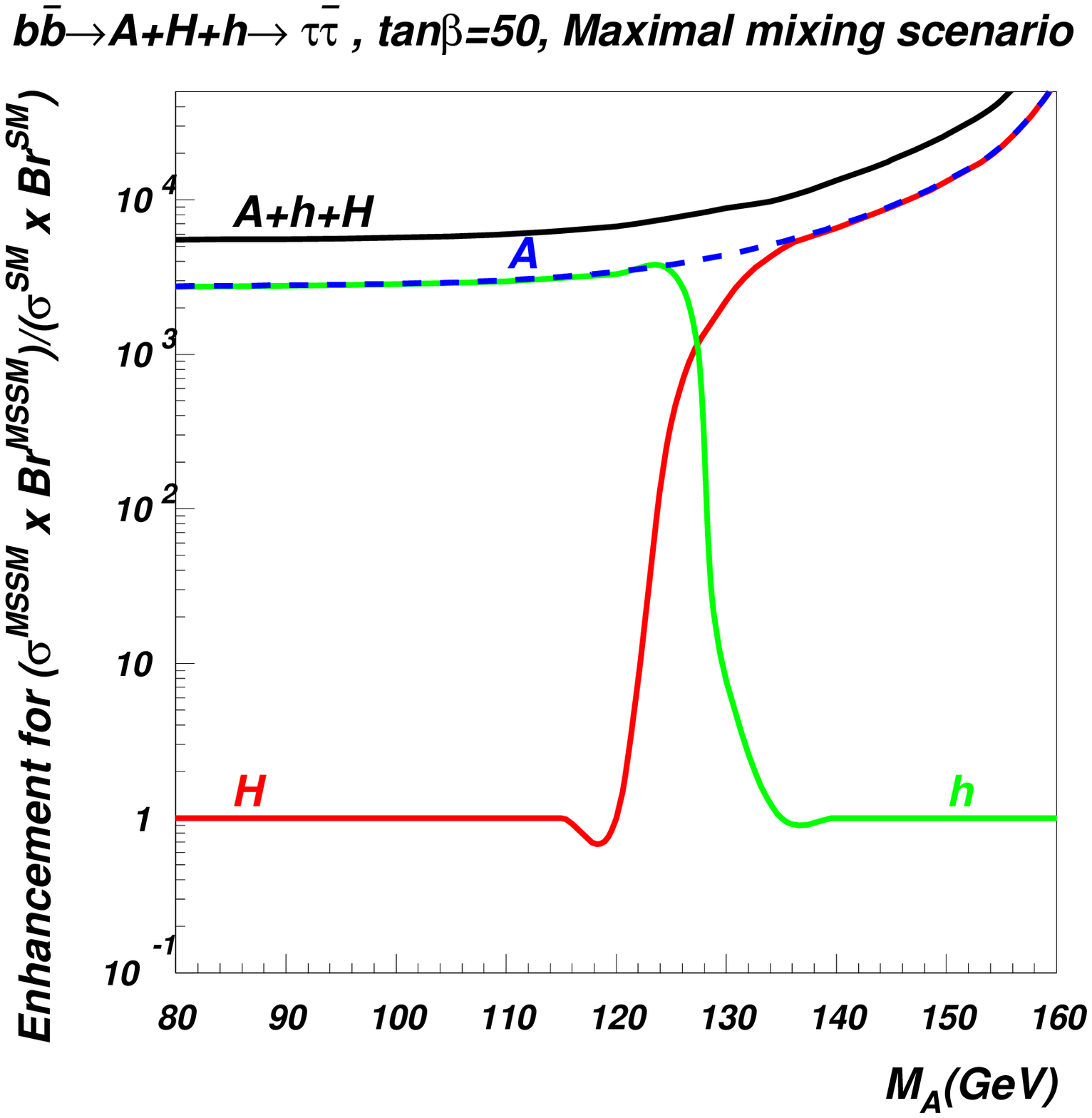}
\caption[fig:min-max]{
Enhancement factor $\kappa^{\cal H}_{bb/\tau\tau}$ (for $b\bar{b}\to H+h+A\to\tau^+{\tau^-}$)
for the minimal (left) and maximal (right) mixing scenarios.
In both scenarios $\kappa^{\cal H}_{bb/\tau\tau}$ is the same 
(within at most a few percent) for fixed $M_A$, $\mu$ and $M_{SUSY}$ parameters:
$\kappa^{\cal H}_{bb/\tau\tau}$ is independent of the collider energy and
essentially independent of the $X_t$ variable which describes top squark mixing.
}
\label{fig:min-max}
\end{figure}

\begin{table}[hbt]
\begin{center}
\let\normalsize=\captsize
\caption {Enhancement factors for a 130 GeV MSSM CP-odd Higgs at the  
Tevatron and LHC, compared to production and decay of a SM Higgs Boson of the  
same mass. The $b$-quark annihilation channel has been combined with
the gluon fusion channel, as described in the text. 
The rightmost column  
shows the cross-section (pb) for NLO \protect{$p\bar{p}/pp \to {\cal H} \to xx$} at  
Tevatron Run II/LHC; for $b\bar{b}$ and $\tau^+\tau^-$ channels 
the production of the $A$ is summed with
that of the $h$ or $H$ if the 
mass gap $\Delta M_A = |M_A-M_h(M_H)|$ is less than $0.3\sqrt{M_A}$. }
\vskip.5pc
\small
\renewcommand\tabcolsep{9pt}
\begin{tabular}{|c|c||c||c||c||c|}
\hline
\multicolumn{6}{|c|}{\bf Tevatron}\\
\hline
Model 			& Decay mode 		&  $\kappa^{\cal H}_{prod}$ 	& $\kappa^A_{dec}$ 	&  $\kappa^{\cal H}_{tot/xx}$ 	& Cross Section  \\\hline
			&$b \overline{b}$ 	& 0.5			& 1.72 			& 0.88			& 0.23 pb		\\
$\tan\beta=5$		&$ \tau^+ \tau^-$ 	& 0.5			&  1.75 		& 0.89			& 0.025 pb		\\
			&$ \gamma \gamma $ 	& 0.5			& $1.2\times 10^{-4}$ 	&$6.2\times 10^{-5}$	& $ 7.0 \times 10^{-8} $ pb \\
\hline\hline
			&$b \overline{b}$ 	& 4.9 			& 1.72  		& 8.5   		& 2.3 pb		\\
$\tan\beta=10$		&$ \tau^+ \tau^-$ 	& 4.9 			&  1.76 		& 8.8   		& 0.24  pb		\\
           		&$ \gamma \gamma $ 	& 4.9 			& $2.4\times 10^{-4}$	&$5.0\times 10^{-3}$	& $ 5.7 \times 10^{-6} $\\
\hline\hline
			&$b \overline{b}$ 	& 42.  			& 1.71  		& 72  			& 19  pb		\\
$\tan\beta=30$		&$ \tau^+ \tau^-$ 	& 42.  			&  1.82 		& 77  			&  2.1  pb		\\
			&$ \gamma \gamma $ 	& 42.  			& $2.7\times 10^{-4}$	&$8.4\times 10^{-3}$	& $ 9.6 \times 10^{-6} $\\
\hline\hline
			&$b \overline{b}$ 	& 115 			& 1.70  		&196    		& 52 pb		\\
$\tan\beta=50$		&$ \tau^+ \tau^-$ 	& 115 			&  1.88 		& 217   		& 6.0 pb		\\
			&$ \gamma \gamma $ 	& 115 			& $2.6\times 10^{-4}$	&$2.6\times 10^{-2}$	& $ 3.0 \times 10^{-5}$	 \\
\hline
\hline
\multicolumn{6}{|c|}{\bf LHC}\\
\hline
Model 			& Decay mode 		&  $\kappa^{\cal H}_{prod}$ 	& $\kappa^A_{dec}$ 	&  $\kappa^{\cal H}_{tot/xx}$ 	& Cross Section  \\\hline
			&$b \overline{b}$ 	& 0.67			&  1.72 		& 1.15			& 19.1 pb		\\
$\tan\beta=5$		&$ \tau^+ \tau^-$ 	& 0.67			&  1.75 		& 1.17			& 2.01 pb		\\
			&$ \gamma \gamma $ 	& 0.67			& $1.2\times 10^{-4}$ 	&$8.1\times 10^{-5}$	& $ 5.7 \times 10^{-6} $ pb \\
\hline\hline
			&$b \overline{b}$ 	& 6.1 			& 1.72  		& 10.5   		& 173 pb		\\
$\tan\beta=10$		&$ \tau^+ \tau^-$ 	& 6.1 			&  1.76 		& 10.8   		& 18.6  pb		\\
           		&$ \gamma \gamma $ 	& 6.1 			& $2.4\times 10^{-4}$	&$5.9\times 10^{-3}$	& $ 4.2 \times 10^{-4} $\\
\hline\hline
			&$b \overline{b}$ 	& 52.9  		& 1.71  		& 90  			& 1500  pb		\\
$\tan\beta=30$		&$ \tau^+ \tau^-$ 	& 52.9  		&  1.82 		& 96  			&  166  pb		\\
			&$ \gamma \gamma $ 	& 52.9  		& $2.7\times 10^{-4}$	&$1.1\times 10^{-2}$	& $ 7.5 \times 10^{-4} $\\
\hline\hline
			&$b \overline{b}$ 	& 144 			& 1.70  		&246    		& 4078 pb		\\
$\tan\beta=50$		&$ \tau^+ \tau^-$ 	& 144 			&  1.88 		&272   			& 467 pb		\\
			&$ \gamma \gamma $ 	& 144 			& $2.6\times 10^{-4}$	&$3.2\times 10^{-2}$	& $ 2.3 \times 10^{-3}$	 \\
\hline
\end{tabular}
\end{center}
\end{table}

Using the Higgs branching fractions from above with these NLO cross sections for
$gg\to H$ and $b\bar{b}\to H$ allows us to derive $\kappa_{total/xx}^{\cal H}$, as
presented in Fig.~\ref{fig:kthtot} for the Tevatron and LHC. 
Several comments  are in order. In  Fig.~\ref{fig:kthtot}(a) one can see a
gap in enhancement factor for $WW$ and $ZZ$ final states at $\tan\beta=5$ for
$M_A$ between 90-130 GeV. This is related to our procedure of combining signals
from $(A,h,H)$ bosons.  The $A$-boson does not couple to $WW$  or $ZZ$, while 
the mass gap between $h$ and $H$ is too big at low values of $\tan\beta$ to satisfy our
combination criterion ($|M_A-M_{H,h}|<0.3\sqrt{M_A/GeV}$), so  one cannot
define an  enhancement factor for  this parameter region. At higher values of
$\tan\beta$  there is no corresponding gap for $WW$ and $ZZ$ final states for $M_A$ between
90-130 GeV, however one can observe artificial peaks
for $M_A$ between 90-130 GeV which are again related to our combination procedure. In
addition, there are
several ``physical" kinks and peaks  in the enhancement factor for various Higgs
boson final states related to  $WW$, $ZZ$ and top-quark thresholds which can be
seen  for the respective  values of $M_A$. At very large  values of $\tan\beta$
the top-quark threshold effect for the $\gamma\gamma$ enhancement factor is almost gone
because the b-quark contribution dominates in the loop.

The enhancement factors and cross sections for a 130 GeV CP-odd
Higgs are listed, for various values of $\tan\beta$, in Table~2.
 From Table~2 one can see that the enhancement factors at the Tevatron
and LHC are very similar.  On the other hand, the values of the total rates at the LHC
are about two orders of magnitude higher than the corresponding rates at the Tevatron.
One should also notice that enhancements of  the $b\bar{b}$ and $\tau^+\tau^-$ signatures
are very similar and they rise swiftly by a factor of 200 as $\tan\beta$ increases from 5 to 50.
In contrast, the  $\gamma\gamma$ signature is always strongly suppressed! This particular feature 
of SUSY models, as we will see below, may be important for distinguishing
supersymmetric models from models with dynamical symmetry breaking.

It is important to note that combining the signal from the neutral Higgs bosons $h, A, H$ in the MSSM
turns out to make our results more broadly applicable across SUSY parameter space.   As discussed earlier,  Figure \ref{fig:degenspec}(left) reveals that at moderate-to-high values of $\tan\beta$ at least two of the
neutral Higgs bosons are degenerate in mass
\footnote{The degenerate pair is either $(h,A)$  for $M_A<M_A^0$
or $(H,A)$ for  $M_A>M_A^0$~, where the value of  $M_A^0$ is related  to the maximal mass of the light
Higgs as a function of $M_A$ with other SUSY parameters held fixed.  See Figure 1.}.  The value of the 
light Higgs mass  $M_h$ also depends on the degree of mixing between
the scalar partners of the top quark; this is parameterized by the variable $A_t$. For a given SUSY scale,
$M_S$, the mass $M_h$ takes its maximum value for $X_t\equiv A_t-\mu\cot\beta=\sqrt{6} M_S$  which corresponds to the
``maximal mixing case" while for $X_t\equiv A_t-\mu\cot\beta=0$ we have the ``minimal mixing case"
and $M_h$ takes on its minimum value.  What is interesting is that although the value of $M_h$ can differ
significantly in the  minimal  and  maximal mixing cases, the combined signal  from  all 3  Higgses  at
high $\tan\beta$ leads to the 
 nearly the same  (within at most few percent) enhancement factor, as shown  in Figure \ref{fig:min-max}.      
Combining the signals from  $A,h,H$ has the virtue of making  the enhancement factor independent of the
degree of top squark mixing (for fixed $M_A$, $\mu$ and $M_{S}$ and medium to high values of $\tan\beta$), 
which greatly reduces the
parameter-dependence of our results.

\subsection{Technicolor}

\subsubsection{PNGB Production via Gluon Fusion}

Single production of a technipion can occur through the axial-vector anomaly which
couples the technipion to pairs of gauge bosons. For an $SU(N_{TC})$ technicolor group
with technipion decay constant $F_P$, the anomalous coupling between the technipion and
a pair of gauge bosons is given, in direct analogy with the coupling of a QCD pion to
photons,\footnote{Note that the normalization used here differs
from that used in \cite{Lynch:2000hi} by a factor of 4.}
 by \cite{Dimopoulos:1980yf,Ellis:1980hz,Holdom:1981bg}
\begin{equation}
N_{TC} {\cal A}_{V_1 V_2} {g_1 g_2 \over 8 \pi^2 F_P} \epsilon_{\mu\nu\lambda\sigma}
k_1^\mu k_2^\nu \epsilon_1^\lambda \epsilon_2^\sigma
\label{eq:anom}
\end{equation}
where ${\cal A}_{V_1 V_2}$ is the anomaly factor, $g_i$ are the gauge boson couplings,
and the $k_i$ and $\epsilon_i$ are the four-momenta and polarizations of the gauge
bosons. The values of the anomaly factors for the lightest PNGB coupling to gluons are
given in Table 3 for each model.

\begin{table}[tb]
\begin{center}
\let\normalsize=\captsize
\caption{Anomaly Factors for the technicolor models under study
\protect{\cite{Lubicz:1995xi,Lynch:2000hi,Casalbuoni:1998fs,Lane:1999uh,Chivukula:1995dt} }}
\vskip.5pc
\small
\renewcommand\tabcolsep{9pt}
\begin{tabular}{|c||c||c||c||c|}\hline
 & 1) one-family & 2) variant one-family & 3) multiscale & 4) low-scale \\
\hline
${\cal A}_{gg}$ & $\frac{1}{\sqrt{3}}$ & $\frac{1}{\sqrt{6}}$ & $\sqrt{2}$ & $\frac{1}{\sqrt{3}}$\\
\hline
${\cal A}_{\gamma\gamma}$ & $- \frac{4}{3\sqrt{3}}$ & $\frac{16}{3 \sqrt{6}}$ & $\frac{4\sqrt{2}}{3}$ & $\frac{34}{9}$\\
\hline
\end{tabular}
\end{center}
\end{table}

The rate of single technipion production in this channel is proportional to the decay width to gluons.  In
the technicolor models, we have 
\begin {equation}
{\Gamma (P \rightarrow gg)} = { m_{P}^3 \over 8 \pi}   \left(\frac {\alpha_s N_{TC}{\cal A}_{gg}}{2 \pi F_P} \right)^2\ .
\end {equation}
while in the  SM, the expression looks like \cite{Gunion:1989we} 
\begin {equation}
{\Gamma (h \rightarrow gg)} ={m_{h}^3 \over 8 \pi}  \left(\frac{ \alpha_s}{3 \pi v}\right)^2~,
\end {equation}
in the heavy top-quark approximation.
Comparing a PNGB to a SM Higgs boson of the same mass, we find the enhancement in the gluon fusion production rate is 
\begin {equation}
\kappa_{gg\ prod} = \frac{ \Gamma (P \rightarrow gg)}{ \Gamma (h \rightarrow gg)} = \frac{9}{4} N_{TC}^2 {\cal A}_{gg}^2 \frac{v^2}{F_P^2}
\label{eq:kappagg}
\end {equation}
The main factors influencing $\kappa_{gg\ prod}$ for a fixed value of $N_{TC}$ are the anomalous coupling to gluons and the technipion decay constant.  The value of $\kappa_{gg\ prod}$ for each model (taking $N_{TC} = 4$) is given in Table 4.

\begin {table} [tb]
\begin{center}
\let\normalsize=\captsize
\caption{Calculated enhancement factors for production at the Tevatron and LHC of a 130 GeV technipion  via $gg$ alone, via
	 $b\bar{b}$ alone, and combined. Note that the small enhancement in the $b\bar{b}$ process slightly reduces the
	 total enhancement relative to that of $gg$ alone. In all cases, $N_{TC}=4$.}
\vskip.5pc
\small
\renewcommand\tabcolsep{9pt}
\begin{tabular}{|c||c||c||c||c|}\hline
& 1) one family & 2) variant one-family & 3) multiscale & 4) low scale \\
\hline
$ \kappa^P_{gg\ prod} $ & 48 & 6 & 1200 & 120 \\
\hline
$ \kappa^P_{bb\ prod} $ & 4  & 0.67 & 16 & 10 \\
\hline\hline
$ \kappa^P_{prod} $ & 47 & 5.9 & 1100 & 120 \\
\hline
\end{tabular}
\end{center}
\end{table}

\subsubsection{Production via $b\bar{b}$ annihilation}

The PNGBs couple to b-quarks courtesy of the extended technicolor interactions
\cite{Dimopoulos:1979es,Eichten:1979ah} responsible for producing masses for
the ordinary quarks and leptons.    The extended technicolor group (of which
$SU(N_{TC})$ is an unbroken subgroup) includes gauge bosons that couple to both
ordinary and technicolored fermions so that the ordinary fermions can interact
with the technicondensates that break the electroweak symmetry.

The rate of technipion production via $b\bar{b}$ annihilation is proportional
to $\Gamma(P\to b\bar{b})$.  In general, the expression for the decay of a
technipion to fermions is

\begin {equation}
{ \Gamma (P \rightarrow f \overline{f})} = {N_C\, \lambda^2_f\, m^2_f\, m_P \over 8 \pi\, F^2_P}\,
\left(1 -  \frac{4m_f^2}{m_P^2}\right)^{\frac{s}{2}}
\label{eq:phasesp}
\end {equation}
where  $N_C$ is 3 for quarks and 1 for leptons.  The phase space exponent,
$s$, is 3 for scalars and 1 for pseudoscalars; the lightest PNGB in models
1 and 4 is a scalar, while in models 2 and 3 it is assumed to be a
pseudoscalar.  For the technipion masses considered here, the value of
the phase space factor in (\ref{eq:phasesp}) is so close to one that the
value of $s$ makes no practical difference.   The factor $\lambda_f$ is a
non-standard Yukawa coupling distinguishing leptons from quarks.  Model
2 has $\lambda_{quark} = \sqrt{\frac{2}{3}}$ and $\lambda_{lepton} =
\sqrt{6}$; model 3 also includes a similar factor, but with average
value 1; $\lambda_f=1$ in models 1 and 4.   Finally, it should be noted that model 2 assumes that the
lightest technipion is composed only of down-type fermions and cannot
decay to $c\bar{c}$; since this decay would usually have a small
branching ratio and $c\bar{c}$ is not a preferred final state for Higgs
searches, this has little impact. 

For comparison, the decay width of the SM Higgs into b-quarks is:
\begin {equation}
{ \Gamma (h \rightarrow b \overline{b})} = {3\,m^2_b\,m_h \over 8\pi\,v^2}
\left(1 - \frac {4m_b^2}{m_h^2}\right)^{\frac{3}{2}}
\end {equation}
The production enhancement  for $b\bar{b}$ annihilation is (again assuming Higgs and technipion have the
same mass): %
\begin {equation}
\kappa_{bb\ prod} = \frac{ \Gamma (P \rightarrow b \overline{b})}
{ \Gamma (h \rightarrow b \overline{b})} = {\lambda^2_b\, v^2 \over F^2_P} 
\left(1 - \frac{4m_b^2}{m_h^2}\right)^{\frac{s-3}{2}}
\label{eq:kappabb}
\end {equation}
The value of $\kappa_{bb\ prod}$ (shown in Table 4) is controlled by the size of the technipion decay
constant.   

We see from Table 4 that $\kappa_{bb\ prod}$ is at least one order of magnitude smaller than $\kappa_{gg\
prod}$ in each model.   Taking the ratio of equations (\ref{eq:kappagg}) and (\ref{eq:kappabb})
\begin{equation}
\frac{\kappa_{gg\ prod}}{\kappa_{bb\ prod}} = \frac{9}{4} N_{TC}^2 {\cal A}_{gg}^2 \lambda_b^{-2}
\left(1 - \frac{4m_b^2}{m_h^2}\right)^{\frac{3-s}{2}}
\end{equation}
we see that the larger size of $\kappa_{gg\ prod}$ is due to the factor of $N_{TC}^2$ coming from the fact
that gluons couple to a technipion via a techniquark loop.  The
extended technicolor (ETC) interactions coupling $b$-quarks to a
technipion have no such enhancement.

In addition, the production cross-section for a SM Higgs boson via $b\bar{b}$ annihilation is 2 to 3 orders of magnitude smaller than that for gluon
fusion at the Tevatron \cite{Carena:2000yx} and LHC \cite{Spira:1996if}.  With a smaller SM cross-section and a smaller enhancement factor, it is
clear that technipion production via $b\bar{b}$ annihilation is essentially  negligible  at these hadron colliders.  
Nonetheless, to be conservative, we include the 
$b\bar{b}$ production channel because it  tends to slightly  reduce the production enhancement factor.
  
Using the combined enhancement factor definition of eqn. (\ref{kappab}), and recalling
that $\kappa_{dec}$ is the same for both colliders,  
we find that the small differences due to the values of $R_{bb:gg}$  do not give a noticable
difference between the values of $\kappa^P_{total/xx}$ at the Tevatron and LHC; the production enhancement factors quoted in Table 4 apply to both
colliders. 

\subsubsection{Decays}

The decay width of a light technipion into gluons or fermion/anti-fermion pairs has been discussed above.
Since the technipions we are studying do not  decay to $W$  bosons and their decay to $Z$
bosons through the axial vector anomaly is negligible in the interesting mass range, 
the remaining possibility is a decay to photons. 
Again, this proceeds through the axial vector anomaly (cf. eqn. (\ref{eq:anom}))  and the anomaly factors
${\cal A}_{\gamma\gamma}$ are shown in Table 3.

We now calculate the technipion branching ratios from the above information, taking $N_{TC} = 4$.  The
values are essentially independent of the size of $M_P$ within the range 120 GeV - 160 GeV; the branching
fractions for $M_P = 130$ GeV are shown in Table 5. The branching ratios for the SM Higgs at NLO are given
for comparison; they were calculated using HDECAY \cite{Djouadi:1997yw}.  Note that, in contrast to the
technipions, a SM Higgs in this mass range already has a noticeable decay rate to off-shell vector
bosons.  
 
\begin{table}[tb]
\begin{center}
\let\normalsize=\captsize
\caption{Branching ratios of Technipions/Higgs of mass 130 GeV}
\vskip.5pc
\small
\renewcommand\tabcolsep{9pt}
\begin{tabular}{|c||c||c||c||c|||c|}\hline
Decay & 1) one family & 2) variant & 3) multiscale & 4) low scale & SM Higgs \\
Channel & &one family & & & \\
\hline
$b \overline{b}$ & 0.60 & 0.53 & 0.23  & 0.60 & 0.53 \\
$c \overline{c}$ & 0.05  & 0  & 0.03  & 0.05  & 0.02  \\
$ \tau^+ \tau^-$ & 0.03  & 0.25  & 0.01 & 0.03 & 0.05  \\
$gg$ & 0.32  & 0.21  & 0.73  & 0.32  & 0.07  \\
$ \gamma \gamma $ & $ 2.7 \times 10^{-4}  $ & $ 2.9 \times 10^{-3}  $ & $ 6.1 \times 10^{-4} $ & $ 6.4 \times 10^{-3}$  & $ 2.2 \times 10^{-3} $ \\
$ W^+ W^-$ & 0 & 0 & 0 & 0 & 0.29 \\
\hline
\end{tabular}
\end{center}
\end{table}

Comparing the technicolor and SM branching ratios in Table 5, 
we see immediately that all decay enhancements, 
except to the $gg$ mode, are generally of order one
and therefore much smaller than the production enhancements.   Decays to
$b\bar{b}$ are slightly enhanced, if at all.  
Decays to $c\bar{c}$ are enhanced in our tree-level
calculations -- but note that it is higher-order corrections that suppress this mode for the SM Higgs; in
any case, this is not a primary discovery channel.  
Decays to $\tau$ leptons  are slightly suppressed  in
general; again, the comparison of tree-level technicolor and loop-level SM Higgs calculations may be a
factor here.  Model 2 is an exception; its unusual Yukawa couplings yield a decay enhancement in the
$\tau^+\tau^-$ channel of order the technipion's (low) production enhancement.  In the $\gamma\gamma$ channel,
the decay enhancement strongly depends on the group-theoretical structure of the model, through the anomaly
factor.  Table 6 includes the decay enhancements $\kappa^P_{dec}$ for the most experimentally promising
search channels.  

\begin{table}[bt]
\begin{center}
\let\normalsize=\captsize
\caption {Enhancement Factors for 130 GeV technipions produced at the Tevatron and LHC,
compared to production and decay of a SM Higgs Boson of the same mass. The
slight suppression of \protect{$\kappa^P_{prod}$} due to the b-quark
annihilation channel has been included. The rightmost column shows the
cross-section (pb) for \protect{$p\bar{p}/pp \to P \to xx$} at Tevatron Run II/LHC. }
\vskip.5pc
\small
\renewcommand\tabcolsep{9pt}
\begin{tabular}{|c|c||c||c||c||c|}\hline
\multicolumn{6}{|c|}{\bf Tevatron}\\
\hline
Model & Decay mode &  $\kappa^P_{prod}$ & $\kappa^P_{dec}$ & $\kappa^P_{tot/xx}$ & Cross Section  \\
\hline
&$b \overline{b}$ & 47 & 1.1 & 52 & 14 pb \\
1) one family&$ \tau^+ \tau^-$ & 47 & 0.6 & 28 & 0.77 pb\\
&$ \gamma \gamma $ & 47 & 0.12 & 5.6 & $ 6.4 \times 10^{-3} $ pb \\
\hline
\hline
&$b \overline{b}$ & 5.9 & 1 & 5.9 & 1.8 pb\\
2) variant&$ \tau^+ \tau^-$ & 5.9 & 5 & 30 & 0.84 pb\\
one family&$ \gamma \gamma $ & 5.9 & 1.3 & 7.7 & $ 8.7 \times 10^{-3}$ pb \\
\hline
\hline
&$b \overline{b}$ & 1100 & 0.43 & 470 & 130 pb\\
3) multiscale &$ \tau^+ \tau^-$ & 1100 & 0.2 & 220 & 6.1 pb\\
&$ \gamma \gamma $ & 1100 & 0.27 & 300 & 0.34 pb \\
\hline
\hline
&$b \overline{b}$ & 120 & 1.1 & 130 & 36 pb\\
4) low scale&$ \tau^+ \tau^-$ & 120 & 0.6 & 72 & 2 pb\\
&$ \gamma \gamma $ & 120 & 2.9 & 350 & 0.4 pb \\
\hline
\hline
\multicolumn{6}{|c|}{\bf LHC}\\
\hline
\hline
Model & Decay mode &  $\kappa^P_{prod}$ & $\kappa^P_{dec}$ & $\kappa^P_{tot/xx}$ & Cross Section  \\
\hline
&$b \overline{b}$ & 47 & 1.1 & 52 & 890 pb \\
1) one family&$ \tau^+ \tau^-$ & 47 & 0.6 & 28 & 48 pb\\
&$ \gamma \gamma $ & 47 & 0.12 & 5.6 & 0.4 pb \\
\hline
\hline
&$b \overline{b}$ & 5.9 & 1 & 5.9 & 100 pb\\
2) variant&$ \tau^+ \tau^-$ & 5.9 & 5 & 30 & 52 pb\\
one family&$ \gamma \gamma $ & 5.9 & 1.3 & 7.7 & 0.55 pb \\
\hline
\hline
&$b \overline{b}$ & 1100 & 0.43 & 473 & 8000 pb\\
3) multiscale &$ \tau^+ \tau^-$ & 1100 & 0.2 & 220 & 380 pb\\
&$ \gamma \gamma $ & 1100 & 0.27 & 300 & 22 pb \\
\hline
\hline
&$b \overline{b}$ & 120 & 1.1 & 130 & 2200 pb\\
4) low scale&$ \tau^+ \tau^-$ & 120 & 0.6 & 72 & 120 pb\\
&$ \gamma \gamma $ & 120 & 2.9 & 350 & 25 pb \\
\hline
\end{tabular}
\end{center}
\end{table}

\subsubsection{Enhancement Factors and Cross-Sections}

Our results for the Tevatron Run II and LHC production enhancements (including both $gg$
fusion and $b\bar{b}$ annihilation), decay enhancements, and overall enhancements
of each technicolor model relative to the SM are shown in Table~6 for a technipion
or Higgs mass of 130 GeV.  Multiplying $\kappa^P_{tot/xx}$ by the cross-section for
SM Higgs production via gluon fusion \cite{Spira:1996if}
yields an approximate technipion production cross-section, as shown in the right-most column of Table 6.   

In each technicolor model, the main enhancement of the possible technipion signal
relative to that of an SM Higgs arises at production, making the size of the
technipion decay constant the most critical factor in determining the degree of
enhancement for fixed $N_{TC}$.

Each decay enhancement is in general of order 1, making it significantly smaller
than the typical production enhancement. In model 3 where the decay ``enhancement"
is actually a suppression, the decay factor is 3 orders of magnitude smaller than
the production enhancement.   We find that $P\to b\bar{b}$ is very similar to
$h_{SM}\to b\bar{b}$.  The decay $P\to \tau^+\tau^-$
generally has a suppressed rate relative to SM expectations; again, this may relate
to comparing leading technicolor and NLO SM results.   An exception is model 2,
where the special structure of the Yukawa coupling leads to a $\tau^+\tau^-$ decay
enhancement of the same order as the production enhancement.   The $P \to
\gamma\gamma$ decay enhancement factor depends strongly on the group-theoretic
structure of the model through the anomaly factor, ranging from a distinct
enhancement in model 4 to a factor-of-10 suppression in model 1.


\section{Interpretation}

We are ready to put our results in context.  The large QCD background for $q\bar{q}$ states of any flavor makes the tau-lepton-pair and di-photon final states the most promising for exclusion or discovery of the Higgs-like states of the MSSM or technicolor.  We now illustrate how the size of the enhancement factors for these two final states vary over the parameter spaces of these theories at the Tevatron and LHC.  We use this information to display the likely reach of each experiment in each of these standard Higgs search channels.  
Then, we compare the signatures of the MSSM Higgs bosons and the various technipions to see how one might tell these states apart from one another.

\subsection{Visibility of MSSM Higgs Bosons}

\begin{figure}
\includegraphics[width=7cm]{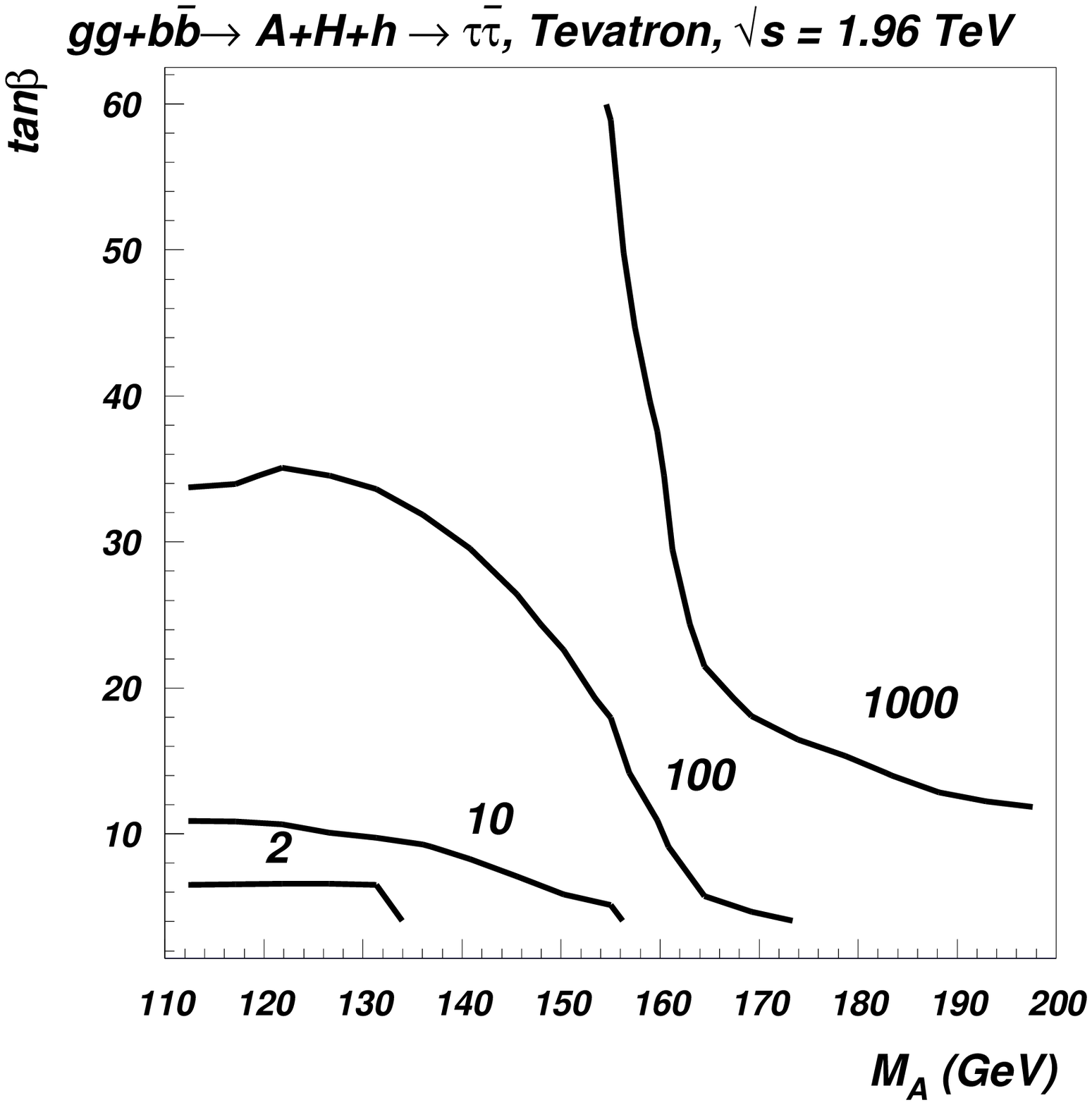}
\includegraphics[width=7cm]{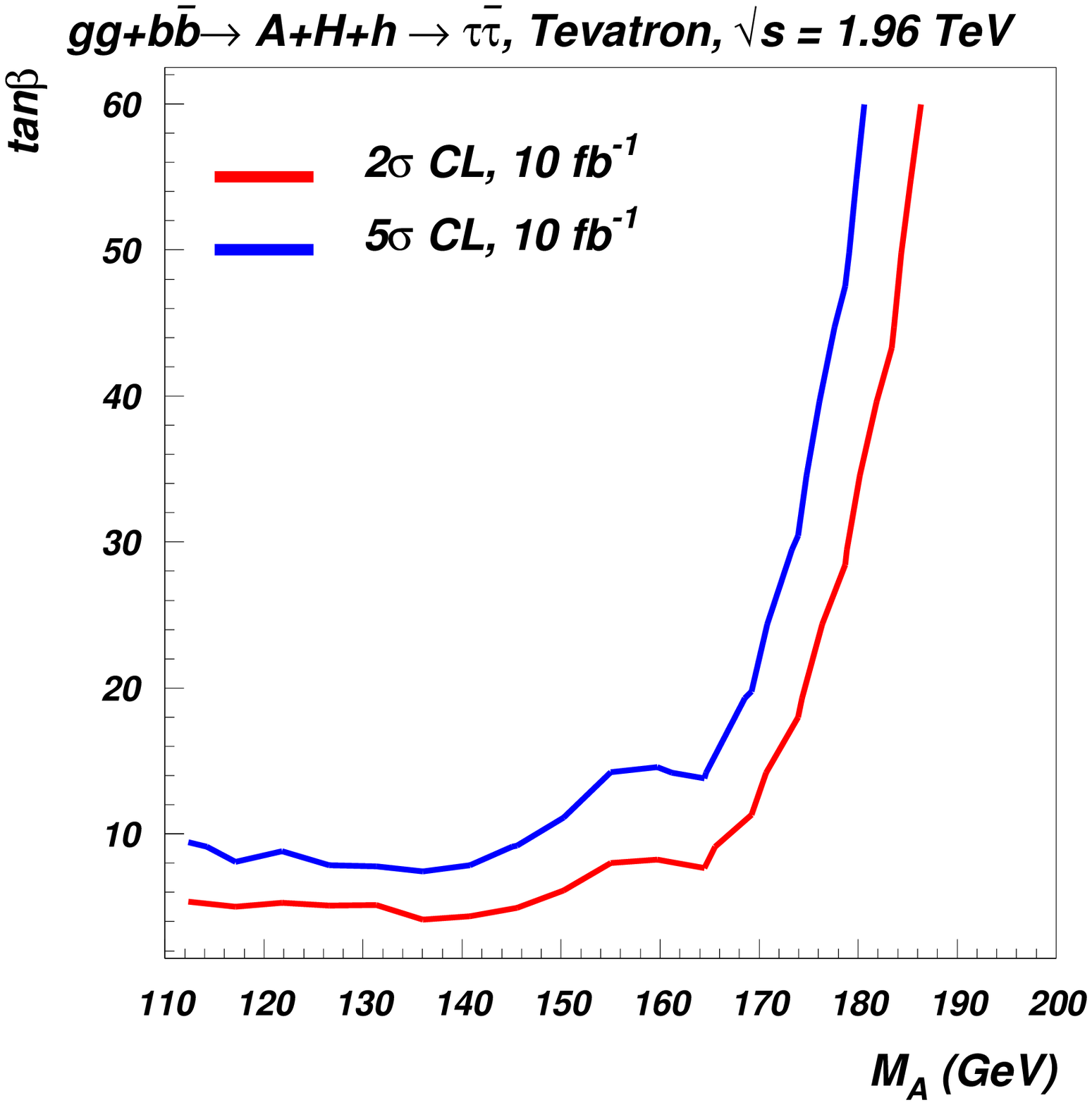}
\caption[mssm-plots]{
Results for  \protect{$gg + b\bar{b} \to h + H + A \to \tau^+\tau^-$} at Tevatron
Run II.   Left frame: Selected contours of given enhancement factor values
\protect{$\kappa^{\cal H}_{total/\tau\tau}$} in the MSSM.  Right frame: 
Predicted Tevatron reach, based on the $h_{SM} \to \tau^+ \tau^-$ studies of \cite{Belyaev:2002zz},
 in the MSSM parameter space.  Lower contour is a
\protect{$2\sigma$} exclusion contour; upper contour is a \protect{$5\sigma$}
discovery contour.}
\label{fig:mssm-contours}
\end{figure}
\begin{figure}
\includegraphics[width=7cm]{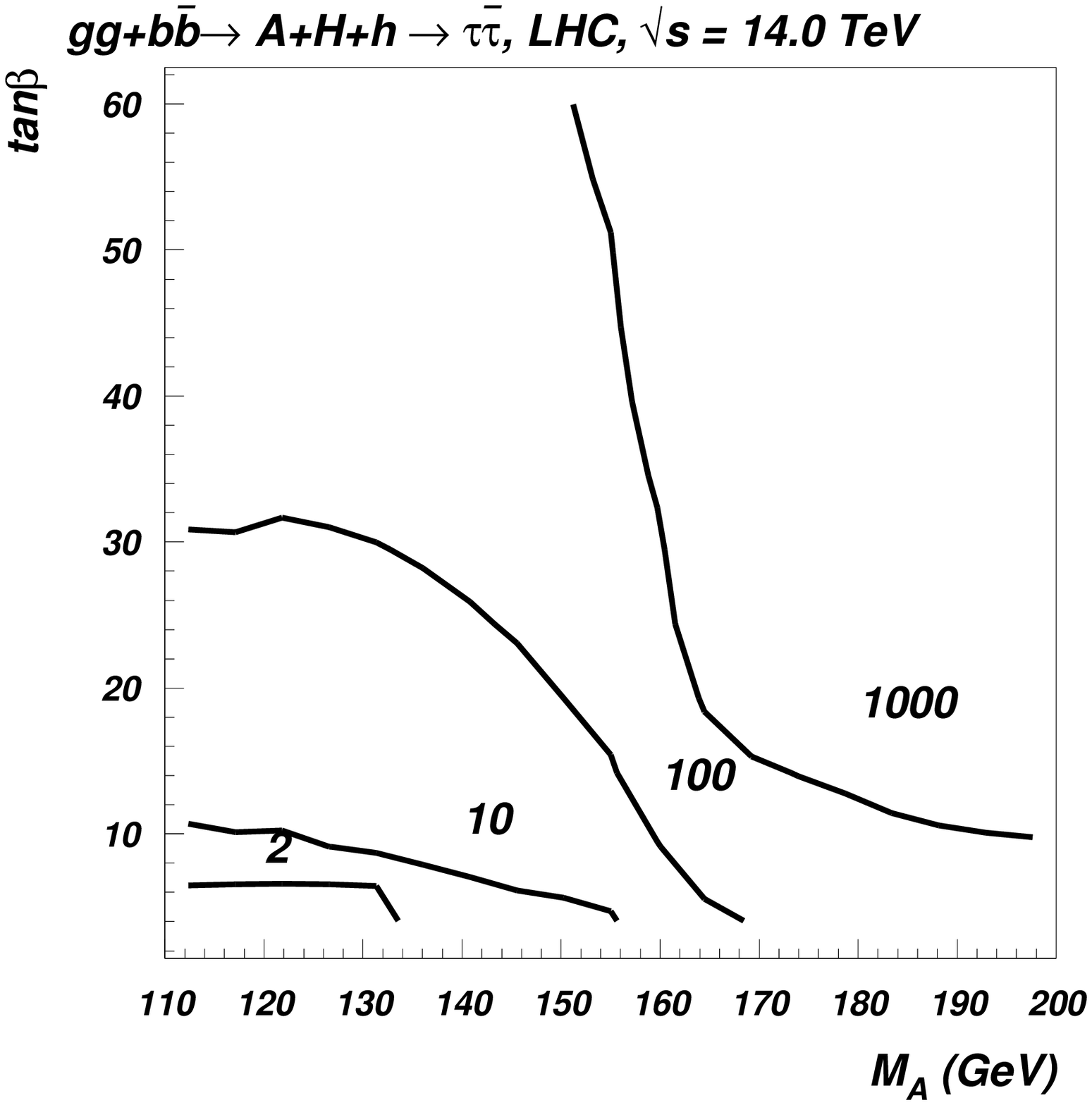}
\includegraphics[width=7cm]{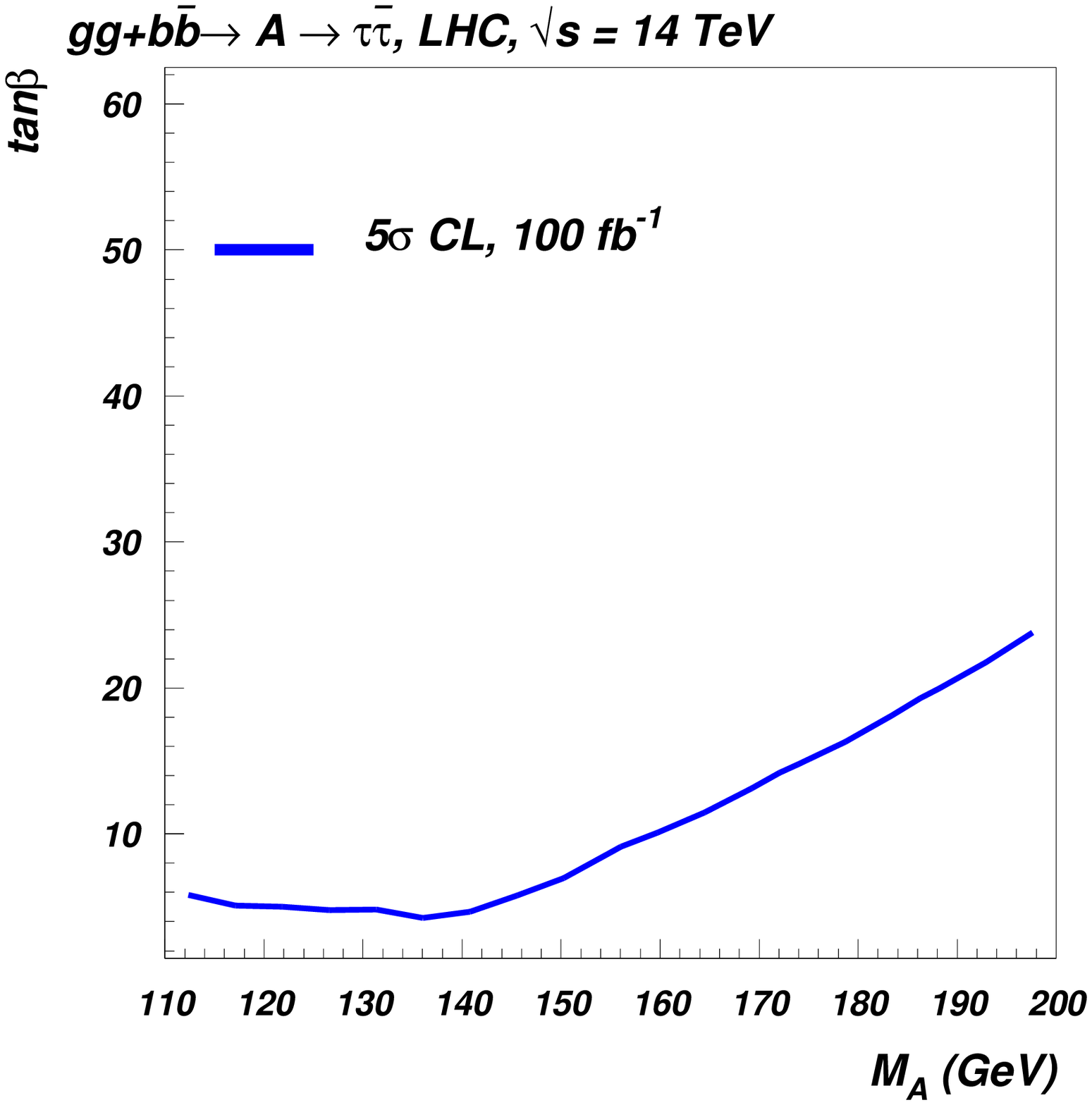}
\caption[mssm-plots]{
Results for  \protect{$gg + b\bar{b} \to h + H + A \to \tau^+\tau^-$} at LHC.  
Left frame: Selected contours of given enhancement factor values
\protect{$\kappa^{\cal H}_{total/\tau\tau}$} in the MSSM.  Right frame: 
Predicted LHC reach, based on the $h_{SM} \to \tau^+\tau^-$ studies of  \cite{Cavalli:2002vs},
 in the MSSM parameter space.}
\label{fig:mssm-contours-lhc}
\end{figure}

The left-hand frame of Figure~\ref{fig:mssm-contours}  displays contours of enhancement factors of 2,
10, 100 and 1000 for the process $gg + b\bar{b} \to h + A + H \to \tau^+ \tau^-$ in the MSSM at the
Tevatron.  
We see that the enhancement factors grow dramatically as either $\tan\beta$ or $M_A$ becomes
large.  These results are consistent with those of \cite{Belyaev:2002zz}.
The large  increase in the  enhancement factor
for large  values of  $M_A$ takes place
because the Standard Model  $Br[H\to\tau\tau(b\bar{b})]$ decreases sharply
when the  $WW$ and $ZZ$ decay channels open,  while in the MSSM the
$Br[A\to\tau\tau(b\bar{b})]$ in the high $\tan\beta$ regime
hardly  changes.

In the right-hand frame of the same figure, we summarize the Tevatron's ability to explore the MSSM 
parameter space (in terms of both a $2\sigma$ exclusion curve and a $5\sigma$ discovery curve) using the
process $gg + b\bar{b} \to h + A + H \to \tau^+ \tau^-$.  Translating the enhancement factors above into
this reach plot draws on the results of \cite{Belyaev:2002zz}.   
As the $M_A$ mass increases up to
about 140 GeV, the opening of the $W^+W^-$ decay channel drives the $\tau^+\tau^-$ branching fraction down, and
increases the $\tan\beta$  value required to make  Higgses  visible in the $\tau^+\tau^-$ channel.  
At still larger $M_A$, a very steep drop in the gluon luminosity (and the related $b$-quark luminosity) at
large $x$ reduces the phase space for ${\cal H}$ production.  Therefore for $M_A>$170 GeV, 
Higgs bosons would
only be visible at very high values of $\tan\beta$. 

Figure~\ref{fig:mssm-contours-lhc}
presents a qualitatively similar picture for LHC, based on the studies
of $h_{SM} \to \tau^+ \tau^-$ of \cite{Cavalli:2002vs}.
The main differences compared to the Tevatron are that the 
required value of  $\tan\beta$ at the LHC 
is lower for a given $M_A$ and it does not climb steeply for  $M_A>$170 GeV because there is much less phase space
suppression.

It is important to notice that both, Tevatron and LHC, 
could observe MSSM Higgs bosons in the $\tau^+\tau^-$ 
channel even for moderate values of $\tan\beta$
for $M_A\lesssim 200$~GeV, because of significant enhancement of this channel.
However the    $\gamma\gamma$ channel is so suppressed that
even the LHC will not be able to observe it in any point of the $M_A < 200$ GeV parameter space studied in this paper!
{\footnote{ In the decoupling limit with large values of $M_A$ and low values of $\tan\beta$,
the lightest MSSM Higgs could be dicovered in the $\gamma\gamma$ mode just like the 
SM model Higgs boson, see e.g. ref.~\cite{djouadi-higgs}}}

\subsection{Visibility of Technipions}

\begin{figure}
\includegraphics[width=7cm]{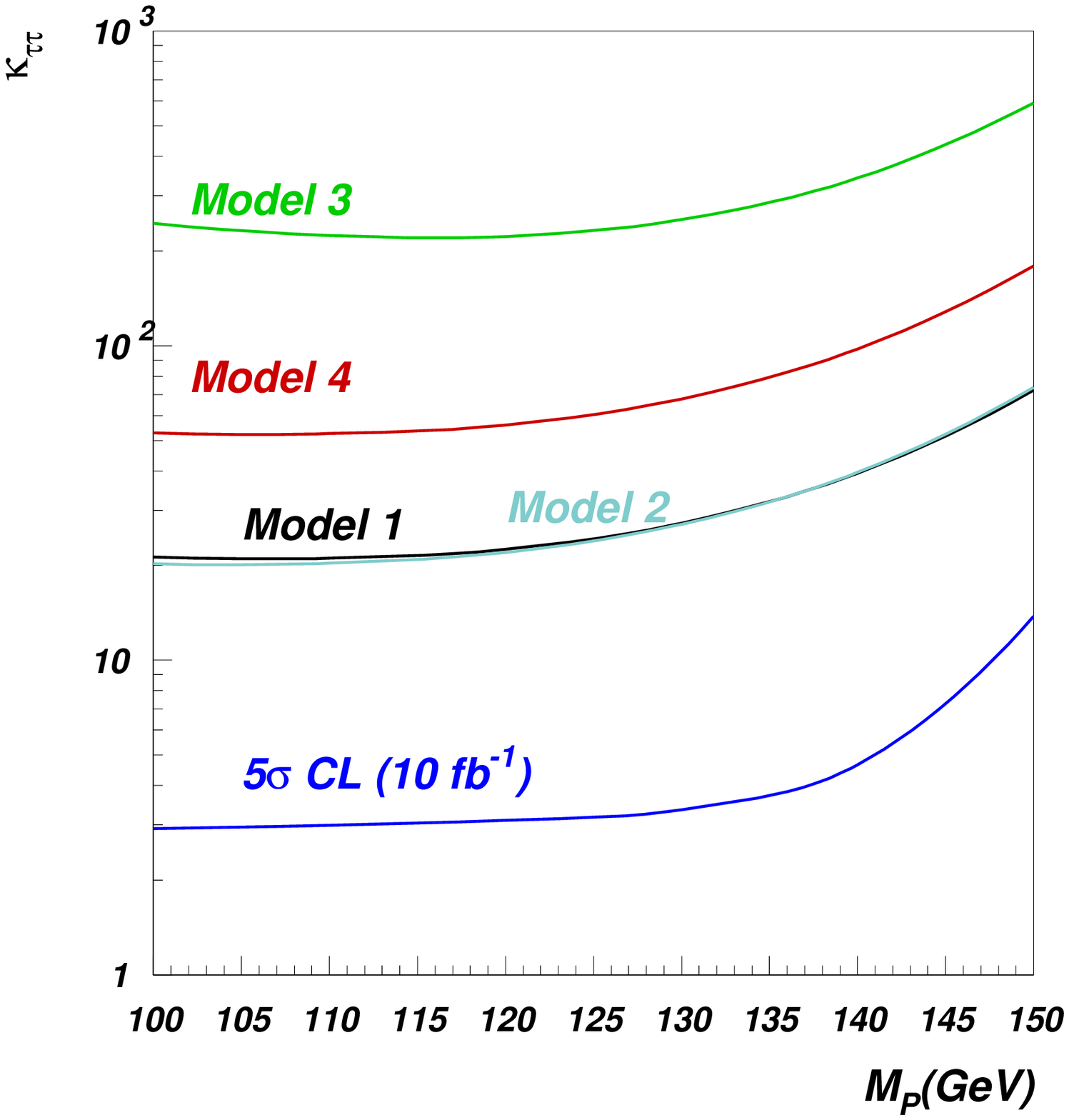}
\includegraphics[width=7cm]{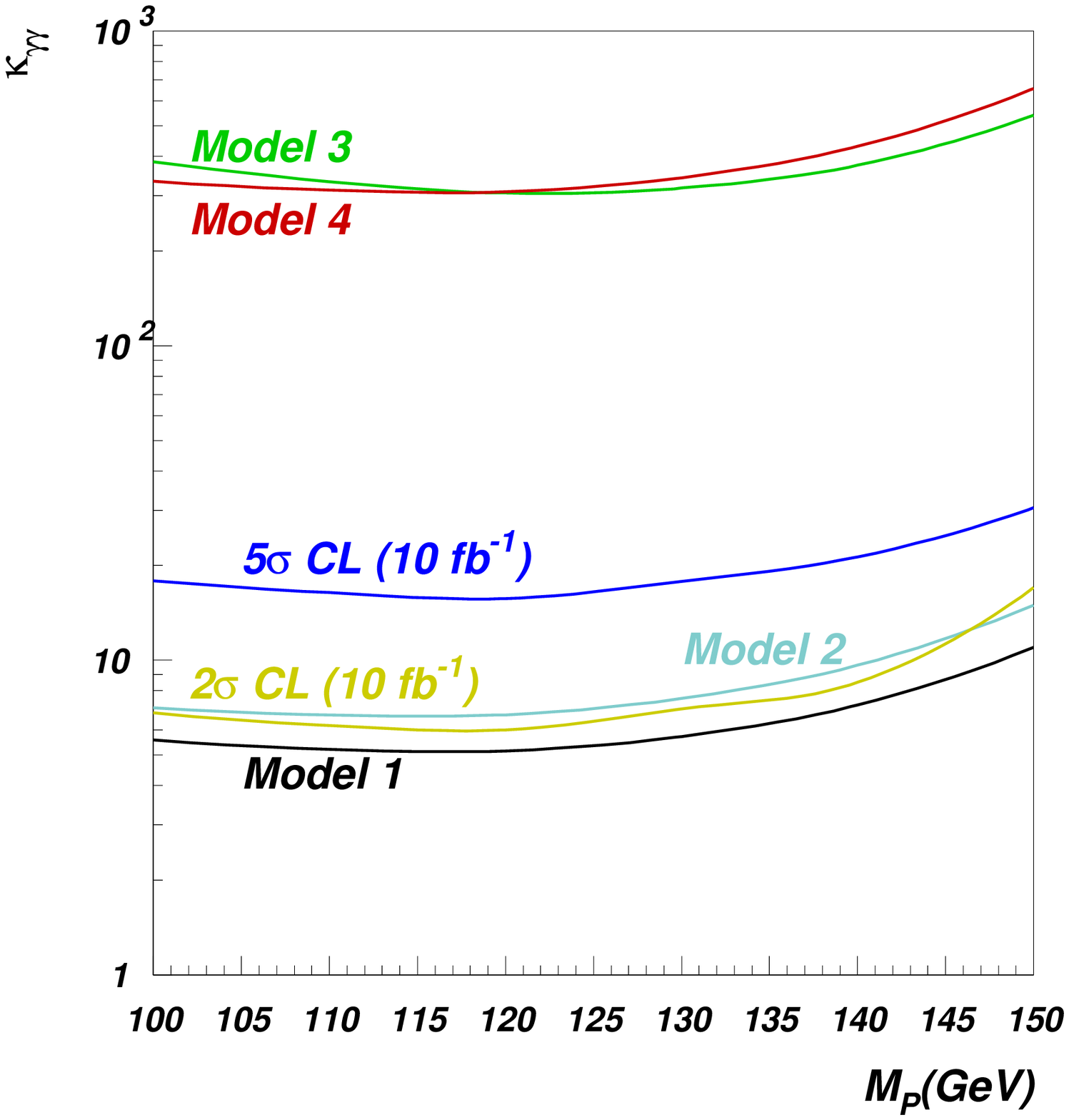}
\caption[sm-h-br]{
Total enhancement factor for each technicolor model  as a function of
technipion mass and assuming the final state is a tau pair (left frame) or photon pair (right frame).  
The $5\sigma$ discovery and $2\sigma$ exclusion curves indicate the required enhancement
factor for a Higgs-like particle at Tevatron Run II when the final state is $\tau^+ \tau^-$
 \protect{\cite{Belyaev:2002zz}} (left frame) or $\gamma \gamma$
 \protect{\cite{Mrenna:2000qh}}(right frame).
\label{fig:techni-limits}}
\end{figure}

\begin{figure}
\includegraphics[width=7cm]{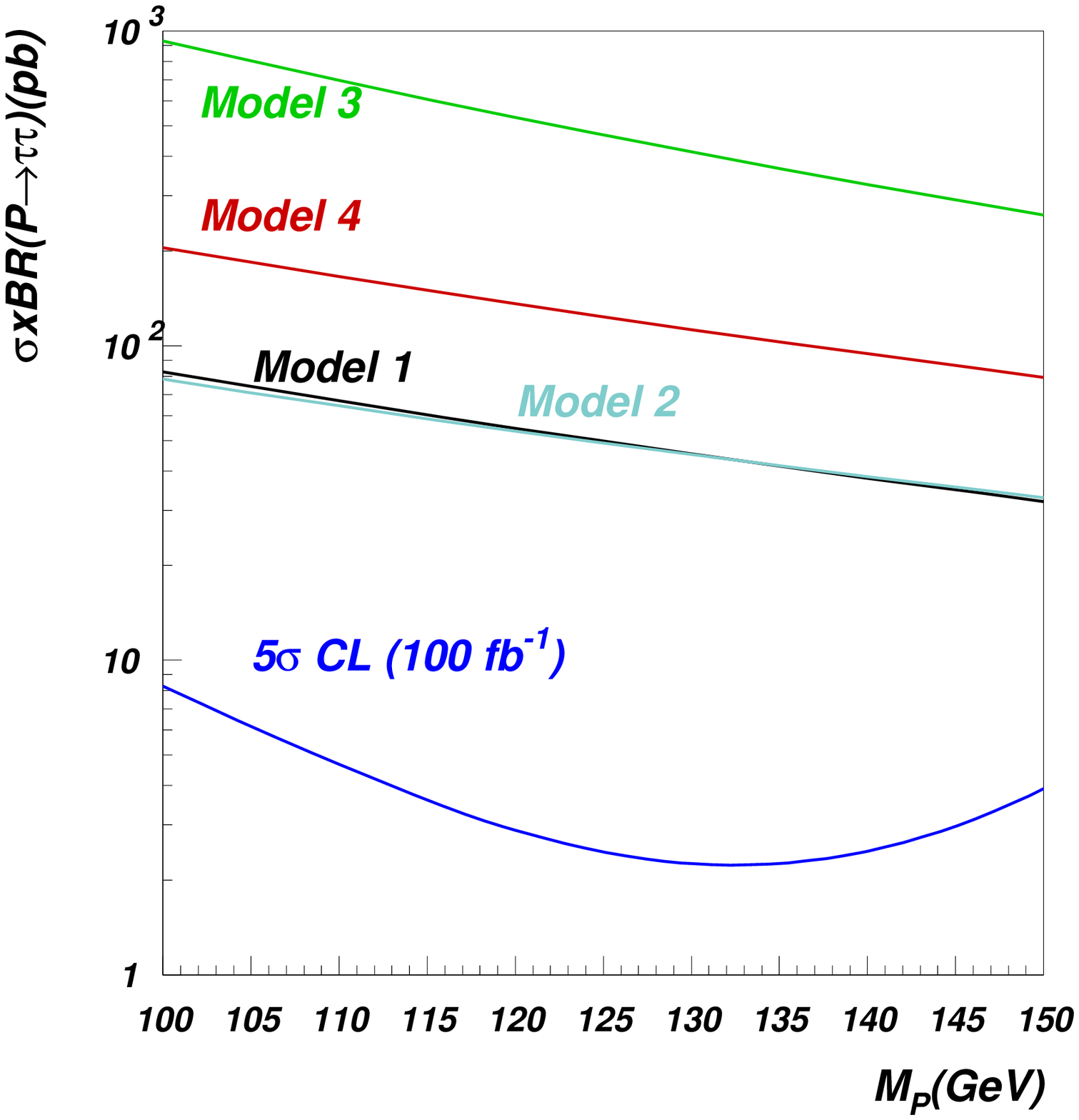}
\includegraphics[width=7cm]{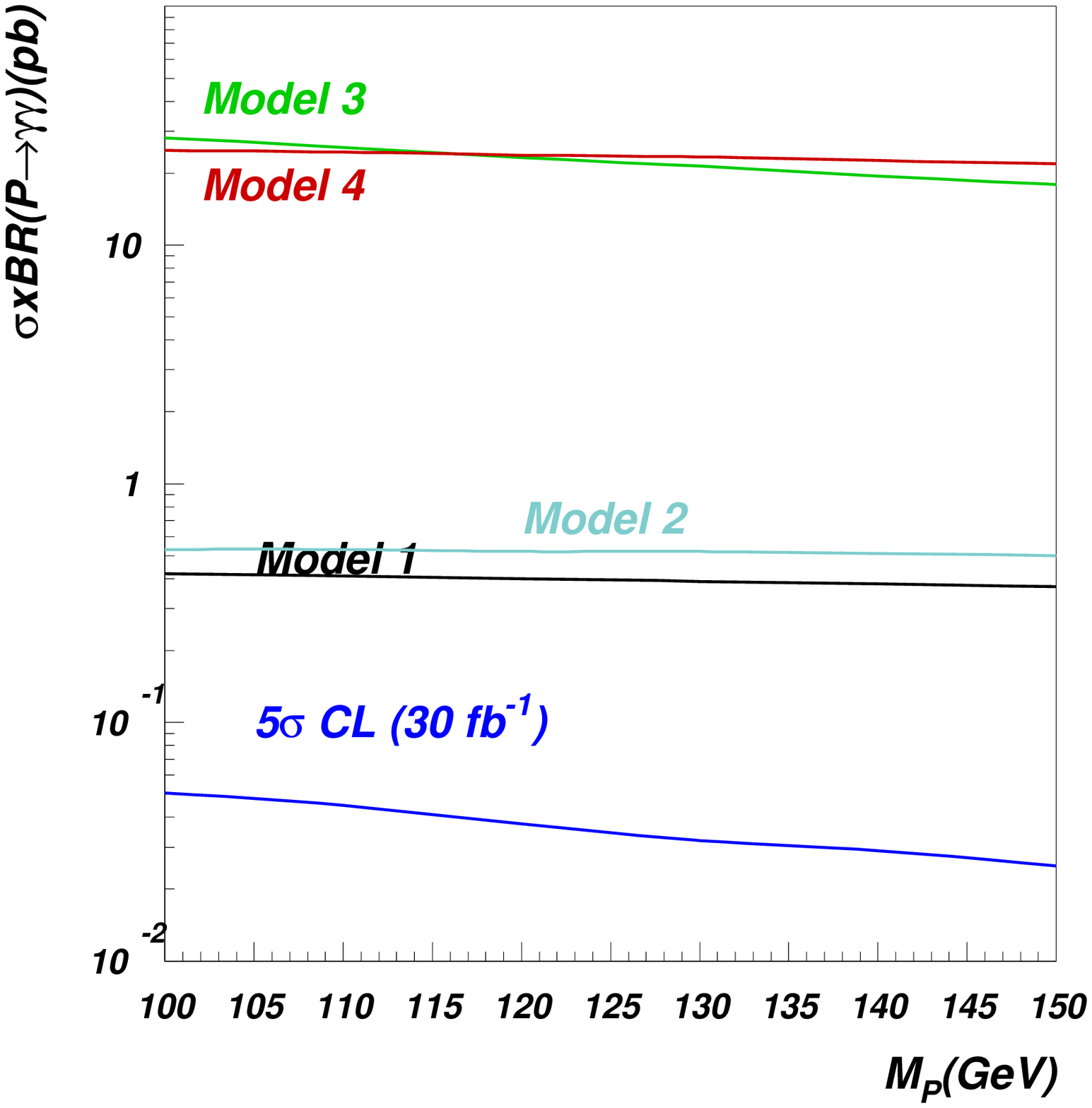}
\caption[mssm-plots]{
$\sigma\times Br$ for each technicolor model plotted as a function of technipion mass and assuming
the final state is a tau($\gamma$) pair -- left(right) at LHC.
The lowest curve is the $\sigma\times Br$ required to make a
Higgs-like particle visible (\protect{$5\sigma$} discovery) in $\tau^+\tau^-$ \protect{\cite{Cavalli:2002vs}}
or in $\gamma \gamma$  \protect{\cite{Kinnunen:2005aq}}  at LHC.
\label{fig:techni-limits-lhc}}
\end{figure}

In Section 3.2 we found a distinct enhancement of the $P$
signal in both the $\tau^+\tau^-$ and $\gamma\gamma$ search channels for each of the technicolor models studied.  As illustrated in the left frame of Figure~\ref{fig:techni-limits}, the available enhancement is
well above what is required to render the $P$ of any of these models visible in the
$\tau^+\tau^-$ channel at the Tevatron.  
Likewise, the right frame of that figure shows that in the $\gamma\gamma$ channel at the Tevatron
the technipions of models 3 and 4 will be observable at the $5\sigma$ level while model 2 is subject
to exclusion at the $2\sigma$ level.  The situation at the LHC is even more promising:
Figure~\ref{fig:techni-limits-lhc} shows that  all four models could be observable at the $5\sigma$
level in both the $\tau^+\tau^-$ (left frame) and $\gamma\gamma$ (right frame)  channels.


\subsection{Distinguishing the MSSM from Technicolor}

In the previous section we have shown that
that the Tevatron and LHC have the potential 
to observe the light (pseudo) scalar states characteristic of both 
supersymmetry and models  of dynamical symmetry breaking.
For both classes of models, the $\tau^+ \tau^-$ channel
is enhanced and could be used for discovery of the light Higgs-like states.

Once a supposed light ``Higgs boson'' is observed in a collider experiment, 
 an immediate  important task will be  to identify the new state 
 more precisely, i.e. to discern 
``the meaning of Higgs'' in this context.  Comparison of the enhancement
factors for different channels will aid in this task.  Our study has shown that comparison
of the  $\tau^+\tau^-$ and $\gamma\gamma$ channels can be particularly informative in
distinguishing supersymmetric from dynamical models.  In the case of supersymmetry, when the
  $\tau^+\tau^-$ channel is enhanced, the  $\gamma\gamma$ channel is suppressed,
   and this suppression is strong enough that even the LHC would not observe the 
   $\gamma\gamma$ signature.  In contrast, for  the dynamical symmetry breaking models studied
   we expect {\it simultaneous} enhancement of both
   the $\tau^+\tau^-$  and  $\gamma\gamma$ channels.
   The enhancement of  the $\gamma\gamma$ channel
   is so significant, that even at the Tevatron
   we may observe technipions via  this signature at the  $5\sigma$
   level for Models 3 and 4, while 
   Model 2 could be excluded at 95\% CL at the Tevatron.
   The LHC collider, which will have better sensitivity
   to the signatures under study,
   will be able to observe all four models of  dynamical symmetry breaking
   studied here  in the $\gamma\gamma$ channel, and can therefore distinguish 
   more conclusively between the supersymmetric and dynamical models.

We also would like to stress
  an important difference between 
   two class of models in their production mechanisms.
   In supersymmetry the $b\bar{b}$ fusion process is likely to be
   as important as the $gg$ fusion mechanism (see Figure 6) in
   contributing to the total
   production cross section. In technicolor models, however, the
   $b\bar{b}$ fusion contribution to technipion production is likely to be negligible.
   This difference could be revealed, in principle,
   by looking at other (exclusive or semi-exclusive) processes:
   in case of supersymmetry, for example,  one would expect 
   significant enhancement
   of Higgs boson production associated with $b$-quarks.

\section{Conclusions}

In this paper we have shown that searches for a light Standard Model Higgs boson
at Tevatron Run II and CERN LHC have the power to provide significant information
about important classes of physics beyond the Standard Model.  
We demonstrated that the new scalar and pseudo-scalar states
predicted in both supersymmetric and dynamical models can have enhanced visibility in
standard $\tau^+\tau^-$ and $\gamma\gamma$ search channels, 
making them potentially discoverable at both the Tevatron Run II and the CERN LHC.
The enhancement arises largely from increases in the production rate; 
we showed that the model parameters exerting the largest influence on the enhancement size are $\tan\beta$ in the case of the MSSM and $N_{TC}$ and $F_P$ in the case of dynamical symmetry breaking.  
At the same time, the  ${\cal H}\to W^+W^-$  decay pathway is suppressed  
     in the models studied here by at least an order of magnitude, compared to Standard Model
     expectations.   
In comparing the key signals for the non-standard scalars across models, we were able to show how one could start to identify which state has actually been found by a standard Higgs search.  
In particular, we investigated the likely mass reach of the Higgs search in $pp/p\bar{p} \to {\cal
H} \to \tau^+\tau^-$ for each kind of non-standard scalar state, and we demonstrated that 
$pp\, \ \, p\bar{p} \to {\cal H} \to \gamma\gamma$ may
cleanly distinguish the scalars of supersymmetric models from those of dynamical models.

\section{Acknowledgments}

We thank M.~Spira and R.~Harlander  for discussions, C.-P. Yuan for providing the code from Ref. \cite{Balazs:1998sb}, and a very thorough referee for a close reading of the manuscript.
This work was supported in part by the U.S. National Science Foundation under awards
PHY-0354838 (A.~Belyaev) and PHY-0354226 (R.~S.~Chivukula and E.~H.~Simmons).
A.~Blum is supported in part by a scholarship from the German National Academic Foundation
(Studienstiftung des deutschen Volkes).

\newpage



\begin{thebibliography}{999}

\bibitem{Rupak:1995kg}
  G.~Rupak and E.~H.~Simmons,
  Phys.\ Lett.\ B {\bf 362}, 155 (1995)
  [arXiv:hep-ph/9507438].
  
\bibitem{Lubicz:1995xi}
  V.~Lubicz and P.~Santorelli,
  Nucl.\ Phys.\ B {\bf 460}, 3 (1996)
  [arXiv:hep-ph/9505336].
  
\bibitem{Lynch:2000hi}
  K.~R.~Lynch and E.~H.~Simmons,
  Phys.\ Rev.\ D {\bf 64}, 035008 (2001)
  [arXiv:hep-ph/0012256].

\bibitem{higgs-lep1}
 R.~Barate {\it et al.}  [ALEPH Collaboration],
  Phys.\ Lett.\ B {\bf 565}, 61 (2003)
  [arXiv:hep-ex/0306033].
  
\bibitem{higgs-lep2}
   P.~Bechtle  [LEP Collaboration],
  Eur.\ Phys.\ J.\ C {\bf 33} (2004) S723
  [arXiv:hep-ex/0401007].
  
\bibitem{higgs-lep3}
  G.~L.~Kane, T.~T.~Wang, B.~D.~Nelson and L.~T.~Wang,
  Phys.\ Rev.\ D {\bf 71}, 035006 (2005)
  [arXiv:hep-ph/0407001].
  
\bibitem{higgs-lep4}
  G.~Abbiendi {\it et al.}  [OPAL Collaboration],
  Eur.\ Phys.\ J.\ C {\bf 40}, 317 (2005)
  [arXiv:hep-ex/0408097].
  
\bibitem{higgs-lep5}
  G.~Sguazzoni,
  Acta Phys.\ Slov.\  {\bf 55}, 93 (2005)
  [arXiv:hep-ph/0411096].


\bibitem{Belyaev:2002zz}
  A.~Belyaev, T.~Han and R.~Rosenfeld,
  JHEP {\bf 0307}, 021 (2003)
  [arXiv:hep-ph/0204210].

\bibitem{Cavalli:2002vs}
D.~Cavalli {\it et al.},
arXiv:hep-ph/0203056.

\bibitem{Mrenna:2000qh}
  S.~Mrenna and J.~D.~Wells,
  Phys.\ Rev.\ D {\bf 63} (2001) 015006
  [arXiv:hep-ph/0001226].



\bibitem{Kinnunen:2005aq}
R.~Kinnunen, S.~Lehti, A.~Nikitenko and P.~Salmi,
J.\ Phys.\ G {\bf 31}, 71 (2005)
[arXiv:hep-ph/0503067].

\bibitem{Gunion:1989we}
  J.~F.~Gunion, H.~E.~Haber, G.~L.~Kane and S.~Dawson,
SCIPP-89/13


\bibitem{Gunion:1992hs}
  J.~F.~Gunion, H.~E.~Haber, G.~L.~Kane and S.~Dawson,
  arXiv:hep-ph/9302272.

\bibitem{Djouadi:1997yw}
A.~Djouadi, J.~Kalinowski and M.~Spira,
Comput.\ Phys.\ Commun.\  {\bf 108}, 56 (1998)
[arXiv:hep-ph/9704448].


\bibitem{'tHooft:1980xb}
  G..~'t Hooft in G.'t Hooft, {\it et. al.},   ``Recent Developments In Gauge Theories. 
  Proceedings, Nato Advanced Study Institute, Cargese, France, August 26 - September 8, 1979,''
 New York, USA Plenum (1980). 

\bibitem{Witten:1981nf}
  E.~Witten,
  Nucl.\ Phys.\ B {\bf 188}, 513 (1981).

\bibitem{Dimopoulos:1981zb}
  S.~Dimopoulos and H.~Georgi,
  Nucl.\ Phys.\ B {\bf 193}, 150 (1981).

\bibitem{Wilson:1971bg}
  K.~G.~Wilson,
  Phys.\ Rev.\ B {\bf 4}, 3174 (1971).

\bibitem{Wilson:1973jj}
  K.~G.~Wilson and J.~B.~Kogut,
  Phys.\ Rept.\  {\bf 12}, 75 (1974).
  
\bibitem{Dawson:1996cq}
  S.~Dawson,
  arXiv:hep-ph/9612229.
 
 

\bibitem{Murayama:2000fm}
  H.~Murayama,
{\it Prepared for Theoretical Advanced Study Institute in Elementary Particle Physics (TASI 2000): Flavor Physics for the Millennium, Boulder,
Colorado, 4-30 Jun 2000}

\bibitem{Boos:2002ze}
  E.~Boos, A.~Djouadi, M.~Muhlleitner and A.~Vologdin,
  Phys.\ Rev.\ D {\bf 66}, 055004 (2002)
  [arXiv:hep-ph/0205160].

\bibitem{Hill:2002ap}
  C.~T.~Hill and E.~H.~Simmons,
  Phys.\ Rept.\  {\bf 381}, 235 (2003)
  [Erratum-ibid.\  {\bf 390}, 553 (2004)]
  [arXiv:hep-ph/0203079].

\bibitem{Anderson:1996ew}
  G.~W.~Anderson, D.~J.~Castano and A.~Riotto,
  Phys.\ Rev.\ D {\bf 55}, 2950 (1997)
  [arXiv:hep-ph/9609463].

\bibitem{Murayama:1996ec}
  H.~Murayama and M.~E.~Peskin,
  Ann.\ Rev.\ Nucl.\ Part.\ Sci.\  {\bf 46}, 533 (1996)
  [arXiv:hep-ex/9606003].

\bibitem{Susskind:1978ms}
  L.~Susskind,
  Phys.\ Rev.\ D {\bf 20}, 2619 (1979).
  
\bibitem{Weinberg:1975gm}
  S.~Weinberg,
  Phys.\ Rev.\ D {\bf 13}, 974 (1976).

\bibitem{Weinberg:1979bn}
  S.~Weinberg,
  Phys.\ Rev.\ D {\bf 19}, 1277 (1979).

\bibitem{Dimopoulos:1979es}
  S.~Dimopoulos and L.~Susskind,
  Nucl.\ Phys.\ B {\bf 155}, 237 (1979).

\bibitem{Eichten:1979ah}
  E.~Eichten and K.~D.~Lane,
  Phys.\ Lett.\ B {\bf 90}, 125 (1980).

\bibitem{Farhi:1980xs}
  E.~Farhi and L.~Susskind,
  Phys.\ Rept.\  {\bf 74}, 277 (1981).

\bibitem{Casalbuoni:1998fs}
  R.~Casalbuoni, A.~Deandrea, S.~De Curtis, D.~Dominici, R.~Gatto and J.~F.~Gunion,
  Nucl.\ Phys.\ B {\bf 555}, 3 (1999)
  [arXiv:hep-ph/9809523].

\bibitem{Lane:1991qh}
  K.~D.~Lane and M.~V.~Ramana,
  Phys.\ Rev.\ D {\bf 44}, 2678 (1991).

\bibitem{Lane:1999uh}
  K.~D.~Lane,
  Phys.\ Rev.\ D {\bf 60}, 075007 (1999)
  [arXiv:hep-ph/9903369].



\bibitem{Carena:1998gk}
M.~Carena, S.~Mrenna and C.~E.~M.~Wagner,
Phys.\ Rev.\ D {\bf 60}, 075010 (1999)
[arXiv:hep-ph/9808312].

\bibitem{Carena:1999bh}
M.~Carena, S.~Mrenna and C.~E.~M.~Wagner,
Phys.\ Rev.\ D {\bf 62}, 055008 (2000)
[arXiv:hep-ph/9907422].

\bibitem{Dawson:1996xz}
  S.~Dawson, A.~Djouadi and M.~Spira,
  Phys.\ Rev.\ Lett.\  {\bf 77}, 16 (1996)
  [arXiv:hep-ph/9603423].

\bibitem{Harlander:2003bb}
  R.~V.~Harlander and M.~Steinhauser,
  next-to-leading
  Phys.\ Lett.\ B {\bf 574}, 258 (2003)
  [arXiv:hep-ph/0307346].

\bibitem{Harlander:2003kf}
  R.~Harlander and M.~Steinhauser,
  Phys.\ Rev.\ D {\bf 68}, 111701 (2003)
  [arXiv:hep-ph/0308210].


\bibitem{Harlander:2004tp}
  R.~V.~Harlander and M.~Steinhauser,
  JHEP {\bf 0409}, 066 (2004)
  [arXiv:hep-ph/0409010].



\bibitem{Balazs:1998sb}
  C.~Balazs, H.~J.~He and C.~P.~Yuan,
  Phys.\ Rev.\ D {\bf 60}, 114001 (1999)
  [arXiv:hep-ph/9812263].
  
\bibitem{Harlander:2003ai}
  R.~V.~Harlander and W.~B.~Kilgore,
  Phys.\ Rev.\ D {\bf 68}, 013001 (2003)
  [arXiv:hep-ph/0304035].

  
\bibitem{Spira:1996if}
  M.~Spira,
  Nucl.\ Instrum.\ Meth.\ A {\bf 389}, 357 (1997)
  [arXiv:hep-ph/9610350].


\bibitem{Dimopoulos:1980yf}
  S.~Dimopoulos, S.~Raby and G.~L.~Kane,
  Nucl.\ Phys.\ B {\bf 182}, 77 (1981).

\bibitem{Ellis:1980hz}
  J.~R.~Ellis, M.~K.~Gaillard, D.~V.~Nanopoulos and P.~Sikivie,
  Nucl.\ Phys.\ B {\bf 182}, 529 (1981).

\bibitem{Holdom:1981bg}
  B.~Holdom,
  Phys.\ Rev.\ D {\bf 24}, 157 (1981).

\bibitem{Chivukula:1995dt}
  R.~S.~Chivukula, R.~Rosenfeld, E.~H.~Simmons and J.~Terning,
  arXiv:hep-ph/9503202.

\bibitem{Carena:2000yx}
  M.~Carena {\it et al.}  [Higgs Working Group Collaboration],
  arXiv:hep-ph/0010338.



\bibitem{djouadi-higgs}
  A.~Djouadi,
  arXiv:hep-ph/0412238.


\bibitem{Spira:1998wh}
M.~Spira,
arXiv:hep-ph/9810289.

\end{thebibliography}
\end{document}